\documentclass[12pt]{article}
\usepackage{amsthm,amsmath,amssymb}
\usepackage{psfrag,graphicx}
\usepackage{bm}

\newcommand{\dio}{R_*}

\newcommand{\Rzt}{R_{3,\theta}}
\newcommand{\Fz}{S_3}

\newcommand{\zunit}{\hat{e}_3}
\newcommand{\speed}{\tilde{\vel}}

\newcommand{\pv}{\tau}
\newcommand{\pf}{\Nnum_{,\vel}}
\newcommand{\pfD}{\Nnum_{,\vel}^{(3)}}
\newcommand{\pfdD}{\Nnum_{,\vel_3}^{(3)}}
\newcommand{\pfj}{\Nnum_{,\vel_j}}
\newcommand{\pfk}{\Nnum_{,\vel_k}}
\newcommand{\pn}{\Nnum_{,\freq}}
\newcommand{\pvD}{\tau^{(3)}}
\newcommand{\ipv}{g}
\newcommand{\iN}{\gamma}
\newcommand{\ipf}{q}
\newcommand{\NnD}{\pn^{(3)}}
\newcommand{\Nnum}{n}

\newcommand{\vu}{\TheNewVec{u}}
\newcommand{\vv}{\TheNewVec{w}}
\newcommand{\Dsol}{\beta}
\newcommand{\vDsol}{\TheNewVec{\Dsol}} 

\newcommand{\phase}{\vartheta}
\newcommand{\indR}[1]{#1^{(1)}}
\newcommand{\indI}[1]{#1^{(2)}}

\newcommand{\TheNewVec}[1]{\bm{#1}}
\newcommand{\wwww}{\xi}
\newcommand{\vw}{\TheNewVec{\xi}}
\newcommand{\vzeta}{\TheNewVec{\zeta}}
\newcommand{\rw}{\indR{\wwww}}
\newcommand{\iw}{\indI{\wwww}}
\newcommand{\vpsi}{\TheNewVec{\psi}}

\newcommand{\tangent}{z}
\newcommand{\vz}{\TheNewVec{\tangent}}

\newcommand{\symp}[2]{\omega( #1,#2)}
\newcommand{\OM}{\Omega}
\newcommand{\OMpos}{\OM_{\vsolp}}
\newcommand{\OMvos}{\OM_{\vsolvn}}
\newcommand{\OMd}{\OM^{(3)}}
\newcommand{\iOM}{\OM^{(c)}}
\newcommand{\jOM}{\OM_{\vsol_{\iniPar}}}

\newcommand{\fe}{f_{\delta_0}}
\newcommand{\pos}{\mathcal{Q}_{\delta_0}}
\newcommand{\po}{\mathcal{Q}}
\newcommand{\VR}{R_V}
\newcommand{\NL}{\mathcal{M}_{\vsolvn}}
\newcommand{\Rest}{\mathcal{R}_{\freq,\vel}}
\newcommand{\hL}{\mathcal{U}_{\freq,\vel,y}}
\newcommand{\Lyp}{\mathcal{S}}
\newcommand{\Csym}{$v$-symmetric}

\newcommand{\vel}{v}
\newcommand{\maxspeed}{r_0}
\newcommand{\Par}{\zeta}
\newcommand{\iPar}{\varsigma}
\newcommand{\iniPar}{\Par^{(c)}}
\newcommand{\zPar}{\Par^{(0)}}
\newcommand{\tPar}[1]{\Par^{(#1)}}
\newcommand{\Pdom}{\set{Z}}
\newcommand{\zPdom}{\Pdom_{0}}
\newcommand{\oPdom}{\Pdom_{1}}
\newcommand{\tPdom}{\Pdom_{2}}
\newcommand{\freq}{\mu}

\newcommand{\iu}{{\rm i}}
\newcommand{\iy}{y^{(c)}}
\newcommand{\iv}{\vel^{(c)}}
\newcommand{\ifreq}{\freq^{(c)}}
\newcommand{\iphase}{\phase^{(c)}}

\newcommand{\sol}{\varphi}
\newcommand{\solw}{\sol_\freq}
\newcommand{\solv}{\sol_{\vel}}
\newcommand{\solvn}{\sol_{\vel,\freq}}
\newcommand{\soli}{\vsol_{\iPar(\psi)}}
\newcommand{\vsol}{\TheNewVec{\sol}}
\newcommand{\vsolw}{\vsol_{\freq}}
\newcommand{\vsolp}{\vsol_{\Par}}
\newcommand{\zsol}{\vsol_{\zPar}}
\newcommand{\Isol}{\vsol_{\iniPar}}
\newcommand{\vsolvn}{\vsol_{\vel,\freq}}
\newcommand{\vsolz}{\vsol_{\vel_0,\freq_0}}
\newcommand{\rsol}{\indR{\sol_{\vel,\freq}}}
\newcommand{\isol}{\indI{\sol_{\vel,\freq}}}
\newcommand{\solD}{\vsolvn^{(3)}}
\newcommand{\vsols}{\vsol_{\speed\zunit,\freq}}

\newcommand{\dotp}[2]{\big( #1,#2\big)_{2}}
\newcommand{\dotP}[2]{\big( #1,#2\big)_{\Ltwo(\mathbb{R}^3,\mathbb{R}^2),\vpsi}}

\newcommand{\dotX}[2]{\big( #1,#2\big)_{\tilde{\Gamma}}}

\newcommand{\set}[1]{\mathrm{#1}}
\newcommand{\lexp}[1]{\mathrm{e}^{#1}}
\newcommand{\RE}{\mathop{\set{Re}}}
\newcommand{\IM}{\mathop{\set{Im}}}
\newcommand{\diag}{\mathop{\mathrm{diag}}}
\newcommand{\Ker}[1]{\mathop{\set{Ker}}(#1)}
\newcommand{\ie}{{\it i.e.\/}, }
\newcommand{\eg}{{\it e.g.\/}, }
\newcommand{\diff}{\mathop{\mathrm{\mathstrut{d}}}\!}

\newcommand{\Half}{\set{H}^{\sind{\frac{1}{2}}}}
\newcommand{\nHalf}{\set{H}^{\sind{-\frac{1}{2}}}}
\newcommand{\Sob}[1]{\set{H}^{#1}}
\newcommand{\C}[1]{\set{C}^{#1}}
\newcommand{\Ltwo}{\set{L}^2}
\newcommand{\Lp}[1]{\set{L}^{#1}}
\newcommand{\Hrhalf}{\Half_{\mathrm{rad}}}
\newcommand{\Hrone}{\Sob{1}_{\mathrm{rad}}}
\newcommand{\sSob}[1]{\Sob{#1}_{\zunit}}
\newcommand{\Lrp}[1]{\Lp{#1}_{\set{rad}}}

\newcommand{\sind}[1]{{\text{{\tiny $#1$}}}}
\newcommand{\eps}{\varepsilon}
\newcommand{\lengthSol}{\ell_{\mathrm{sol}}}
\newcommand{\lengthExp}{\ell_{\mathrm{exp}}}

\newtheorem{proposition}{Proposition}[section]
\newtheorem{remark}[proposition]{Remark}
\newtheorem{theorem}{Theorem}[section]
\newtheorem{lemma}[proposition]{Lemma}
\newtheorem{corollary}[proposition]{Corollary}
\newtheorem{ass}[proposition]{Assumption}
\newtheorem{Def}[proposition]{Definition}

\newcommand{\Pn}{\mathcal{P}}

\newcommand{\LL}{L}
\newcommand{\Lv}{\LL_{\vel,\freq}}
\newcommand{\Lvz}{\LL_{\vel_0,\freq_0}}
\newcommand{\Lz}{\LL_{\freq}}
\newcommand{\Lzz}{\LL_{\freq_0}}
\newcommand{\LzO}{\LL_{11,\freq}}
\newcommand{\LzOz}{\LL_{11,\freq_0}}
\newcommand{\LzT}{\LL_{22,\freq}}
\newcommand{\LzTz}{\LL_{22,\freq_0}}

\newcommand{\En}{\mathcal{E}}
\newcommand{\Es}{\En_{\speed\zunit,\freq}}
\newcommand{\EE}{\En_{\vel,\freq}}
\newcommand{\Nn}{\mathcal{N}}
\newcommand{\Hn}{\mathcal{H}}
\newcommand{\HV}{\Hn_V}
\newcommand{\Nc}{N_{\mathrm{c}}}

\newcommand{\MfL}{\set{M}_0}
\newcommand{\MfS}{\set{M}_1}
\newcommand{\MfT}{\set{M}_2}
\newcommand{\TMLp}{\set{T}_{\vsolp}\MfL}
\newcommand{\Mf}{\set{M}}
\newcommand{\TMp}{\set{T}_{\vsolp}\Mf}
\newcommand{\vTM}{\set{T}_{\vsolvn}\Mf}
\newcommand{\vTMp}{\set{T}_{\vsolp}\Mf}
\newcommand{\TMo}{\set{T}_{\vsolp}\MfS}

\newcommand{\Span}{\mathop{\mathrm{span}}}
\newcommand{\Oh}{\mathcal{O}}
\newcommand{\nrmLp}[2]{\|#2\|_{#1}}
\newcommand{\nrmHp}[2]{\|#2\|_{\Sob{#1}}}
\newcommand{\nrmH}[1]{\|#1\|_{\Half}}
\newcommand{\Laplace}{\Delta}
\newcommand{\nrmX}[1]{\|#1\|_{\tilde{\Gamma}}}

\newcommand{\nrm}[1]{\|#1\|_2}

\newcommand{\Rnrm}[1]{|#1|}
\newcommand{\Inrm}[1]{|#1|_{\infty}}
\newcommand{\Dnrm}[1]{|#1|_{1,\infty}}
\newcommand{\zNrm}[1]{\|#1\|_{\infty,w}}
\newcommand{\dNrm}[1]{\|#1\|_{1,\infty,w}}

\newcommand{\Tm}{\big ( \sqrt{-\Laplace+m^2} - m \big ) }
\newcommand{\RR}{\mathbb{R}}
\newcommand{\dd}{\set{d}}
\newcommand{\Vpot}[1]{ \int_{\RR^3} \big ( \frac{1}{|x|} \ast |#1|^2 \big ) |#1|^2 \diff x }

\usepackage{fancyhdr}

\makeatletter 
\def\fixNumberingInArticle{
\@addtoreset{figure}{section}
\@addtoreset{equation}{section}
\renewcommand{\thefigure}{\thesection.\arabic{figure}}  
\renewcommand{\theequation}{\thesection.\arabic{equation}}  
}

\makeatother

\newcommand{\ArticleNowFoot}{
\fancypagestyle{plain}{
\fancyhf{}
\renewcommand{\headrulewidth}{0pt}
\fancyfoot[R]{{\scriptsize May 17, 2006}}
\fancyfoot[C]{\thepage}
}
\pagestyle{plain}
}

\title{Effective Dynamics for Boson Stars}
\author{J\"urg Fr\"ohlich\footnote{Addresses are given at the end.},\ \ \  B. Lars G. Jonsson,\ \ \ Enno Lenzmann}
\date{May 17, 2006}

\begin{document}
\ArticleNowFoot
\fixNumberingInArticle
\maketitle
\begin{abstract}
  We study solutions close to solitary waves of the
  pseudo-relativistic Hartree equation describing boson stars under
  the influence of an external gravitational field. In particular, we
  analyze the long-time {\it effective dynamics} of such solutions.
  In essence, we establish a (long-time) stability
  result for solutions describing boson stars that move
  under the influence of an external gravitational field.
\end{abstract}

\section{Introduction}\label{sec:intro}

In this paper, we study boson stars described as 
solutions of the pseudo-relativistic Hartree equation
which, initially, are close to a solitary wave. The {\it
  pseudo-relativistic Hartree equation} is the nonlinear evolution equation
\begin{equation}\label{eq:SRH}
  \iu \partial_t \psi = (\sqrt{-\Laplace + m^2}-m)\psi +V\psi 
  - (\frac{1}{|x|}*|\psi|^2)\psi,
\end{equation}
where $\psi=\psi(x,t)$ is a complex wave field on space-time,
$x\in\mathbb{R}^3$ is a point in space, and $t\in\mathbb{R}$ is time.
Here the symbol $*$ denotes spatial convolution. The kinetic energy
operator $\sqrt{-\Laplace + m^2}-m$ is appropriate to describe
relativistic quantum particles of mass $m>0$.  This operator is
defined by its symbol $\sqrt{k^2+m^2}-m$ in momentum space.  The
convolution kernel, $|x|^{-1}$, represents the Newtonian potential of
gravitational 2-body interactions.  We use units such that the
speed of light and Planck's constant are equal to unity. By 
rescaling $\psi$ we may set Newton's gravitational constant times
$m^{2}$ equal to unity.  

Equation~\eqref{eq:SRH} describes a system of
gravitating bosonic particles in a regime where effects of special
relativity are important, because the particles have velocities close
to the speed of light, but retardation effects and space-time
curvature can be neglected. As recently shown in \cite{Elgart+Schlein2005}, equation~\eqref{eq:SRH} emerges as the correct evolution equation for the mean-field dynamics of many-body quantum systems modelling pseudo-relativistic boson stars. The external potential, $V=V(x)$,
accounts for gravitational fields from other stars. $V$ is a smooth,
bounded, slowly varying real function; (precise assumptions on $V$ are
stated in Section~\ref{sec:THM}).

Equation~\eqref{eq:SRH} admits some important conserved quantities.
Namely, the {\it mass} of the system (proportional to the number of
particles), and its {\it energy}. These quantities are given by\footnote{Note that in \cite{Frohlich+Jonsson+Lenzmann2005} we used $\Nn(\psi) = \int |\psi|^2 \diff x$.}
\begin{equation}
  \Nn(\psi):=\frac{1}{2}\int_{\mathbb{R}^3} |\psi|^2 \diff x,  
  \label{eq:NC} 
\end{equation}
and
\begin{equation}\label{eq:HC}
 \HV(\psi):=\frac{1}{2}\int_{\RR^3} (|(-\Laplace+m^2)^{1/4}\psi|^2 -m|\psi|^2 
  +V|\psi|^2)\diff x 
 - \frac{1}{4}\int_{\RR^3} \big(\frac{1}{|x|}*|\psi|^2\big)|\psi|^2\diff x,
\end{equation}
respectively.  The momentum, $\Pn$, also plays an important role. It
is defined by
\begin{equation}\label{eq:PC}
  \Pn(\psi):=\frac{-\iu}{2}\int_{\RR^3} \bar{\psi}\nabla\psi \diff x
\end{equation}
and is conserved when the external potential $V$ is constant. 

In \cite{Lenzmann2005a} it was shown for equation \eqref{eq:SRH} that initial data $\psi_0 \in \Half$ with $\Nn(\psi) < \Nc$, where $\Nc > 2/\pi$ is a universal constant, lead to global-in-time solutions $\psi \in \C{0}\big([0,\infty);\Half(\mathbb{R}^3)\big) \cap
\C{1}\big([0,\infty);\nHalf\big)$. Furthermore, if $V\equiv 0$ holds then we have {\it solitary wave solutions} of~\eqref{eq:SRH} given by
\begin{equation}\label{eq:swave}
  \psi(x,t)=\lexp{\iu t \freq}\solvn(x-\vel t) .
\end{equation}
Here the parameter $v \in \RR^3$ corresponds to the travelling velocity, and the function $\solvn \in \Half$ is a minimizer of the functional
\begin{equation}\label{eq:Evz}
\En_{\vel,0}:=\Hn_{V\equiv 0}(\psi)-\vel\cdot\Pn(\psi),
\end{equation} 
subject to the constraint $\Nn(\psi)=N$, with $N< \Nc(v)$, where
$\Nc(\vel)< \Nc$ is a finite constant. The minimizers
$\solvn$ are referred to as {\it boosted ground states}. They solve the
Euler-Lagrange equation
\begin{equation}\label{eq:solvn}
  (\sqrt{-\Laplace + m^2}-m)\sol + \freq\sol + \iu \vel\cdot \nabla \sol
  - (\frac{1}{|x|}*|\sol|^2)\sol=0,
\end{equation}
where the frequency $\freq$ is a Lagrange multiplier for the
constraint $\Nn(\psi)=N$. As shown in our companion paper \cite{Frohlich+Jonsson+Lenzmann2005}, such boosted ground state exists for all $|\vel|< 1$, \ie the travelling velocity is below the speed of light, and the constant
$\Nc(v)$ satisfies the bounds $(1-|\vel|)\Nc\leq \Nc(\vel)\leq
\Nc(0)\equiv \Nc$. In addition, we remark that the ground states, $\solvn$, decay exponentially, with rate~$\delta=\delta(\freq,\vel)$. Further properties of $\solvn$ are
established in Section~\ref{sec:sol}; see Proposition~\ref{prop:sol}.

The main goal of the present paper is to provide a detailed description of
solutions of~\eqref{eq:SRH}, initially close to a manifold of boosted
ground states, over a long interval of time. In this study, two
length scales will play a crucial role. The first one is determined by
the external potential and is given by
\begin{equation}
\lengthExp:=\nrmLp{\infty}{\nabla V}^{-1}.
\end{equation}
The second length scale is inferred from the requirement that the
initial condition, $\psi_0$, of~\eqref{eq:SRH} be close to a ground
state, $\solvn$. We will project, see Section~\ref{sec:skew}, such an
initial condition to a point on a manifold of boosted ground states.
All ground states are exponentially localized with rate
$\delta=\delta(\freq,\vel)$. The projection singles out one ground
state, with an associated length scale given by its exponential decay
rate. Thus, the second length scale is defined by
\begin{equation}
\lengthSol := \delta^{-1}.
\end{equation}
In the regime where
\begin{equation}\label{eq:epsdef}
\eps:=\frac{\lengthSol}{\lengthExp} \ll 1,
\end{equation}
we expect that solutions of~\eqref{eq:SRH} with initial condition
close to a ground state $\solvn$ behave like relativistic point particles.

We now sketch our {\it Main Result}.  Let $(y^{(0)},v^{(0)}, \phase^{(0)},
\mu^{(0)})$ be a point in $\RR^3\times \RR^3\times [0,2\pi)\times
\RR_+$, with $|v|^{(0)}\leq r<1$ for some small $r>0$ and
$\freq^{(0)}\in I\subset \RR_+$, where $I$ is some open interval.  We
consider an initial condition, $\psi_0 \in \tilde{\Gamma}$, such that
\begin{equation}
\nrmX{\psi_0 - \lexp{\iu \phase^{(0)}}\sol_{\vel^{(0)},\freq^{(0)}}(\cdot-y^{(0)})}\leq \eps,
\end{equation}
where $\tilde{\Gamma} \subset \Half$ is a weighted Sobolev space with norm $\nrmX{\cdot}$. We then show that the
solution of~\eqref{eq:SRH} with initial condition $\psi_0$ remains
close to a ground state, for times of order $\eps^{-1}$. More
explicitly, we prove that
\begin{equation}
\psi(x,t) = \lexp{\iu \phase(t)}\big(\sol_{\vel(t),\freq(t)}(x-y(t)) + \xi(x-y(t),t)\big),
\end{equation}
with $\nrmX{\xi}\leq C\eps$, for times $0\leq t\leq C\eps^{-1}$. Here
the time-dependent functions $(y,\vel,\phase,\freq)$ satisfy the
{\it Equations of Motion},
\begin{equation}\label{d1}
\dot y = \vel + \Oh(\eps^2), \ \ \gamma(\vel,\freq)\dot \vel = -\nabla V(y) + \Oh(\eps^2),
\end{equation}
where the factor $\gamma(\vel,\freq)$ is a relativistic correction, and 
\begin{equation}\label{d2}
\frac{\dd}{\dd t} \Nn(\vsolvn) = 
\Oh(\eps^2), \ \ \dot\phase = V(y)-\freq + \Oh(\eps^2).
\end{equation}
These results yield a fairly detailed description of the solution,
$\psi(x,t)$, up to times of order $\eps^{-1}$. For a precise statement of
our main result, see Theorem~\ref{thm:main} and its hypotheses in Section~\ref{sec:THM} below.

We remark that similar results for the Nonlinear Schr\"odinger
Equation (NLS) can be found
in~\cite{Frohlich+Tsai+Yau2002,FGJS-I,FGJS-II}, and, for the
Korteweg-de~Vries equation,
in~\cite{Dejak+Sigal2006,Dejak+Jonsson2005}.

Next, we review some previous results for systems of gravitating
relativistic bosons.  One of the first studies of self-gravitating
scalar bosons can be found in~\cite{Ruffini+Bonazzola1969}. Important
properties of a Hamiltonian describing a relativistic particle in an
external potential proportional to $|x|^{-1}$ are proven
in~\cite{Herbst1977}; see also \cite{Weder1974}. Bosonic matter is analyzed in~\cite{Messer1981};
and bosonic black holes are discussed in~\cite{Thirring1983}. In these
papers, it is argued that the Chandrasekhar limit for boson stars
(with $m\approx 1-50$ GeV) is approximately the mass of a mountain.
Moreover, the intuitive scaling ideas used
in~\cite{Messer1981,Thirring1983} are turned into rigorous statements
in~\cite{Lieb+Thirring1984}. There, it is conjectured that the
ground-state energy of $n$ bosonic particles with relativistic kinetic
energy is given by the minimum of the pseudo-relativistic Hartree
energy functional, $\Hn_{V\equiv 0}$, in the `mean-field limit'. This
has subsequently been shown in~\cite{Lieb+Yau1987}, where it is also
proven that there exist minimizers, $\solw$, for $\Hn_{V\equiv
  0}(\psi)$ subject to the constraint $\Nn(\psi)=N<\Nc$, and that
these minimizers can be chosen to be spherically symmetric.  The
constant $\Nc$ satisfies the bounds $2/\pi<\Nc<1.4$, and is
interpreted as the critical mass for stability of a boson star;
(bosonic Chandrasekhar limit mass).

A recent review paper on the mean-field limit of quantum Bose gases is
paper~\cite{Frohlich+Lenzmann2004}, which contains rigorous results
and a survey of open problems for Bose gases. It is shown
in~\cite{Lenzmann2005a} that the initial value-problem for
equation~\eqref{eq:SRH}, is locally well-posed, satisfies a blow-up
alternative and has global solutions for initial conditions $\psi_0\in
\Half(\RR^3)$ with $\Nn(\psi_0)<\Nc$.  The mean-field limit for
systems of gravitating relativistic bosons is discussed
in~\cite{Elgart+Schlein2005}.  The existence of blow-up solutions
of~\eqref{eq:SRH} with spherically symmetric initial conditions is
shown in~\cite{Frohlich+Lenzmann2005}, using a virial-type argument. This blow-up result is indicative of ``gravitational collapse'' of Boson stars with
mass beyond the (boson) Chandrasekhar limit.  In a companion paper,
\cite{Frohlich+Jonsson+Lenzmann2005}, we show existence of boosted ground
states $\solvn$, see~\eqref{eq:swave}, \ie of minimizers of the
functional $\En_{\vel,0}(\psi)$, subject to the constraint
$\Nn(\psi)=N<\Nc(v)$. Here $\En_{\vel,0}$ is the functional defined
in~\eqref{eq:Evz}, above.  We also prove exponential decay of these
ground states and orbital stability of solutions of~\eqref{eq:SRH}
with vanishing external potential.  In paper
\cite{Frohlich+Jonsson+Lenzmann2006b}, we present numerical evidence
for the unproven (kernel) assumption used in this paper.

Equations~\eqref{d1}, \eqref{d2} can be seen as modulation equations.
For previous work on modulation equations, see
\cite{Kaup1976,McLaughlin+Scott1978,Kodama+Ablowitz1981,Weinstein1985,Stuart1992,BP92,Frohlich+Tsai+Yau2002,Stuart2001,Buslaev+Sulem2002,FGJS-I,
  Stuart2004b,Merle2005,Dejak+Sigal2006,FGJS-II}.

The organization of our paper is as follows.  In
Section~\ref{sec:Ham}, we rephrase equation~\eqref{eq:SRH} as a
Hamiltonian evolution equation and discuss its Hamiltonian structure.
We also state a fundamental assumption.  In Section~\ref{sec:THM}, we
state our main theorem and sketch its proof.
Sections~\ref{sec:sol}--\ref{sec:up} contain numerous auxiliary
results used in the proof of our main theorem.  The main theorem is
proven in Sect.~\ref{sec:main}.  The appendices contain proofs of
spectral properties and positivity of a certain linear operator, as
well as the proof of Corollary~\ref{cor:uniform}.

\paragraph{Notation.} The space of measurable functions, $f$, with
$|f|^p$ integrable, is denoted by $\Lp{p}$, and its norm by
$\nrmLp{p}{\cdot}$. For $p=2$, this space is the Hilbert space of 
square-integrable functions. The space of $n$ times continuously
differentiable functions is denoted by $\C{n}$. The usual inhomogeneous
Sobolev space is denoted by $\Sob{s}$, its norm by $\nrmHp{s}{\cdot}$.
In particular,
\begin{equation}
  \nrmH{u} := \nrm{(1-\Laplace)^{1/4}u},
\end{equation}
for $u\in\Half$.
We define the weighted norm $\nrmX{\cdot}$ by
\begin{equation}
\nrmX{u}^2 := \nrmH{u}^2 + \eps \nrm{|x|^{1/2}u}^2,
\end{equation}
for $\eps>0$ given as in~\eqref{eq:epsdef}. We also use the notation $d_t:=\frac{\dd}{\dd t}$.

\section{The Hamiltonian Nature of Equation~\eqref{eq:SRH}}
\label{sec:Ham}

Equation~\eqref{eq:SRH} is a Hamiltonian evolution equation on an
infinite-dimensional phase space, $\Gamma$. In this paper we make
extensive use of this fact and of the symplectic structure of the phase
space.  We therefore begin with a brief review of some basic notions in
Hamiltonian dynamics.

The phase space, $\Gamma$, where~\eqref{eq:SRH} is well defined, for
bounded $V$, is the complex Sobolev space
$\Half(\mathbb{R}^3,\mathbb{C})$. A point in the phase space is
identified with a complex-valued function
$\psi\in\Half(\RR^3,\mathbb{C})$. The decomposition of $\psi$ into
real and imaginary parts,
\begin{equation}
\psi=\indR{\psi}+\iu\indI{\psi}
\end{equation}
where $\indR{\psi}$ and $\indI{\psi}$ are real-valued functions in 
$\Half(\RR^3,\RR)$, corresponds to the identification 
\begin{equation}
\Gamma\cong \Half(\RR^3,\RR^2).
\end{equation}
Note that $\indR{\psi}$ and $\indI{\psi}$ are canonically conjugate
variables.  In this paper, we use $\Half(\RR^3,\RR^2)$ as the phase
space, and, to distinguish the wave fields in this representation from
$\psi\in \Half(\RR^3,\mathbb{C})$, we use the boldface notation
\begin{equation}
(\indR{\psi},\indI{\psi})=\vpsi\in\Half(\RR^3,\RR^2).
\end{equation}

The tangent space, $\set{T}_{\vpsi}\Gamma$, to $\Gamma$ at a point
$\vpsi\in\Gamma$ is given by
\begin{equation}
\{\vz(x):\vz(\cdot)\in\Half(\RR^3,\RR^2), \vpsi+\vz\in\Gamma\}.
\end{equation}
Hence 
\begin{equation}
  \set{T}_{\vpsi}\Gamma = \Half(\mathbb{R}^3,\mathbb{R}^2).
\end{equation}

A section of the tangent bundle $\set{T}\Gamma$ is a {\em vector
  field}, \ie an assignment of a vector $\vz_{\vpsi}\in
\set{T}_{\vpsi}\Gamma$ to each point $\vpsi\in\Gamma$ that depends
continuously on $\vpsi$.

The phase space carries a natural metric,
$\dotP{\cdot}{\cdot}$: For $\vu=(\indR{u}_{\vpsi},\indI{u}_{\vpsi})$, 
$\vv=(\indR{w}_{\vpsi},\indI{w}_{\vpsi})$ in $\set{T}_{\vpsi}\Gamma$,
$\vpsi\in \Gamma$, we set
\begin{equation}\label{eq:inp}
\dotp{\vu}{\vv}\equiv \dotP{\vu}{\vv}:=\int_{\RR^3} 
(\indR{u}_{\vpsi} \indR{w}_{\vpsi} + \indI{u}_{\vpsi} \indI{w}_{\vpsi}) \diff x.
\end{equation}
Furthermore, $\Gamma$ carries a complex structure
denoted by $J$: For $\vz\in\set{T}_{\vpsi}\Gamma$, we set
\begin{equation}
J\vz = (\indI{\tangent},-\indR{\tangent}).
\end{equation}

Of course, $\Gamma$ is {\em symplectic} with symplectic 2-form
given by
\begin{equation}\label{eq:symp}
\symp{\vu}{\vv}:=\int_{\RR^3} (\indI{u} \indR{w} 
- \indR{u} \indI{w}) \diff x,
\end{equation}
for $\vu$, $\vv$ in $\set{T}_{\vpsi}\Gamma$.
We observe that 
\begin{equation}\label{eq:rel}
\symp{\vu}{\vv}=-
\dotp{\vu}{J\vv}.
\end{equation}

In what follows we also consider a sub-space, $\tilde{\Gamma}\subset
\Gamma$, given by
\begin{equation}
\tilde{\Gamma}:=\{\vpsi\in \Gamma :|x|^{1/2}\vpsi\in \Ltwo\}
\end{equation}
and equipped with the norm
\begin{equation}\label{eq:nrmX}
  \nrmX{\vu}^2:=\nrmH{\vu}^2 + \eps \nrm{|x|^{1/2}\vu}^2,
\end{equation}
where $\eps$ is defined in~\eqref{eq:epsdef}.  By the proof of Lemma~3
in~\cite{Frohlich+Lenzmann2005}, we find that if $\vpsi_0\in
\tilde{\Gamma}$, then $\vpsi(\cdot,t)\in \tilde{\Gamma}$. See also
Proposition~\ref{prop:gamma}.

The Hartree energy functional $\HV$, the mass $\Nn$, and the momentum
functional $\Pn$ (see \eqref{eq:NC}--\eqref{eq:PC}) have the form
\begin{equation}\label{eq:Ham}
\HV(\vpsi):= \frac{1}{2}
\dotp{\vpsi}{\Tm\vpsi} 
+ \frac{1}{2}\dotp{\vpsi}{V\vpsi}
-\frac{1}{4}\dotp{\frac{1}{|x|}*|\vpsi|^2}{|\vpsi|^2},
\end{equation}
\begin{equation}\label{eq:Nn}
  \Nn(\vpsi):= \frac{1}{2}\nrm{\vpsi}^2,
\end{equation}
\begin{equation}\label{eq:Pn}
\Pn(\vpsi):=\frac{1}{2}\dotp{\vpsi}{J\nabla\vpsi}.
\end{equation}
These functionals are well defined on $\Gamma=\Half(\RR^3;\RR^2)$.

We claim that~\eqref{eq:SRH} is the Hamiltonian equation of motion
corresponding to the Hamiltonian $\HV(\vpsi)$, given
in~\eqref{eq:Ham}.  This is verified by noticing that the equation
\begin{equation}\label{eq:srh}
\partial_t \vpsi = J\HV'(\vpsi)
\end{equation}
is equivalent to~\eqref{eq:SRH}.

If $V\equiv 0$, then the Hamiltonian is invariant under spatial
translations $x\mapsto x+a$. The corresponding conserved quantity is
the momentum $\Pn(\vpsi)$ defined above. For bounded $V$, $\Pn$
satisfies an Ehrenfest identity
\begin{equation}\label{eq:Ehr}
d_t \Pn(\vpsi) = -\frac{1}{2}\dotp{\vpsi\nabla V}{\vpsi}.
\end{equation}
This identity was shown in~\cite[App. A]{FGJS-I} for the nonlinear
Schr\"odinger equation, but the proof carries over to~\eqref{eq:SRH}.

We define a functional $\EE$ as
\begin{equation}\label{eq:EE}
\EE(\vpsi) := \Hn_{V=0}(\vpsi) + \freq\Nn(\vpsi) - \vel \cdot \Pn(\vpsi),
\end{equation}
which is given, more explicitly, by
\begin{equation}
\EE(\vpsi) = \frac{1}{2}\dotp{\vpsi}{\Tm\vpsi} + 
\frac{\freq}{2}\nrm{\vpsi}^2 - \frac{1}{2}\vel\cdot\dotp{\vpsi}{J\nabla \vpsi}
-\frac{1}{4}\dotp{|\vpsi|^2}{\frac{1}{|x|}*|\vpsi|^2}.
\end{equation}
This functional plays a key role in this paper, and we briefly discuss
its properties. The ground states $\vsolvn = (\RE\solvn,\IM\solvn)$,
\ie solutions to eq.~\eqref{eq:solvn} are solutions to
\begin{equation}\label{eq:EEp}
\EE'(\vsolvn) = 0.
\end{equation}
Its Hessian, $\Lv:=\EE''(\vsolvn)$, is given by the linear symmetric
operator:
\begin{equation}
\Lv:=\begin{pmatrix}L_{11} & L_{12} \\ L_{21} & L_{22}\end{pmatrix}
\end{equation}
where, for $\xi\in\Half(\RR^3,\RR)$, 
\begin{align}\label{eq:Lv}
  L_{11,\vel,\freq}\xi&:=(\sqrt{-\Laplace + m^2}-m+\freq -
  \frac{1}{|x|}*|\vsolvn|^2) \xi -
  \big(\frac{2}{|x|}*(\xi \rsol)\big)\rsol, \\
  L_{12,\vel,\freq}\xi&:=-v\cdot \nabla \xi-(\frac{2}{|x|}*(\xi \isol))\rsol, \\
  L_{21,\vel,\freq}\xi&:=v\cdot \nabla \xi -(\frac{2}{|x|}*(\xi \rsol))\isol, \\
  L_{22,\vel,\freq}\xi&:=(\sqrt{-\Laplace + m^2} -m + \freq -
  \frac{1}{|x|}*|\vsolvn|^2)\xi -\big( \frac{2}{|x|}*(\xi \isol)\big)\isol.
\end{align}

We find $\nabla \vsolvn$ and $J\vsolvn$ to be elements of the kernel
of $\Lv$, $\Ker{\Lv}$, because $\EE$ is invariant under gauge
transformations and translations.  For $\vel=0$, $\Lv$ reduces to
$\Lz:=\diag(\LzO,\LzT)$.  A key assumption underlying our analysis is
\begin{ass}\label{ass:ker}
  Let $\Lz=\diag(\LzO,\LzT)$ be defined as above.  We assume that the
  dimension of the null space of $\LzO$ is three, \ie
\begin{equation}\label{eq:ass}
  \dim \Ker{\LzO} = 3, \ \text{for}\ \freq>0.
\end{equation}
\end{ass}
In~\cite{Frohlich+Jonsson+Lenzmann2006b} this assumption is verified
numerically for some $\freq>0$, following~\cite{Demanet+Schlag2005}.

\section{The main theorem}
\label{sec:THM}

In this section we state our assumptions and the main theorem. We then
present an outline of the proof, which is implemented in the remaining
sections of this paper.

Given a number $\eps>0$, we require that the external potential $V\in
\C{3}$ satisfies
\begin{equation}\label{eq:Vass}
\sup_{x} |\partial_x^{\alpha}V|\leq C\eps^{|\alpha|},\ \text{for}\ 
|\alpha|\leq 3,
\end{equation}
where $\alpha$ is a multi-index and $C$ is a constant.

The ground state $\vsolvn$ is {\it not} known to be unique modulo
phase transformations and translations. We therefore single out a
particular class of solutions to~\eqref{eq:EEp}, also denoted by
$\vsolvn$, near a spherically symmetric minimizer
$\vsolw:=\vsol_{\vel=0,\freq}$ by the use of an implicit function
theorem and Assumption~\ref{ass:ker}. As we will see, there is a
maximal number $r_0$ and a maximal open interval $I_0$, with $0<r_0<1$,
such that for $|\vel|<r_0$ and $\freq\in I_0$, $\vsolvn$
solves~\eqref{eq:EEp}.  The construction, as well as several
properties of these functions are given in
Proposition~\ref{prop:sol}.  For any $r<r_0$ and an open non-empty
interval $I\subset I_0$, let $\Pdom(r,I)$ be defined by
\begin{equation}\label{eq:Z}
\Pdom(r,I):=\RR^3\times \set{B}_{r}(0)
\times [0,2\pi)\times I,
\end{equation}
where $\set{B}_{r}(0) = \{ \vel\in\RR^3: |\vel|< r \}$.  Consider the
manifold
\begin{equation}\label{eq:aMf}
  \Mf(\Pdom) := \{\lexp{-\phase J}\vsolvn(\cdot-y): (y,\vel,\phase,\freq)\in 
  \Pdom \},
\end{equation}
The {\it soliton manifold} is defined by $\MfL:=\Mf(\zPdom)$ with
$\zPdom:=\Pdom(\maxspeed,I_0)$, where $r_0$ is the
maximal speed and $I_0$ the maximal frequency interval such that
$\vsolvn$ is well defined in the sense of Proposition~\ref{prop:sol}.
Thus, $\zPdom$ is the parameter space for $\MfL$. Furthermore, we
introduce a symbol, $\Par$, for a point in $\zPdom$ by
\begin{equation}\label{eq:Par}
\Par:=(y,\vel,\phase,\freq),
\end{equation}
and note that each point in $\MfL$ can be described by $\vsolp$, where 
\begin{equation}\label{eq:spar}
\vsolp(x):=\lexp{-\phase J}\vsolvn(x-y).
\end{equation}
The tangent space to $\MfL$ at $\vsolp$ is given by
\begin{equation}\label{eq:TM}
\TMLp := \Span(\nabla_y \vsolp, \nabla_{\vel} \vsolp, J\vsolp, 
\partial_\freq \vsolp).
\end{equation}

We can now state our main theorem.
\begin{theorem}\label{thm:main}
  Suppose that Assumption~\ref{ass:ker} is satisfied. Let $r_0$,
  $I_0$, and $\vsolvn$ be as above. Let the external potential $V$
  satisfy~\eqref{eq:Vass}.  Then there is an open non-empty interval
  $\tilde{I}$, with\footnote{Here $A \Subset B$ is defined by
    $\bar{A}\subset B$, $\bar{A}$ compact.} $\tilde{I} \Subset I_0$ as
  well as numbers $0<\tilde{r}<r_0$ and $\eps>0$, with $\eps$
  sufficiently small, such that the following holds. If the initial
  condition $\vpsi_0\in\tilde{\Gamma}$ for \eqref{eq:srh} satisfies
\begin{equation}\label{eq:initialZ}
  \nrmX{\vpsi_0 - \vsol_{\zPar}} \leq \eps,
\end{equation} 
for some $\zPar\in \tilde{\Pdom}$, with
$\tilde{\Pdom}:=\Pdom(\tilde{r},\tilde{I})$, and $\nrmX{\cdot}$ as defined
in~\eqref{eq:nrmX}, then the solution $\vpsi$ of \eqref{eq:srh} 
is of the form
\begin{equation}
\vpsi(x,t) = \lexp{-\phase J}(\vsolvn(x-y)+\vw(x-y,t)),
\end{equation}
where $y,\vel,\phase,\freq$ are {\em time-dependent} functions and
\begin{equation}
\nrmX{\vw} \leq C\eps.
\end{equation}
Moreover, we have that
\begin{align}\label{eq:fdd1}
  |d_t \Nn(\vsolvn)| &\leq C\eps^2, &&
  |d_t \Pn(\vsolvn) +\Nn(\vsolvn)\nabla V(y)|\leq C\eps^2, \\
  |\dot y-\vel|&\leq C\eps^2, && |\dot\phase + \freq - V(y)| \leq
  C\eps^2,\label{eq:fdd2}
\end{align}
for times $0\leq t \leq C/\eps$, where $C>0$ is some constant.
\end{theorem}

The requirement that the initial condition be `close' in norm to
$\Mf(\tilde{\Pdom})$ can, as in
\eg~\cite{FGJS-I,FGJS-II,Dejak+Sigal2006,Dejak+Jonsson2005}, be used
to introduce an additional small parameter, $\varepsilon_0$, to
separate the two scales mentioned in the introduction. But to simplify
the exposition in the present paper, we assume this distance also to
be bounded by $\eps$.

\begin{figure}[hbtp]
\psfrag{I0}{$I_1$}
\psfrag{I1}{$I_3$}
\psfrag{fl}{$\freq_{l}(|\vel|)$}
\psfrag{dP}{$\propto \delta_P$}
\psfrag{dM}{$\propto \delta_{\Mf}$}
\psfrag{r0}{$r_3$}
\psfrag{r1}{$r_1$}
\psfrag{1}{1}
\psfrag{v}{$|\vel|$}
\psfrag{f}{$\freq$}
\psfrag{n}{$\vsolp$}
\psfrag{H}{$\Half$}
\psfrag{LJ}{}
\psfrag{M0}{$\MfS$}
\psfrag{M1}{$\Mf_3$}
\psfrag{Z0}{$\oPdom$}
\psfrag{Z1}{$\Pdom_3$}
   \centering
   \centerline{\includegraphics[scale=.85]{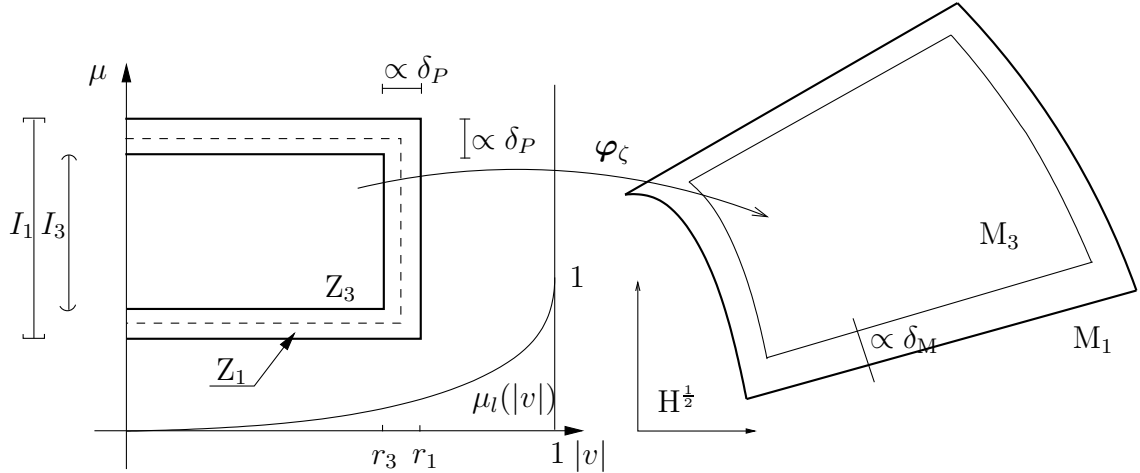}}
   \parbox{\linewidth}{
     \caption{The left figure displays a schematic view of the
       parameter spaces $\oPdom$ and $\Pdom_3$. The function
       $\freq_l(|\vel|)$ is defined in Proposition~\ref{prop:sol} and
       represents the lower bound on $\freq$.  The dotted line
       indicates $\tPdom$;  $\delta_M$ and $\delta_P$ are the distance
       between the manifolds and upper bounds on $\eps$, see
       Remark~\ref{rem:dm}.}
\label{fig:domain}}
\end{figure}

Next, we outline the essential parts of the proof of
Theorem~\ref{thm:main}.  Above, we have introduced $r_0$, $I_0$ and
$\tilde{r}$, $\tilde{I}$. In the process of the proof, we will find a
nested sequence of manifolds: There are numbers $r_j$ and open
non-empty intervals $I_j$, for $j=0,\ldots,3$, with $r_3:=\tilde{r}$
and $I_3:=\tilde{I}$.  Here $0<r_{j+1}<r_j$ and $\bar{I}_{j+1}\subset
I_{j}$ for $j=0,1,2$.  
The corresponding parameter domains are
$\Pdom_j:=\Pdom(r_j,I_j)$ corresponding to soliton submanifolds
$\Mf_j:=\Mf(\Pdom_j)$.  One of the constraints on $\eps$ is that it
has to be smaller than the distance between the boundaries of the
nested sequence of the manifolds. These distances are indicated in
Figure~\ref{fig:domain}; see also Remark~\ref{rem:dm}.

Once we derive the properties of the ground state
(Sect.~\ref{sec:sol}), we need to derive the finite-dimensional
dynamics expressed by~\eqref{eq:fdd1}--\eqref{eq:fdd2}. To do this, we
show that the symplectic form is non-degenerate on $\MfL$ by using
the symmetry properties of $\vsolvn$, and its derivatives; see
Section~\ref{sec:J}.  This non-degeneracy of the symplectic form on
$\MfL$ is the key fact to show the existence of a skew (or
symplectically) orthogonal decomposition of $\vpsi$ in a tubular
neighborhood $U_{\delta}(\Pdom_2)$ around $\Mf_2$. That is, there is a
unique map $\iPar:U_{\delta}(\tPdom)\rightarrow \oPdom$ such that
\begin{equation}\label{dec1}
\vpsi(x) = \vsol_{\iPar(\vpsi)}(x) + \lexp{-\phase(\vpsi)J}\vw(x-y(\vpsi),t)
\end{equation}
and
\begin{equation}\label{dec2}
\symp{\vpsi - \soli}{\vz} = 0, \ \ \forall \vz\in \set{T}_{\soli}\MfS.
\end{equation}
This result is proven in Section~\ref{sec:skew}.

\begin{figure}[htbp]
\psfrag{Z}{$\oPdom$}
\psfrag{z}{$\Par(t)$}
\psfrag{nz}{$\vsol_{\iPar(\vpsi_0)}$}
\psfrag{R}{$\RR^8$}
\psfrag{n}{$\vsolp$}
\psfrag{nt}{$\vsol_{\Par(t)}$}
\psfrag{zz}{$\iPar(\vpsi_0)$}
\psfrag{H}{$\Half$}
\psfrag{M}{$\MfS$}
\psfrag{P}{$\vpsi(\cdot,t)$}
\psfrag{Pz}{$\vpsi_0$}
\psfrag{LJ}{}
   \centering
   \centerline{\includegraphics[scale=.9]{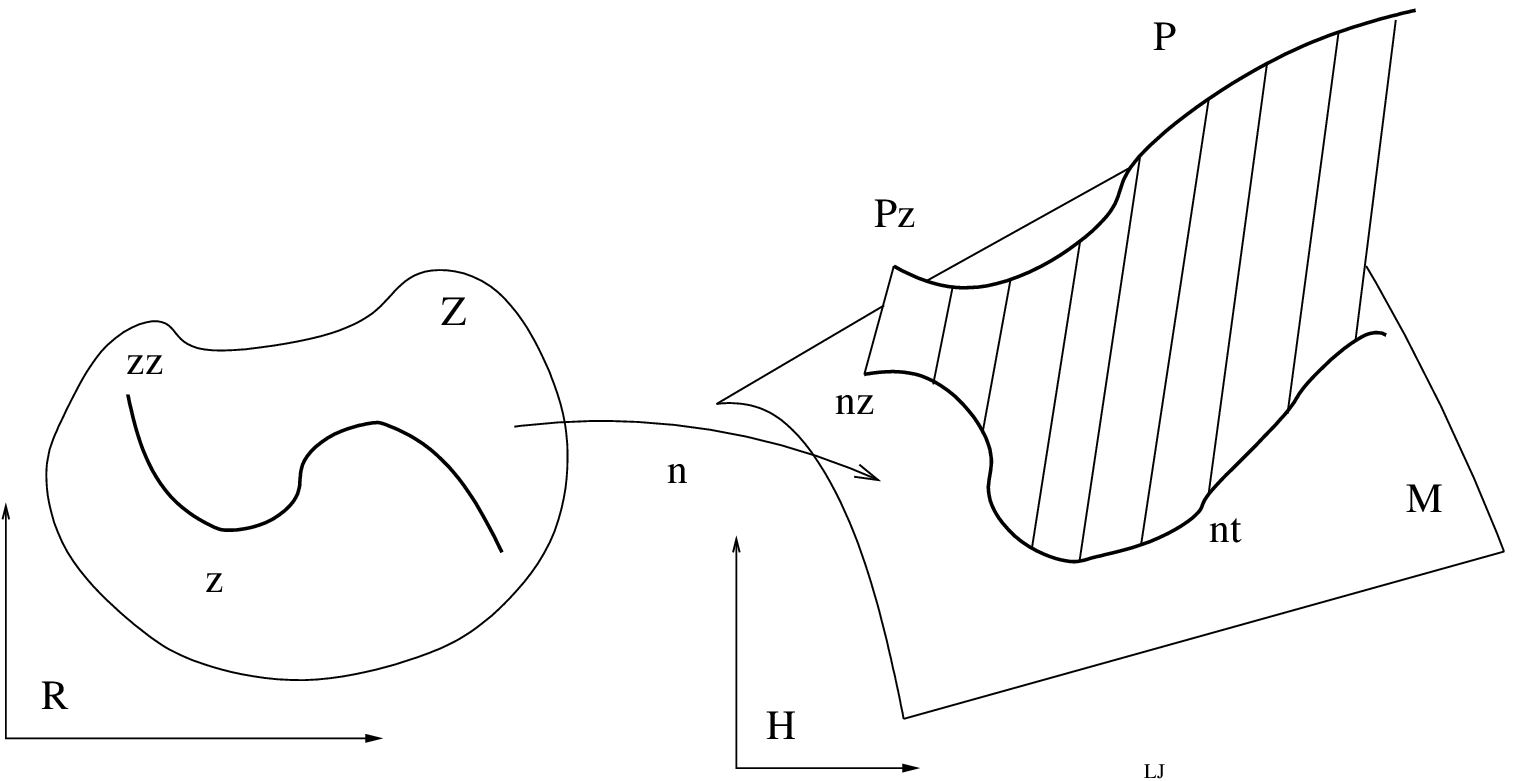}}
   \parbox{\linewidth}{
     \caption{The dynamics of $\vpsi$, $\vsolp$ and
       $\Par(t):=\iPar(\vpsi(\cdot,t))$. The diagonal lines connecting
       the trajectories of $\vsolp$ and $\vpsi$ indicate the
       skew-orthogonal projection.}
\label{fig:2}}
\end{figure}

The existence of the decomposition~\eqref{dec1}--\eqref{dec2} enables
us to `change variables' from $\vpsi$ to $(\Par,\vw)$, where
$\Par(t):=\iPar(\vpsi(\cdot,t))$.  The proof of the uniqueness of the
decomposition also gives estimates for the distance between the
initial parameter point $\zPar$, (see~\eqref{eq:initialZ}) and the
starting point is given by the decomposition $\iPar(\vpsi_0)$ as well
as the distance between $\vsol_{\zPar}$ and $\vsol_{\iPar(\psi_0)}$.
Figure~\ref{fig:2} shows what we have in mind. (See also
Figure~\ref{fig:ift}).

In Section~\ref{sec:dyn}, we insert the symplectically orthogonal
decomposition into the equation of motion~\eqref{eq:srh}. We then use
the symplectic form to project out the finite-dimensional equations
for $\Par$, which schematically take the form
\begin{equation}\label{eq:X}
\dot\Par = X(\Par) + Y(\Par,\vw).
\end{equation}
We then show that
\begin{equation}
|Y(\Par,\vw)|\leq C (\eps^2 + \nrmH{\vw}|\alpha| + \nrmH{\vw}^2 + \nrmH{\vw}^3),
\end{equation}
and $\alpha=(\dot y -\vel,\dot\vel,\freq-\dot\phase -V(y),\dot \freq)$
is bounded by
\begin{equation}
|\alpha|\leq C(\eps + |Y|).
\end{equation}
Finally, $\dot \vel$ and $\dot\freq$ are shown to be of size $C\eps$,
whereas the full dynamics $|\dot\Par - X(\Par)|$ is bounded by $C\eps^2$.
Thus, by~\eqref{eq:X}, we have found the finite-dimensional dynamics,
\ie the equation of motion for $\Par$.

The equation for the perturbation $\vw$ takes the form
\begin{equation}
\partial_t \vw = \Lv \vw + \NL(\vw) + ...
\end{equation}
The procedure in Sections~\ref{sec:J}--\ref{sec:dyn} of the paper is
collected in Figure~\ref{fig:1}.
\setlength{\unitlength}{.09em}
\begin{figure}[htbp]
\centering
\centerline{  
\begin{picture}(100,125)(160,00)
\put (0,100){\framebox(150,20){Symplectic structure, $\symp{\cdot}{\cdot}$}}
\put (75,100){\vector(0,-1){50}}
\put (85,65){\makebox(60,20){$\partial_\freq \Nn(\solw)>0$}}
\put (0,30){\framebox(150,20){$\symp{\cdot}{\cdot}$ non-degenerate on $\MfS$}}
\put (150,40){\vector(1,0){110}}
\put (153,15){\makebox(105,20){\parbox{8em}{
Implicit Function Theorem}
}}
\put (260,8){\framebox(150,50){\parbox{12em}{$\exists !$ decomposition in $U_{\delta}(\tPdom)$\\ $\vpsi = \vsolp + \vw_{\phase,y}$,\\  $\symp{\vw}{\vz}=0$, $\forall \vz\in\set{T}_{\vsolp}\MfS$}}}
\put (322,58){\vector(0,1){25}}
\put (324,60){\makebox(70,20){$\vpsi\rightarrow (\Par,\vw)$}}
\put (260,83){\framebox(120,36){
    \parbox{10em}{$\dot\Par = X(\Par)+Y(\Par,\vw)$, $\dot\vw = J\Lv \vw + g(\vw,\Par) $}}}
\end{picture}
}
   \centerline{}
   \parbox{\linewidth}{
     \caption{The way to the dynamics of $(\Par,\vw)$. A schematic
       representation of Sections~\ref{sec:J}--\ref{sec:dyn}.
       $\vw_{\phase,y}(x,t):=\lexp{-\phase J}\vw(x+y,t)$,
       $g(\vw,\Par)$ is the coupling term. For an explicit form
       see~\eqref{eq:vw}.}
\label{fig:1}}
\end{figure}

The next step in the proof of our main theorem is to control the
$\vw$-term. This is done in three steps as follows. We first show, in
Section~\ref{sec:xdyn}, that the dynamics of $\po:=\dotp{\vw}{|x|\vw}$
is well behaved and satisfies the inequality
\begin{equation}
\po(t)\leq \po(0) + C\nrmH{\vw},
\end{equation}
for times $t\leq T_1$ and such that the symplectic decomposition is valid.

To control the $\Half$-norm of $\vw$, we introduce a Lyapunov
functional, $\Lyp(t)$, which is a linear combination of $\HV$, $\Nn$
and $\Pn$ at $\vpsi$, minus the same quantities at $\vsolp$, that is
\begin{equation}
\Lyp(t):=(\freq-V(y))\big(\Nn(\vpsi)-\Nn(\vsolp)\big)
+v\cdot (\Pn(\vpsi)-\Pn(\vsolp)) + \HV(\vpsi)-\HV(\vsolp).
\end{equation}
In Section~\ref{sec:low}, we bound this functional from below in terms
of $\nrmH{\vw}$, plus small perturbation terms.  To do this, we use the
spectral properties of $\Lv$.

In Section~\ref{sec:up}, we bound this functional from above in terms
of small quantities, (\ie powers of $\nrmH{\vw}$, $|Y|$), for times
$t\leq T_2$ and such that the decomposition is valid. To do this, we
use the that $\Lyp(t)$ is ``almost conserved''.

In the last Section~\ref{sec:main}, we combine our estimates on
$\nrmH{\vw}$, $\po$ and $|Y|$.  The arguments in
Section~\ref{sec:main} prove the main theorem.

\section{Properties of ground states}
\label{sec:sol}

In this section, we derive properties of the elements of $\MfL$.

Let $R_{\vel,\theta}$ be a rotation of angle $\theta$ in $x$ around
the $\vel$-direction. Let $S_{\vel}$ be reflection in $x$ along the
$\vel$-direction.
\begin{Def}\label{def:sym}
  Let $\vel\in\RR^3$ with $\vel\neq 0$. A function
  $\vv=(\indR{w},\indI{w}):\RR^3\rightarrow \RR^2$ is called
  $\vel$-symmetric if
  \begin{equation}
    R_{\vel,\theta}\vv =\vv, \ \text{for all}\ \theta\in[0,2\pi), \ \text{and}\  S_{\vel}\vv = (\indR{w},-\indI{w}).
  \end{equation}
\end{Def}
Analogously for any direction, say $\zunit$, we call a function
$\zunit$-symmetric if it satisfies Definition~\ref{def:sym} with $\vel$
replaced with $\zunit$.
\begin{Def}\label{def:sim}
  Let $f:\RR\rightarrow \RR$. If $f$ is an even function then we write
  $f\sim (e)$ and if it is odd $f\sim (o)$.
\end{Def}
For functions $\vv:\RR^3\rightarrow \RR^2$ we use $\sim$ likewise. For
example, $\vv\sim (eee,eeo)$ means that the first component of $\vv$
is even in its three coordinate directions and the second component is
even in its first two coordinate directions and odd in the last. We
write $\vv\sim\vu$ to indicate that $\vv$ and $\vu$ have the same
reflection symmetries.

We have the result:

\begin{proposition}\label{prop:sol}
  Suppose Assumption~\ref{ass:ker} is satisfied and let $m>0$. Define
  $\freq_l(|\vel|):=(1-\sqrt{1-|\vel|^2})m$. Then there is a number
  $\maxspeed\in(0,1)$ and an open non-empty interval
  $I_0:=(\freq_{l0},\freq_{h0})$, with $\freq_{l0}> \freq_l(|\vel|)$
  such that for all $\freq\in I_0$ and $|\vel|\leq \maxspeed<1$ we
  have:
\renewcommand{\theenumi}{(\roman{enumi})}
\begin{enumerate}
\item\label{sol:-10} there is a function $(\vel,\freq) \mapsto
  \vsolvn\in \C{\infty}(B_{r_0}\times I_0,\Sob{1})$ 
  that solves~\eqref{eq:EEp} and $\vsolvn \in \Sob{s}$ for all $s \geq 1$. Moreover, we have $\vsolw:=\vsol_{0,\freq}$, with
  $\vsolw=(\solw,0)$ and $\solw$ is spherically symmetric and
  positive; 
\item\label{sol:1} $\vsolvn$ and $\partial_\freq \vsolvn$ are \Csym, 
and $\partial_{\vel_j}\vsolvn\sim  J\partial_{x_j}\vsolvn$;
\item\label{sol:3} $\vsolvn$, $\partial_{x_j}\vsolvn$,
  $\partial_{\freq}\vsolvn$, and $\partial_{\vel_k}\vsolvn$ are
  pointwise exponentially localized;
\item\label{sol:2} `Stability condition', $\partial_\freq
  \nrm{\vsolvn}^2>c$, where $c$ is independent of $\vel\in
  \{\vel:|\vel|\leq \maxspeed\}$, $\freq\in I_0$. Furthermore $\vsolw$
  is a minimizer of $\En_{0,0}$ subject to $\Nn$ constant;
\item\label{sol:spec} $\Lv:=\EE''(\vsolvn)$ has one negative
  eigenvalue. Moreover, $\dim\Ker \Lv = 4$; there is a spectral gap
  between zero and its next spectral point; and the essential spectrum
  starts at $\freq-\freq_l(|v|)>0$.
\end{enumerate}
\renewcommand{\theenumi}{\arabic{enumi}}
\end{proposition}
The ground state $\vsolw$ constructed in
Proposition~\ref{prop:sol}\eqref{sol:-10} is used to define the
soliton manifold $\MfL:=\Mf(\Pdom_0)$, where $\Pdom_0=\Pdom(r_0,I_0)$
in~\eqref{eq:aMf}. It is convenient to use
$\Par:=(y,\vel,\phase,\freq)$ as a point in $\Pdom_0$, and
subsequently (as in~\eqref{eq:spar})
\begin{equation}
\vsolp(x):=\lexp{-\phase J}\vsolvn(x-y).
\end{equation}
These ground states $\vsolp$ satisfy the Euler-Lagrange equation
\begin{equation}\label{eq:EEpp}
\EE'(\vsolp)=0,
\end{equation}
and taking derivatives of \eqref{eq:EEpp} with respect to $y$ and
$\phase$ evaluated at $(y,\phase)=(0,0)$ yield
\begin{equation}\label{eq:Lz1}
  \Lv \partial_{x_j} \vsolvn = 0, \ \ \Lv J \vsolvn = 0.
\end{equation}
Thus $\partial_{x_j}\vsolvn$ and $J\vsolvn$ are zero modes to $\Lv$.
The derivatives of Eq.~\eqref{eq:EEpp} with respect to $\vel_j$ and
$\freq$, respectively, at $y=0$, $\phase=0$ lead to
\begin{equation}\label{eq:Lz2}
  \Lv \partial_{\vel_j} \vsolvn = J \partial_{x_j} \vsolvn, 
  \ \ \ \Lv \partial_\freq \vsolvn = -\vsolvn.
\end{equation}
Therefore $\partial_{\vel_j}\vsolvn$ and $\partial_\freq \vsolvn$ are
zero modes to $(J\Lv)^2$. The tangent vectors of $\MfL$ are hence in
the generalized null space of $J\Lv$ (compare with
\eg~\cite{Weinstein1985}).  Below, we prove
part~\ref{sol:-10}--\ref{sol:spec} of Proposition~\ref{prop:sol} and
the completion of the proof can be found in
Subsection~\ref{sub:done}.

\subsection{Proof of Proposition~\ref{prop:sol}\ref{sol:-10},\ref{sol:1}}
\label{sub:e+s}

The functions $\vsolvn$ will be constructed as a class of solutions to
the Euler-Lagrange equation~\eqref{eq:EEp}, starting from an
unboosted, with $\vel=0$, minimizer $\vsol_0:=(\sol_0,0)$ at frequency
$\freq_0$. Here, $\sol_0$ is a spherically symmetric positive
minimizer to $\En_{0,0}$ at constant $\Nn$ from~\cite{Lieb+Yau1987}
and~\cite{Frohlich+Jonsson+Lenzmann2005}.

A non-zero velocity $\vel$ breaks the rotation symmetry of the map
$\EE'$. Without loss of generality we pick a preferred direction,
$\zunit$, the unit vector parallel to the $x_3$-axis, and choose
coordinates so that $\vel=\speed\zunit$.  Let $\Rzt$ be the spatial
rotation around $\zunit$ of angle $\theta$, $\Fz$ the spatial
reflection along $\zunit$ and let $K$ be the matrix
\begin{equation}
  K=\begin{pmatrix} 1 & 0 \\  0 & -1\end{pmatrix}.
\end{equation}
The $\zunit$-symmetric Sobolev space, $\sSob{s}$, of order $s$ is
defined as
\begin{equation}
  \sSob{s} := \{\vpsi\in \Sob{s}(\RR^3,\RR^2):\ \Rzt\vpsi=\vpsi, \forall \theta\in [0,2\pi), \  K\Fz\vpsi=\vpsi\},
\end{equation}
where $\Rzt \vpsi (x) = \vpsi(\Rzt x)$.

\begin{remark}
That $\sSob{s}$ is a closed subspace of $\Sob{s}(\RR^3, \RR^2)$ follows by noting that $\Rzt$ and $K \Fz$ are bounded operators on $\Sob{s}(\RR^3, \RR^2)$ and that $\sSob{s} =  \cap_{\theta \in [0, 2 \pi)} \Ker (1-\Rzt) \cap \Ker (1-K \Fz)$.
\end{remark}

We recast Proposition~\ref{prop:sol}\eqref{sol:-10} and the first
part of \eqref{sol:1} into:
\begin{proposition}\label{prop:exists}
  Suppose that Assumption~\ref{ass:ker} is satisfied.  Let
  $\vsol_{0}$, $\freq_0$ and $\sSob{s}$, be as above, let
  $\vel:=\speed\zunit$, with $\speed\in\RR$. Then there is an open
  neighborhood, $W\subset\RR\times\RR_+$, with $(0,\freq_0)\in W$, and
  a unique function $(\speed,\freq)\mapsto
  \vsol_{\speed\zunit,\freq}\in \C{\infty}(W,\sSob{1})$ such that $\vsol_{0\zunit,\freq_0}=\vsol_0$ and
  $\vsol_{\speed\zunit,\freq}$ solves $\Es'(\vpsi)=0$ for all
  $(\speed,\freq)\in W$. In addition, $\vsol_{\speed\zunit,\freq}$ belongs to $\Sob{s}$ for all $s \geq 1$.
\end{proposition}
\begin{remark}\label{rem:solw}
  (a) A solution to $\EE'(\vpsi)=0$ when $\vel\neq 0$ points in
  arbitrary direction is obtained by rotating
  $\vsol_{\speed\zunit,\freq}$ in $x$ from $\zunit$ to
  $\hat{\vel}:=\vel/|\vel|$. See the proof of Corollary~\ref{cor:sim}
  for details. (b) The Sobolev space of order one of radially
  symmetric functions, $\Hrone(\RR^3,\RR)$, is a scalar subspace of
  $\sSob{1}$. This ensures existence and uniqueness of a solution
  $\vsol_{\vel=0,\freq}=\vsolw=(\solw,0)$ to
  $\En_{0,\freq}'(\vpsi)=0$, where $\solw\in \Hrone$. (c) $\EE$ is
  invariant under translation and change of gauge. Thus, $\lexp{\phase
    J}\vsolvn(\cdot+y)$ for any $y\in \RR^3$, $\phase\in[0,2\pi)$ is
  also a solution to~\eqref{eq:EEp}.
\end{remark}

Proposition~\ref{prop:exists} proves that $\vsol_{\speed\zunit,\freq},
\partial_\freq\vsol_{\speed\zunit,\freq}\in\sSob{s}$. For arbitrary
coordinates this implies that $\vsolvn$ and $\partial_{\freq}\vsolvn$
are $v$-symmetric.  The reflection symmetries of
$\partial_{\vel_j}\vsolvn$ now follows from:
\begin{corollary}\label{cor:sim}
  Suppose that the Assumption~\ref{ass:ker} is satisfied. Then
\begin{equation}
\partial_{\vel_j}\vsolvn\sim J\partial_{x_j}\vsolvn, \ j=1,2,3.
\end{equation}
\end{corollary}
The corollary is proved at the end of this subsection.

\begin{lemma}\label{lem:diffEE}
  We have that
\begin{equation}
  (\speed, \freq, \vpsi) \mapsto \Es'(\vpsi)\in
  \C{\infty}(\RR \times \RR \times \sSob{1}, \sSob{0}).
\end{equation}
\end{lemma}

\begin{proof}[Proof of Proposition~\ref{prop:exists}]
  Let $F(\speed,\freq,\vpsi):=\Es'(\vpsi)$. To find solutions to the
  equation $F(\speed,\freq,\vpsi)=0$, we use the implicit function
  theorem in~\cite{Dieudonne1969}, which has three assumptions, that
  we now verify: That $F$ is $\C{\infty}$ is shown by
  Lemma~\ref{lem:diffEE}. The equation $F=0$ has a solution
  $(0,\freq_0,\vsol_0)$ and $\vsol_0\in\sSob{1}$.  The last condition
  is that $F_{\vpsi}'(0,\freq_0,\vsol_0)=:\Lzz$, where
  $\Lzz=\diag(\LzOz,\LzTz)$ is invertible. We have that $\LzTz\geq 0$
  and $\LzTz\sol_0=0$, the zero eigenvalue is non-degenerate since $\lexp{-t\LzTz}$ is positivity improving. (This follows from the explicit kernel for $\lexp{-t \sqrt{-\Delta + m^2}}$ and Trotter's product formula.) The kernel of $\LzOz$ is
  spanned by $\{\partial_{x_j}\sol_0\}_j$, thanks to the kernel
  assumption.  Thus the kernel of $\Lzz$ is spanned by
  $\{J\vsol_0,\partial_{x_j}\vsol_0, j=1,2,3\}$, neither of these
  functions are $\zunit$-symmetric. Thus, $\Lzz$, as a map
  $\sSob{1}\subset\sSob{0} \rightarrow \sSob{0}$, is invertible.

  We conclude, by the implicit function
  theorem~\cite{Dieudonne1969}, that there is a neighborhood $W\subset
  \RR\times\RR_+$ with $(0,\freq_0)\in W$ and a unique map
  $(\speed,\freq)\mapsto \vsols\in\C{\infty}(W,\sSob{1})$ such that
  $\vsol_{0\zunit,\freq_0}=\vsol_0$ and $\vsols$ solves
  $F(\speed,\freq,\vpsi)=0$.

  That $\vsols \in \Sob{n}$ for any $1\leq n\in\mathbb{N}$ follows from a simple bootstrap argument; see the proof of Theorem~3 in~\cite{Frohlich+Jonsson+Lenzmann2005}.
 
\end{proof}

\begin{proof}[Proof of Lemma \ref{lem:diffEE}]
  The Hardy-Littlewood-Sobolev inequality together with the Sobolev
  embedding theorem (see \eg~\cite{Lieb+LossII}) shows that 
  $\EE'\in\set{C}(\Sob{1},\Ltwo)$. To see that $\Es'$ preserves
  $\zunit$-symmetry, let $U$ be either of $K\Fz$ or $\Rzt$. Both operations 
  leave $\Es$ invariant \ie
\begin{equation}\label{eq:EU} 
\Es(U\vpsi) = \Es(\vpsi), \ \ \vpsi\in \Sob{1}.
\end{equation}
By (Fr\'echet) differentiation of~\eqref{eq:EU} 
\begin{equation}\label{eq:U}
  \Es'(U\vpsi)= U\Es'(\vpsi).
\end{equation}
Let $\vpsi$ satisfy $U\vpsi=\vpsi$. Eqn.~\eqref{eq:U} then states
\begin{equation}\label{eq:UUU}
  \Es'(\vpsi) = \Es'(U\vpsi) = U\Es'(\vpsi), 
  \ \text{where}\ \vpsi\in \Sob{1},
\end{equation}
and hence $\Es':\sSob{1}\rightarrow\sSob{0}$.  Repeating the argument
for higher order derivatives of~\eqref{eq:EU} gives that
$\Es^{(n)}(\vpsi)$ preserves $\zunit$-symmetry.

That $\vpsi\mapsto\EE'(\vpsi)$ is $\C{1}$ follows from the
Hardy-Sobolev-Littlewood inequality and the Sobolev embedding theorems
for $\Sob{1}$. That is, let $\vu,\vv, \vw, \vzeta\in \Sob{1}$, then
\begin{multline}
\sup_{\nrmHp{1}{\vw}=1,\nrmHp{1}{\vzeta}=1}|
\dotp{\EE''(\vu)\vw}{\vzeta}- \dotp{\EE''(\vv)\vw}{\vzeta}| 
= \sup_{\nrmHp{1}{\vw}=1,\nrmHp{1}{\vzeta}=1} 
\\ |\dotp{\frac{1}{|x|}*((\vu+\vv)\cdot(\vu-\vv))}{\vw\cdot\vzeta} 
+ \dotp{\vzeta\cdot(\vu-\vv)}{\frac{2}{|x|}*(\vu\cdot\vw)} + 
 \dotp{\vzeta \cdot\vv}{\frac{2}{|x|}*((\vu-\vv)\cdot\vw)}|\\ 
\leq C(\nrmHp{1}{\vv},\nrmHp{1}{\vu})\nrmHp{1}{\vu-\vv}.
\end{multline}
Analogously one can show that $\vpsi\mapsto\EE'(\vpsi)$ is $\C{2}$.
The polynomial nature of the nonlinearity implies that
$\EE^{(4)}(\vpsi)$ is a (tri-)linear bounded operator independent of
$\vpsi$. Thus $\vpsi\mapsto\EE'(\vpsi)$ is $\C{\infty}$ in $\vpsi$.

The map $(\speed,\freq)\mapsto\Es'(\vpsi)$ is linear and hence smooth. Differentiation with respect to either $\freq$ or
$\speed$ does not change $\Es'(\vpsi)$'s symmetries. \end{proof}

\begin{proof}[Proof of Corollary~\ref{cor:sim}]
  The momentum term is the only term in $\EE$ that breaks the rotation
  symmetry. For an arbitrary rotation
\begin{equation}
\EE(R\vpsi) = \En_{R^{-1}\vel,\freq}(\vpsi),
\end{equation}
talking the derivative gives the relation
\begin{equation}\label{eq:RRR}
R^{-1}\EE'(R\vpsi) =\En'_{R^{-1}\vel,\freq}(\vpsi).
\end{equation}
Given the $\zunit$-symmetric function $\vsols$, we find $\vsolvn$ by any rotation, $R$, that takes $\zunit$ to $\hat{\vel}$ as $\vsolvn=R\vsols$.

This relation between $\vsolvn$ and $\vsols$ is the key to show the
corollary. Let $R_1$ be the rotation from $v_1\hat{e}_1 + v_3\zunit$
to $|\vel|\zunit$ given by
\begin{equation}
R_1(\theta):=\begin{pmatrix} 
\cos \theta & 0 & -\sin\theta \\ 
0 & 0 & 0 \\
\sin \theta & 0 & \cos \theta 
\end{pmatrix},
\end{equation}
where $\theta$ is the angle between $\vel$ and $\zunit$. We find for
$\speed=|\vel|>0$ that
\begin{align}\label{eq:rot}
\partial_{\vel_1}\vsolvn &= \partial_{\vel_1} (R_1\vsols) = 
\frac{\vel_1}{\speed}R_1\partial_{\speed}\vsols + 
(\partial_{\vel_1}\theta)(\partial_{\theta}R_1x)\cdot R_1\nabla_x\vsols \\
 & = R_1 (\frac{\vel_1}{\speed}\partial_{\speed}\vsols + 
(\partial_{\vel_1}\theta)(R_1^{-1}\partial_{\theta}R_1x)\cdot 
\nabla_x\vsols).
\end{align}
At the point $\vel_1=0$, $\vel_3=\speed$ this simplifies to
\begin{equation}
  \left.\partial_{\vel_1}\vsolvn\right|_{\vel=(0,0,\speed)} = -\frac{1}{\speed}
\hat{e}_2 \cdot (x\wedge\nabla_x) \vsols,
\end{equation}
where $\wedge$ is the cross product.  The above expression is $\sim
J\partial_{x_1}\vsols$.  Analogously for $\vel=(0,v_2,v_3)$ we find
\begin{equation}
  \left.\partial_{\vel_2}\vsolvn\right|_{\vel=(0,0,\speed)} = 
  \frac{1}{\speed} \hat{e}_1 \cdot (x\wedge\nabla_x) \vsols \sim J\partial_{x_2}\vsols.
\end{equation}
Recall from Proposition~\ref{prop:exists} that
$\partial_{\speed}\vsolvn\in \sSob{1}$ and thus $\sim
J\partial_{x_3}\vsols$. Therefore we have shown the corollary for a
given coordinate system, \ie coordinates such that $\vsolvn$ is
$\vsols$, rotation $R$, once again, from $\zunit$ to $\hat{\vel}$ of
this case gives the general result.
\end{proof}
  
\subsection{Proof of exponential decay of tangent vectors}
\label{sub:exp}

In this subsection we prove Proposition~\ref{prop:sol}\ref{sol:3}, \ie
the pointwise exponential decay of the tangent vectors
$\{\partial_{x_j}\vsolvn, \partial_{\vel_j}\vsolvn, J\vsolvn,
\partial_{\freq}\vsolvn\}$. In \cite[App.
C]{Frohlich+Jonsson+Lenzmann2005} we showed that $\vsolvn$ satisfies
the bound
\begin{equation}\label{eq:solExp}
|\vsolvn|\leq c_1(\beta)\lexp{-\beta|x|},
\end{equation}
for $0<\beta<\min(m,(\freq-\freq_l)(1-\vel^2)^{-1/2})$, where
$\freq_l$ is defined in Proposition~\ref{prop:sol}. The remaining
tangent vectors all satisfy an equation of the type
\begin{equation}\label{eq:FF}
  \vu=F(\vu,\vv),
\end{equation}
with
\begin{equation}\label{eq:Fest}
   F(\vu,\vv):=(H_\vel+\freq)^{-1}(W_1\vu + W_2(\vu)+\vv)
\end{equation}
and
\begin{equation}
H_\vel = \sqrt{-\Laplace+m^2}-m-J\vel\cdot\nabla, \ W_1:=\frac{1}{|x|}*|\vsolvn|^2, \
W_2(\vu):=\big(\frac{2}{|x|}*(\vu\cdot\vsolvn)\big)\vsolvn,
\end{equation}
for $\freq>\freq_{l}$ and some $\vv$ depending on the tangent vector,
see~\eqref{eq:Lz1}--\eqref{eq:Lz2}. 

We have the following result.
\begin{lemma}\label{lem:Fexp}
  Let $m>0$, $\freq>\freq_l$ and let $\vu$ be a solution
  to~\eqref{eq:FF}, for some $\vv$ with $|\vv|\leq
  c_2\lexp{-\beta_2|x|}$ where $c_2,\beta_2$ are some positive
  constants.  Then, there is $\theta>0$ and a constant $C(\theta)>0$
  such that
\begin{equation}
  |\vu|\leq C(\theta)\lexp{-\theta|x|}.
\end{equation}
\end{lemma}
We now return to equations~\eqref{eq:Lz1}--\eqref{eq:Lz2} to find
pairs $(\vu,\vv)$ that solve~\eqref{eq:FF}. These are
$(\partial_{x_j}\vsolvn,0)$, $(\partial_{\freq}\vsolvn,-\vsolvn)$ and
$(\partial_{\vel_j}\vsolvn,J\partial_{x_j}\vsolvn)$.  The first pair,
inserted in Lemma~\ref{lem:Fexp} ensures that $\partial_{x_j}\vsolvn$
is pointwise exponentially decaying. Thus, we know that the last two
pairs also satisfy the assumptions of Lemma~\ref{lem:Fexp} and hence,
both $\partial_{\freq}\vsolvn$ and $\partial_{\vel_j}\vsolvn$ are
pointwise exponentially decaying. It remains to prove
Lemma~\ref{lem:Fexp}.
\begin{proof}[Proof of Lemma~\ref{lem:Fexp}]
  The proof is based on~\cite{Slaggie+Wichmann1962} as presented
  in~\cite{Hislop2000} and we extend the result to include the source
  terms. That the integral kernel of $(H_\vel+\freq)^{-1}$,
  $G_{\freq,\vel}$, satisfies the bound, \cite[App.
  C]{Frohlich+Jonsson+Lenzmann2005}
\begin{equation}\label{eq:Gker}
|G_{\freq,\vel}(x)|\leq c_3\frac{\lexp{-\delta|x|}}{|x|^2},
\end{equation}
for some $\delta>0$ depending on $m>0$, $|v| < 1$, and $\freq>\freq_l$.  

Our first goal is to use \eqref{eq:FF} to bound $|\vu|$ as
\begin{equation}\label{eq:mest}
  |\vu(x)|\leq h_{\theta}(x)M(x) + C_1\lexp{-\gamma|x|},
\end{equation}
where $\theta>0$ remains to be chosen later, and
\begin{equation}
  M(x):=\sup_{x'} |\vu(x')|\lexp{-\theta|x-x'|}
\end{equation}
To this end, we need estimates on each term in~\eqref{eq:Fest} and we
begin with the $W_2(\vu)$ term
\begin{equation}\label{eq:W2}
  W_2(\vu)(x') = \vsolvn(x')\int_{\RR^3}\frac{1}{|x'-y|}\vu(y)\cdot\vsolvn(y)\diff y.
\end{equation}
The identity $\vu(y)=\vu(y)\lexp{-\theta|y-x|}\lexp{\theta|y-x|}$, the
inequality $|y-x|\leq |y-x'|+|x'-x|$, and the upper
bound~\eqref{eq:solExp} lead to
\begin{equation}\label{eq:W2p}
  |W_2(\vu)(x')| \leq C_2'M(x)\lexp{-\beta|x'|+\theta|x-x'|} \int_{\RR^3} 
  \frac{\lexp{\theta|x'-y|}}{|x'-y|}\lexp{-\beta|y|} \diff y.
\end{equation}
We evaluate the integral, with $0<\theta<\beta$, to find for some
$C_2=C_2(\theta)>0$ that
\begin{equation}
  |W_2(\vu)(x')| 
  \leq C_2\frac{\lexp{\theta|x-x'|}}{1+|x'|}M(x).
\end{equation}
The estimate for $W_1$ follows similarly, by once again integrating an
integral of the type that appears in~\eqref{eq:W2p}:
\begin{equation}
  |W_1(x')|\leq C_3\frac{1}{1+|x'|}.
\end{equation}
The `potentials' $W_1$ and $W_2$ are hence bounded and decaying and we
can choose $h_\theta$ to be
\begin{equation}\label{eq:HT}
  h_\theta(x):=C_4 \int_{\RR^3} \frac{\lexp{-(\delta-\theta)|x-x'|}}{|x-x'|^2}\frac{1}{1+|x'|} \diff x',
\end{equation}
where $C_4$ is composed of the constants $c_3$, $C_2$ and $C_3$.

We use the integral kernel $G_{\freq,\vel}$ of $(H_{\vel}+\freq)^{-1}$
to express the last term of~\eqref{eq:Fest}. By~\eqref{eq:Gker}, we
have
\begin{equation}
  |G_{\freq,\vel}*\vv| \leq c_3\int_{\RR^3}\frac{\lexp{-\delta|x'|}}{|x'|^2}|\vv(x-x')| \diff x'.
\end{equation}
The assumed, pointwise exponential decay of $\vv$ together with
the inequality $|x-x'|\geq \big||x|-|x'|\big|$ yields
\begin{equation}
  |G_{\freq,\vel}*\vv| \leq C_1\lexp{-\gamma|x|},
\end{equation}
where $\gamma=\min(\delta,\beta_2)$ and $C_1>0$ are suitable constants.
We have thus established~\eqref{eq:mest}. 

To proceed, we show that $h_{\theta}$ is bounded and that it decays
pointwise as $|x|\rightarrow \infty$.  The first of these properties
follows from Young's inequality, since for $\theta<\delta$,
$\lexp{\theta|\cdot|}G_{\freq,\vel}(\cdot)\in \Lp{1}$ and
$(1+|\cdot|)^{-1}<1\in\Lp{\infty}$
\begin{equation}\label{eq:hsup}
  \sup_{x}|h_{\theta}|\leq C_4\nrmLp{1}{\lexp{\theta|\cdot|}G_{\freq,\vel}(\cdot)}
  \sup_{x} |(1+|x|)^{-1}|=C_5(\theta)<\infty.
\end{equation}
To show the decay of $h_\theta$ as $|x|\rightarrow \infty$, let
$\alpha:=\delta-\theta$, $\alpha>0$, we use~\eqref{eq:HT} and split
the region of integration into two parts $|x-x'|\leq\kappa$,
$|x-x'|>\kappa$.  In the outer region we use the uniform bound of
$(1+|x'|)^{-1}<1$ to find
\begin{equation}
  \int_{|x-x'|>\kappa} \frac{\lexp{-\alpha|x-x'|}}{|x-x'|^2}\frac{1}{1+|x'|}
  \diff x' \leq \frac{1}{\kappa^2} \int_{\RR^3} \lexp{-\alpha|x'|}\diff x',
\end{equation}
and in the inner region,
\begin{equation}
  \int_{|x-x'|\leq\kappa} \frac{\lexp{-\alpha|x-x'|}}{|x-x'|^2}\frac{1}{1+|x'|}
  \diff x' \leq \int_{|y|\leq\kappa} \frac{\diff y}{|y|^2(1+|x-y|)} 
  \leq \frac{4\pi \kappa}{1+\big||x|-\kappa\big|},
\end{equation}
The choice of $\kappa=|x|^{1/2}$ ensures that $h_{\theta}\leq
C_6|x|^{-1/2}$ as $|x|\rightarrow \infty$ and $\theta\leq\delta$.

The following two identities will be used repeatedly in the next step
of the proof, let $\theta>0$, $\gamma>0$,
\begin{equation}\label{eq:expexp}
\sup_{y}\ \lexp{-\theta|x-y|-\theta|y-x'|} = \lexp{-\theta|x-x'|}, \ \
\sup_{y}\ \lexp{-\theta|y|-\gamma|y-x|} = \lexp{-\min(\theta,\gamma)|x|}.
\end{equation}

The exponential decay of $\vu$ now follows from the properties of
$h_{\theta}$, through two inequalities. Since $h_{\theta}$ decay, for
a fixed small $\theta<\delta$, there is a radius $R$, such that for
$|x|>R$ we have that $h_{\theta}\leq C_6R^{-1/2}$.  In this exterior
region, we use~\eqref{eq:mest} together with \eqref{eq:expexp} to
obtain
\begin{multline}\label{eq:first}
\sup_{|x'|> R} |\vu(x')|\lexp{-\theta|x-x'|} \leq 
\sup_{|x'|> R} \big(h_{\theta}(x')M(x')+C_1\lexp{-\gamma|x'|}\big)
\lexp{-\theta|x-x'|} \\ \leq C_6R^{-1/2}M(x)
+\sup_{|x'|> R}C_1\lexp{-\gamma|x'|-\theta|x-x'|}.
\end{multline}
In the interior we have
\begin{multline}\label{eq:second}
\sup_{|x'|\leq R} |\vu(x')|\lexp{-\theta|x-x'|} \leq \\ 
\sup_{|x'|\leq R} \Big(h_{\theta}(x')\big(
\sup_{|y|\leq R} |\vu(y)|\lexp{-\theta|x'-y|}+
\sup_{|y|> R}|\vu(y)|\lexp{-\theta|x'-y|}\big)+C_1\lexp{-\gamma|x'|}\Big)
\lexp{-\theta|x-x'|} .
\end{multline}
Insert the result~\eqref{eq:first} into~\eqref{eq:second}.  The upper
bound~\eqref{eq:hsup} ensures that $h_{\theta}<C_5$; the exterior term
$\sup_{|y|\geq R}|\vu(y)|\lexp{-\theta|x'-y|}$ is estimated
by~\eqref{eq:first}; for the interior term $\sup_{|y|\leq
  R}|\vu(y)|\lexp{-\theta|x'-y|}$ we have by continuity and boundedness
of $\vu$ that $|\vu|\leq C_7'(R)$ and hence $\sup_{|y|\leq
  R}|\vu(y)|\lexp{-\theta|x'-y|}\leq C_7(R,\theta)\lexp{-\theta|x'|}$;
yields~\eqref{eq:second} to become
\begin{multline}\label{eq:secondP}
  \sup_{|x'|\leq R} |\vu(x')|\lexp{-\theta|x-x'|} \leq  
  C_1\sup_{|x'|\leq R}\lexp{-\gamma|x'|}\lexp{-\theta|x-x'|} \\
  +C_5 \sup_{|x'|\leq R} \big(C_7\lexp{-\theta|x'|}+
  C_6R^{-1/2}M(x') + C_1\sup_{|y|>
    R}\lexp{-\gamma|y|-\theta|x-y|}\big)
  \lexp{-\theta|x-x'|}.
\end{multline}
Adding~\eqref{eq:first} to~\eqref{eq:secondP}, rewriting and the use
of~\eqref{eq:expexp} give
\begin{equation}
M(x)\leq 
C_8(R^{-1/2}M(x)+\lexp{-\min(\gamma,\theta)|x|})
+   C_9(R,\theta)\lexp{-\theta|x|},
\end{equation}
for suitable constants $C_8=C_8(\theta)$, $C_9$. By choice of $R=R^*$,
sufficiently large and $\theta>0$, sufficiently small, we find
\begin{equation}
M(x)\leq C'(R^*,\theta) \lexp{-\theta|x|},
\end{equation}
This upper bound inserted into~\eqref{eq:mest} together
with~\eqref{eq:hsup} yields $|\vu|\leq C(\theta)\lexp{-\theta|x|}$ and
we have proved the lemma.
\end{proof}

\subsection{Proof of the Stability Condition}
\label{sub:stab}

We now derive the {\em ``stability condition''} stated in
Proposition~\ref{prop:sol}\ref{sol:2} for unboosted ground states,
$\vsol_{v=0,\freq}(x)=(\solw,0)$. As mentioned in
Remark~\ref{rem:solw} (see
also~\cite{Lieb+Yau1987,Frohlich+Jonsson+Lenzmann2005}), these
functions can be assumed to be real-valued and spherically symmetric.
In view of this, we introduce the subspace
\begin{equation}
\Sob{s}_{\mathrm{rad}}(\RR^3,\RR) = \left \{ \psi \in \Sob{s}(\RR^3;\RR) 
   : \psi\ \text{is spherically symmetric} \right \} ,
\end{equation}
for $s \geq 0$.

\begin{lemma} \label{lem-stab} Suppose that Assumption \ref{ass:ker}
  holds. Then, for almost every $0 < N < \Nc$, there exists an
  unboosted ground state, $\sol_* = \sol_{v=0,\freq_{*}}$, with
  $\Nn((\sol_*,0)) = N$ and Lagrange multiplier, $\mu_*$, satisfying
  the following properties. For every sufficiently small $\delta > 0$,
  there exists a $\C{\infty}$-map
  \begin{equation} \label{def-solmap} (\mu_* - \delta, \mu_* + \delta)
    \rightarrow \Hrone(\RR^3,\RR), \quad \mu \mapsto \solw,
\end{equation}
where $(\solw,0)$ solves \eqref{eq:EEp} with $\vel=0$ and we have that
$\sol_{\freq_*} = \sol_*$. In addition, there exists a non-empty
interval $I \subset (\mu_* - \delta, \mu_* + \delta)$, such that
\begin{equation} \label{eq-stabcond} \frac{\dd}{\dd \mu} \Nn \big (
  (\sol_{\freq},0) \big ) > 0
\end{equation}
holds for all $\mu \in I$.
\end{lemma}

\begin{proof}[Proof of Lemma \ref{lem-stab}]
  By Remark \ref{rem:solw}(b), we can assume that unboosted ground
  states $\sol(x)=\sol_{\vel=0,\freq}$ are spherical symmetric and
  real-valued. Let $E(N):=\inf\{\En_{0,0}(\vpsi):
  \vpsi\in\Half,\Nn(\vpsi)=N\}$. It is known that the function $E:(0,\Nc)\rightarrow
  \RR$ is strictly
  concave~\cite[Lemma~2.3]{Frohlich+Jonsson+Lenzmann2005}.  This
  implies in particular the following properties.
\begin{itemize}
\item $E(N)$ is continuous on $(0,\Nc)$.
\item $E'_-(N)$ and $E_+'(N)$ (which denote the left and right derivative, respectively) exist for all $N \in (0,\Nc)$.
\item $E'(N) = E'_-(N) = E'_+(N)$ for all $N \in (0,\Nc) \setminus \Sigma$, where $\Sigma$ is some countable set.
\end{itemize}
For convenience, we denote the set where $E'(N)$ exists by
\begin{equation}
\Sigma^c := (0,\Nc) \setminus \Sigma.
\end{equation}
Let us now pick $N_* \in \Sigma^c$ and a strictly decreasing sequence, $(N_k)$, in $\Sigma^c$ such that
\begin{equation}
N_k \searrow N_*, \quad \mbox{as $k \rightarrow \infty$}.
\end{equation} 
By density $\Sigma^c \subset (0,\Nc)$, this always possible.
Correspondingly, let $(\sol_k) \subset \Half$ be a sequence of minimizers with
$\En_{0,0}((\sol_k,0)) = E(N_k)$ and $\Nn((\sol_k,0)) = N_k$, which, by
continuity of $E(N)$, implies that
\begin{equation}
\En_{0,0}((\sol_k,0)) \rightarrow E(N_*) \quad \mbox{and} \quad \mathcal{N}((\sol_k,0)) \rightarrow N_*, \quad \mbox{as $k \rightarrow \infty$}.
\end{equation}
By arguments similar to those in the proof of \cite[Theorem
2]{Frohlich+Jonsson+Lenzmann2005} and the relative compactness
property stated in \cite[Theorem 1]{Frohlich+Jonsson+Lenzmann2005}, we
see that $(\sol_k)$, after passing to a subsequence, converges
strongly in $\Hrhalf$ to some minimizer $\sol_*$ with $\Nn((\sol_*,0))
= N_*$ and Lagrange multiplier $-\mu_*$.  (Note that due to $v=0$, we
can restrict our attention to radial functions and translations do not
have to be taken into account.)

Next, we observe that any $\sol_{k}$ satisfies the identity
\begin{equation} \label{eq-ident}
E(N_k) - \frac{1}{4} \Vpot{\sol_{n_k}} = - \mu_k N_k, 
\end{equation}
where $-\mu_k$ is the Lagrange multiplier for the minimizer
$\sol_{k}$. This identity follows from multiplication of the
Euler-Lagrange equation \eqref{eq:EEp} with $(\sol_k(x),0)$ and
integration. Now we claim that
\begin{equation} \label{eq-E-mu}
E'(N_k) = - \mu_k
\end{equation}  
holds for all $k$. Note that $E'(N_k)$ exists due to $N_k \in
\Sigma^c$ for all $k$. To prove (\ref{eq-E-mu}), we observe that
$\En_{0,0}(\sqrt{\tau} (\sol_k,0)) \geq E(\tau N_k)$ holds for all
$\tau \geq 0$ with equality for $\tau = 1$. Hence it is
straightforward to see that the right derivative, $E'_+(N_k)$, obeys
the following estimate
\begin{align}
  E'_+(N_k) & = \frac{1}{N_k} \lim_{N \searrow N_k} \frac{E(N) -
    E(N_k)}{N/N_k - 1} \leq \frac{1}{N_k} \lim_{\tau \searrow 1}
  \frac{\En_{0,0}(\sqrt{\tau} (\sol_k,0)) -
    \En_{0,0}((\sol_k,0))}{\tau -
    1}  \nonumber  \\
 & = \frac{1}{N_{k}}\lim_{\tau\searrow 1} \frac{d}{d\tau}
  \En_{0,0}(\sqrt{\tau}(\sol_k,0)) =
  \left.\frac{1}{2N_l\sqrt{\tau}}\right|_{\tau=1}
  \dotp{\En_{0,0}'((\sol_k,0))}{(\sol_k,0)} \\
&= \frac{-\freq_k}{N_k}\frac{1}{2}\int_{\RR^3} |\sol_k|^2\diff x = -\freq_k,
\end{align} 
using the Euler-Lagrange equation $\mathcal{E}_{0,0}'((\sol_k,0)) = -
\mu_k (\sol_k,0)$. Similarly, we obtain $-\mu_k \leq E'_-(N_k)$. Since
$E'(N_k)$ exists for $N_k \in \Sigma^c$, we have equality and we
conclude that (\ref{eq-E-mu}) holds.

Next, let us define the map
\begin{equation}
G(\psi, \mu) := \Tm \psi - \big ( \frac{1}{|x|} \ast |\psi|^2 \big ) \psi  + \mu \psi,
\end{equation}     
which is seen to a $\set{C}^\infty$-map $G : \Hrone \times \RR
\rightarrow \Lrp{2}$, see the proof of Proposition~\ref{prop:exists}
and Remark~\ref{rem:solw}(b). Moreover, we note that $G(\sol_*, \mu_*)
= 0$ holds and we have that $\partial_\psi G (\sol_*, \mu_*)$ equals
$L_{11,\freq_*}$ restricted on $\Hrone$. But Assumption~\ref{ass:ker}
implies that $L_{11,\freq_*}$ restricted to $\Hrone$ has trivial
kernel (since $\partial_{x_i} \phi_* \not \in \Hrone$). Thus, we can apply the implicit function theorem to find a
unique $\C{\infty}$-map
\begin{equation}\label{eq-C1}
  (\mu_* - \delta, \mu_* + \delta )
  \longrightarrow U, \quad \mu \longmapsto \sol_\mu,
\end{equation}
for every sufficiently small $\delta > 0$, where $\sol_{\mu_*} =
\sol_*$ and $U$ is some open $\Hrone$-neighborhood around $\sol_*$.

We now show that strong convergence of $\sol_k$ to $\sol_*$ in
$\Hrhalf$ implies strong convergence in $\Hrone$. This can be seen as
follows. Each $\sol_k$ satisfies the equation
\begin{equation}
\sol_k = R_{\mu_k} F(\sol_k),
\end{equation}
where $R_\mu:= (H_0 + \mu)^{-1}$ with $H_0:= \sqrt{-\Delta + m^2} -m$, and $F(\sol) := (|x|^{-1} \ast |\sol|^2 ) \sol$. Therefore we have
\begin{align}
\nrmHp{1}{\sol_k - \sol_\ast } & 
=  \nrmHp{1}{ R_{\mu_k} F(\sol_k) - R_{\mu_*} F(\sol_*) \big } \nonumber \\
& \leq \nrmHp{1}{ (R_{\mu_k} - R_{\mu_*}) (F(\sol_k) + F(\sol_*)) } \nonumber \\
& \quad + \nrmHp{1}{(R_{\mu_k} + R_{\mu_*}) (F(\sol_k) - F(\sol_*))} .
\end{align}
By \eqref{eq-ident}, the fact that $\sol_k \rightarrow \sol_*$ in $\Hrhalf$, and $N_k \searrow N_*$, we see that $\mu_k \searrow \mu_*$ (note that \eqref{eq-E-mu} holds and that $E'(N)$ is strictly decreasing on $\Sigma^c$).  Using now the resolvent identity $R_{\mu_k} - R_{\mu_*} = (\mu_* - \mu_k) R_{\mu_k} R_{\mu_*}$, as well as $\|R_{\mu_k} \|_{\set{L}^2 \rightarrow \set{H}^1} \leq C/\mu_k$, we deduce that
\begin{equation} \label{ineq-H1-strong}
\nrmHp{1}{ \sol_k - \sol_\ast } \leq C \big ( |\mu_k - \mu_*| + 
\nrmH{ \sol_k - \sol_*} \big ) \rightarrow 0 \quad \mbox{as $k \rightarrow \infty$},
\end{equation}
where we also used the local Lipschitz estimate 
\begin{equation}
\nrm{ F(u) - F(v) } \leq \nrmH{ F(u)-F(v) } \leq C ( \nrmH{ u }^2 + \nrmH{v}^2) \nrmH{u-v},
\end{equation}
see \cite[\S3 Lemma 1]{Lenzmann2005a}. In estimate (\ref{ineq-H1-strong}), $C=C(M,\mu_*)$ denotes a suitable constant with $M = \sup_k \nrmH{\sol_k}$.

By the strong convergence of $\sol_k$ to $\sol_*$ in $\Hrone$ shown
above, we thus obtain that $\sol_k \in U$ whenever $k \geq k_0$, where
$k_0$ is sufficiently large.
Moreover, since the left-hand
side of (\ref{eq-ident}) converges to its value at $N_*$, we conclude
that $\mu_k$ converges to $\mu_*$. In addition, by (\ref{eq-E-mu}) and
the strict concavity, we deduce that $\mu_k \searrow \mu_*$ (note
$E'(N)$ has to be strictly decreasing on $\Sigma^c$). In summary, we
find that $\sol_{k_0} \in U$ and $\mu_{k_0} \in (\mu_* -\delta, \mu_*
+ \delta)$ for some $k_0$ and $\mu_{k_0} > \mu_*$. By uniqueness of
the map (\ref{eq-C1}), we see that $\sol_{k_0} = \sol_{\mu_{k_0}}$,
where $\sol_{k_0}$ belongs to the sequence $(\sol_k)$ and
$\sol_{\mu_{k_0}}$ is constructed via the map (\ref{eq-C1}). Hence we
have that the $\C{\infty}$-function
\begin{equation}
f(\mu) := \Nn(\sol_\mu)
\end{equation}
satisfies $f(\mu_*) < f(\mu_{k_0})$. By the mean-value theorem, there
exists some $\xi \in (\mu_* - \delta, \mu_* + \delta)$ such that
$f'(\xi) > 0$. By continuity of $f'$, we conclude that $f'(\mu) > 0$
for all $\mu \in I$ with some open interval $I$ containing $\xi$. This
completes the proof of Lemma \ref{lem-stab}. \end{proof}

\subsection{Completion of the proof of Proposition~\ref{prop:sol}}
\label{sub:done}

In Appendix~\ref{app:spec} we prove
Proposition~\ref{prop:sol}\ref{sol:spec}. Each part is shown for some
small open neighborhood in $\RR^2$ around the point $(\freq_0,0)$.  We
can now complete the proof of Proposition~\ref{prop:sol}.

\begin{proof}[Proof of Proposition~\ref{prop:sol}]
  Let $\freq_0>0$, with minimizer, $\vsol_{\freq_0}$, be a point where
  Assumption~\ref{ass:ker} holds. Above, in Sec.~\ref{sub:stab}, we
  showed that for almost all $N>0$ with corresponding $\freq_*$ there
  is an open non-empty interval $I$ around $\freq_*$ and a unique
  spherically symmetric, real function $\vsolw$ such that
  $\partial_{\freq}\Nn(\vsolw)\geq c$, and that $c$ is independent of
  $\freq$. We thus have singled out an `admissible' $\vsol_{\freq_*}$
  around which we construct our ground states. 

  In Sec.~\ref{sub:e+s} we constructed $\vsolvn$ and its symmetries
  around $\vsol_{\freq_*}$ for $(\speed,\freq)\subset W$, where $W$ is
  some open, non-empty neighborhood in $\RR^2$ and
  $\vel=\speed\hat{\vel}$, and $\hat{\vel}=\vel/|\vel|$. That
  $\freq_*>0$ and $\freq_l(0)=0$ ensure the existence of a, possibly
  smaller, open non-empty set, also denoted $W$ with points so that
  $\freq>\freq_{|\vel|}$ is satisfied.  Thus for all
  $(\speed,\freq)\in W$ we have shown
  Proposition~\ref{prop:sol}\ref{sol:-10}, \ref{sol:1}.

  The proof of the exponential decay \ref{sol:3} (Sec.~\ref{sub:exp})
  does not constraint further the set $W$.

  We now show Proposition~\ref{prop:sol}\ref{sol:2}, \ie the stability
  condition for non-zero velocities. At $\vel=0$ the stability
  condition holds, and since the above constructed $\vsolvn$ depends
  continuously on $\speed$ and $\freq$ so does $\Nn(\vsolvn)$ and there
  exists an open, non-empty, possibly smaller, region $W_1\subset W$
  such that $\partial_{\freq}\Nn(\vsolvn)>c/2$ for all
  $(\speed,\freq)\in W_1$.

  The spectral properties of $\Lv$,
  Proposition~\ref{prop:sol}\ref{sol:spec} (see
  Appendix~\ref{app:spec}) is shown for $(\speed,\freq)\in W_2$, 
  where $W_2$ is some open non-empty neighborhood around
  $(0,\freq_*)$ such that $W_2\subset W_1$.

  Finally, we have this small open and non-empty set $W_2$, where
  \ref{sol:-10}--\ref{sol:spec} hold, we now choose $r_0>0$, and
  an open non-empty interval $I_0$ such that $W_3:=(-r_0,r_0)\times
  I_0$ and $W_3\Subset W_2$.  This concludes the proof of 
  Proposition~\ref{prop:sol}.
\end{proof}

\section{The symplectic form reduced to the soliton manifold}
\label{sec:J}

The purpose of this section is to show that the symplectic form
$\symp{\cdot}{\cdot}$ reduced to a subset of the soliton manifold
$\MfL=\Mf(\zPdom)$ is non-degenerate. The result follows if the matrix
$(\OMpos)_{jk}:=\symp{\vz_{j,\Par}}{\vz_{k,\Par}}$ is invertible. Here
$\vz_{j,\Par}$ are elements in the tangent space $\TMp$, for some
$\Mf\subset\MfL$, defined by (see also \eqref{eq:TM})
\begin{equation}\label{eq:zj}
\{\vz_{1,\Par},\ldots,\vz_{8,\Par}\} := 
\{\partial_{x_1}\vsolp,\partial_{x_2}\vsolp,\partial_{x_3}\vsolp,
\partial_{\vel_1}\vsolp,\partial_{\vel_2}\vsolp,\partial_{\vel_3}\vsolp,
\partial_\phase\vsolp,\partial_\freq\vsolp\}.
\end{equation}

\begin{proposition}\label{prop:J}
  Let $\Pdom(r,I)$ be as in~\eqref{eq:Z} and let $\OMpos$ be defined
  as above.  Under Assumption~\ref{ass:ker}, there are numbers $0<r_1<
  \maxspeed$, $\kappa>0$ and an open non-empty interval $I_1\Subset
  I_0$ such that
\begin{equation}
  \det\OMpos \geq \kappa>0.
\end{equation}
for all $\Par\in\Pdom(r_1,I_1)$.  The constant $\kappa$ depends only
on $r_1$ and $I_1$.
\end{proposition}
Using the definitions~\eqref{eq:Z} and \eqref{eq:aMf} we set
$\oPdom:=\Pdom(r_1,I_1)$ and $\MfS:=\Mf(\oPdom)$. We have now defined
the first number and non-empty interval in the sequence mentioned in
the main theorem. The size of $|\OMpos^{-1}|$ may depend on $r_1$ and
$I_1$, thus the first requirement on $\eps$ is that $|\OMpos^{-1}|
= \Oh(1)$. This is a natural requirement, as we will see in
Proposition~\ref{prop:dyn}.  As a direct consequence of the
non-degeneracy of $\OMpos$ we have the following corollary.
\begin{corollary}\label{cor:nondeg}
  For all $0\neq \vz\in \vTMp_1$, there is at least one element,
  $\tilde{\vz} \in\vTMp$, such that $\symp{\vz}{\tilde\vz}\neq 0$
\end{corollary}

\begin{proof}[Proof of Proposition~\ref{prop:J}]
  By the explicit form of $\symp{\cdot}{\cdot}$ we have with $\Par = (y,v,\phase,\freq)$
\begin{equation}
\OMpos = \OMvos, \  \text{and} \ (\OMvos)_{jk} = -(\OMvos)_{kj}.
\end{equation}
Thus, it suffices to consider elements in the tangent space with
$y=0$, $\phase=0$. By the anti-symmetry of $\OMvos$, it is sufficient
to calculate the upper half triangle of the matrix.  Without loss of
generality we may choose coordinates so that $\vel$ is parallel to
$x_3$-axis, where $x=(x_1,x_2,x_3)$. That is, $\vel = |\vel|\zunit$
and for such $\vel$ we use the notation $\solD$ and $\OMd$.

The determinant of $\OMvos$ will be expressed in terms of $\pv$, $\pf$
and $\pn$, where
\begin{align}\label{eq:a}
  \pv_{jk} := \symp{\partial_{x_j}
    \vsolvn}{\partial_{\vel_k}\vsolvn}, 
  \   &
  \ (\pf)_j := -\symp{\partial_{x_j} \vsolvn}{\partial_\freq
    \vsolvn}, \ j,k=1,2,3, \\ 
  \pn(\solD) & := \frac{1}{2}\partial_{\freq}\nrm{\vsolvn}^2.
  \label{eq:Np}
\end{align}
Here $\Nnum(\freq,\vel):=\Nn(\vsolvn)$.  The relations~\eqref{eq:Lz2}
yield the identities
\begin{align}\label{eq:aprim}
  \pv_{jk} & = \dotp{\Lv
    \partial_{\vel_j}\vsolvn}{\partial_{\vel_k}\vsolvn},
  \ &&
  (\pf)_j =
  \dotp{\partial_{\vel_j}\vsolvn}{\vsolvn}, \ j,k=1,2,3.
\end{align}
The last equation yields $(\pf)_j=\pfj$.  Once the coordinates $\vel=|\vel|\zunit$ are chosen, we obtain $\pvD$, $\pfD$ and $\NnD$, the corresponding notation for $\vsolvn$ is $\solD$.

Each element in $\OMpos$ is an integral of a product between a pair of
tangent vectors.  The reflection symmetry of the tangent vectors,
shown under Assumption~\ref{ass:ker} in Proposition~\ref{prop:sol}, is
the key to this proposition.  We have
\begin{align}\label{eq:sym}
\partial_{x_1}\solD\sim (oee,oeo), \
\partial_{x_2}\solD\sim (eoe,eoo), \
\partial_{x_3}\solD\sim (eeo,eee)\sim J\solD, \\ \nonumber
\partial_{v_1}\solD \sim (oeo,oee),\ \
\partial_{v_2}\solD \sim (eoo,eoe),\ \
\partial_{v_3}\solD \sim (eee,eeo)\sim \partial_\freq\solD .
\end{align}

Let us calculate the cross term $\symp{
  \partial_{x_1} \solD}{ \partial_{x_2} \solD}$. It is an integral
over a product of functions with symmetries
$J\partial_{x_1}\solD\sim (oeo,oee)$ and $(eoe,eoo)\sim
\partial_{x_2}\solD$. Thus both components of $J\partial_{x_1}\solD$
are odd in the first variable whereas $\partial_{x_2}\solD$ is even,
hence the integral over this product vanish. Analogously, most of the
other integrals vanish and by repeated use of~\eqref{eq:sym} and
\eqref{eq:a}--\eqref{eq:aprim} we find the matrix
\begin{equation}\label{eq:vos}
\OMd = \begin{pmatrix} 
0&0&0&\pvD_{11}&0&0&0&0\\ 
0&0&0&0&\pvD_{22}&0&0&0\\ 
0&0&0&0&0&\pvD_{33}&0&-\pfdD\\ 
-\pvD_{11}&0&0&0&0&0&0&0\\ 
0&-\pvD_{22}&0&0&0&0&0&0\\ 
0&0&-\pvD_{33}&0&0&0&\pfdD&0\\ 
0&0&0&0&0&-\pfdD&0&-\NnD\\ 
0&0&\pfdD&0&0&0&\NnD&0\\ 
\end{pmatrix}.
\end{equation}
Its determinant is $\tilde{\kappa}(\vel,\freq):=\det \OMd =
(\pvD_{11}\pvD_{22})^2 (\pvD_{33} \NnD + (\pfdD)^2)^2$. By
Lemma~\ref{lem:pos} below there is an $0<\tilde{r}_1\leq
\maxspeed\leq 1$ such that $\pvD_{jj}>0$, and by Part~\eqref{sol:2} of
Proposition~\ref{prop:sol} $\NnD>0$ for all $|\vel|\leq \maxspeed$ and
$\freq\in I_0$. Thus $\tilde{\kappa}>0$. Now, let $r_{1}<\tilde{r}_1$
and $I_1$ be an open non-empty subinterval of $I_0$ such that
$\bar{I}_1\subset I_0$, then on the closed set
$[0,r_1]\times\bar{I_1}$, $\tilde{\kappa}$ attains its minimum
$\kappa>0$.
\end{proof}
\begin{corollary}\label{cor:OM}
The matrices $\OMvos$ and $\OMvos^{-1}$ have the form
\begin{equation}\label{eq:explOM}
\OMvos = 
\begin{pmatrix} 
0 & \pv & 0 & -\pf \\
-\pv & 0 & \pf & 0 \\
0 & -\pf^T & 0 & -\pn \\
\pf^T & 0 & \pn & 0
\end{pmatrix},
\ 
\OMvos^{-1} = 
\begin{pmatrix} 
0 & -\ipv & 0 & \ipf \\
\ipv & 0 & -\ipf & 0 \\
0 & \ipf^T & 0 & -\iN \\
-\ipf^T & 0 & \iN & 0
\end{pmatrix},
\end{equation}
with $\pv$, $\pf$ and $\pn$ as in~\eqref{eq:a}--\eqref{eq:Np} and
where 
\begin{equation}
\ipv = (\tau+\pn^{-1}\pf \pf^{T})^{-1},\  \
\ipf = (\tau \pn + \pf\pf^{T})^{-1}\pf
\end{equation}
and
\begin{equation}
\iN = \pn^{-1}(-1+\pf^{T}(\tau\pn+\pf\pf^T)^{-1}\pf).
\end{equation}
\end{corollary}
\begin{proof}
  To obtain $\OMvos$, we observe that each block matrix \eg $\pv$,
  $\pf$ $\pn$, is related to the corresponding matrix block in
  Eq.~\eqref{eq:vos} by a change of coordinates. Thus, matrix blocks
  that in~\eqref{eq:vos} are identically zero remain so, and $\pv$,
  $\pf$ and $\pn$ remain as in the general form
  from~\eqref{eq:a}--\eqref{eq:Np}.
\end{proof} 

\begin{lemma}\label{lem:pos}
  Let $\pvD_{jj}$ be as in~\eqref{eq:a}.  There is a number
  $0<\tilde{r}_1\leq\maxspeed$ such that $\pvD_{jj}>0$ for $j=1,2,3$.
\end{lemma}
\begin{proof}
  At $\vel=0$, $\pvD$ reduces to
\begin{equation}
  \left.\pv_{jj}\right|_{\vel=0}=\dotp{\Dsol_j}{\LzT\Dsol_j},
\end{equation}
where $\Dsol_j$ is defined through
$\vDsol_j:=\left.\partial_{\vel_j}\vsolvn\right|_{\vel=0}$ and
$\vDsol_j=(0,\Dsol_j)$. The linear operator $\LzT\geq 0$ has a
non-degenerate zero eigenvalue, with corresponding eigenfunction is
$\solw$, see the proof of Proposition~\ref{prop:specEE}. But, since
$\Dsol_j\sim
\partial_{x_j} \solw$ we have $\Dsol_j\bot \solw$ and hence
$\left.\pv_{jj}\right|_{\vel=0}>0$. By the continuity of $\pvD_{jj}$ in
$\vel$, there is a number $\tilde{r}_{1j}>0$ such that for all
$|\vel|<\tilde{r}_{1j}$ we have that $\pvD_{jj}>0$.  Now let
$\tilde{r}_1=\min_j\tilde{r}_{1j}$.
\end{proof}

\section{Symplectically orthogonal decomposition}
\label{sec:skew}

In this section we introduce the symplectically (or skew) orthogonal
decomposition of a function $\vpsi$ close to the soliton manifold. The
decomposition has two components, one on the manifold,
$\vsol_{\iPar(\vpsi)}$, and one in the symplectically orthogonal
direction, $\vw$. We show that the decomposition uniquely defines the
modulation parameter $\Par=(y,\vel,\phase,\freq)$ and a perturbation
$\vw$.

Recall from Proposition~\ref{prop:J} that the modulation parameter
$\Par$ is a point in the parameter space $\oPdom=\Pdom(r_1,I_1)$ and
\begin{equation}
  \Pdom(r_1,I_1) := \RR^3\times B^3_{r_1}(0)\times [0,2\pi)\times I_1.
\end{equation}
All ground states described by the modulation parameters in $\oPdom$
define the soliton manifold $\MfS:=\Mf(\oPdom)$.  Above,
$B^n_r(0)\subset \mathbb{R}^n$ denotes an open ball of radius $r$ and
$I_1$ is an open interval on $\RR$.  The element $\phase$ in $\Par$ is
a phase, and we can replace its domain $[0,2\pi)$ with $\set{S}^{1}$.
With this replacement we note that only the velocity, $\vel$, and the
frequency, $\freq$, parameter have the constrained domains $B_{r}^3$
and $I_1$ respectively.  The dependence of the solitary waves on the
parameters $\vel$ and $\freq$ requires our attention (see
Corollary~\ref{cor:dep} below) when constructing a `uniform' tubular
neighborhood, $U_\delta$, of the soliton manifold, where the
decomposition exists and is unique.  For subsets
$\Pdom_3\subset\tPdom\subset \oPdom$, to be introduced below, we
define the tubular neighborhood $U_{\delta}=U_{\delta}(\Pdom_j)$ of
$\Mf(\Pdom_j)$ by
\begin{equation}\label{eq:Ud}
  U_\delta(\Pdom_j) := \{\vpsi\in \tilde{\Gamma} 
  : \inf_{\Par\in \Pdom_j}\nrmX{\vpsi-\vsolp} 
  < \delta\}, \ j=2,3.
\end{equation}
Let $b_{R}$ be an open ball around $\vsolp$ in the phase space
$\tilde{\Gamma}$ with radius $R$, defined by
\begin{equation}
  b_R(\vsolp) := \{\vpsi\in \tilde{\Gamma} : \nrmX{\vpsi-\vsolp} < R\},
\end{equation}
where $\tilde{\Gamma}:=\{\vpsi\in \Half: \nrmX{\vpsi}<\infty\}$, see also \eqref{eq:nrmX}.

We have the result:
\begin{proposition}\label{prop:decomp}
  Suppose Assumption~\ref{ass:ker} is satisfied and let $\oPdom$ be
  defined as above. Given $\delta>0$ sufficiently small, and let
  $r_2$ and $r_3$ be such that $0<r_3<r_2<r_1$ and let $I_2$, $I_3$ be
  open non-empty intervals with $\bar{I}_{j+1}\subset I_j$, $j=1,2$,
  with corresponding parameter domains $\tPdom$, $\Pdom_3$ and soliton
  manifolds $\MfT$, $\Mf_3$.  Then for every $\vpsi\in
  U_{\delta}(\Pdom_{k+1})$ and $k=1,2$ there is a unique
  $\C{1}(U_{\delta}(\Pdom_{k+1}),\Pdom_{k})$-map $\iPar$ such that 
\begin{itemize}
\item [(i)]
  For each $\vpsi\in U_\delta(\Pdom_{k+1})$, we have
\begin{equation}\label{eq:decomp}
  \symp{\vpsi - \vsol_{\iPar(\vpsi)}}{\vz} = 0, \ \forall \vz\in 
  \set{T}_{\vsol_{\iPar(\vpsi)}}\Mf_k.
\end{equation}
Furthermore, for each $\vpsi\in U_{\delta}(\Pdom_{k+1})$ there exists
a $\zPar\in\Pdom_k$ such that $\vsol_{\zPar}$ is the orthogonal
projection of $\vpsi$ onto $\Mf_k$, and
\item[(ii)] $\nrmX{\vsol_{\iPar(\vpsi)}-\vsol_{\zPar}} \leq C\delta$,
\item[(iii)] $|\zPar-\iPar(\vpsi)|\leq C\delta$,
\end{itemize}
for some positive constant $C>0$. 
\end{proposition}
The above proposition defines a unique function
$\iPar:U_{\Pdom_{k+1}}\rightarrow \Pdom_{k}$. Consequently,
$\vpsi\mapsto (\iPar,\vw)$ with $\vw:=\vpsi-\vsol_{\iPar(\vpsi)}$
defines a unique decomposition of $\vpsi\in U_{\delta}(\tPdom)$.
\begin{remark}\label{rem:dm}
  Given $r_j$ and $I_j$ for $j=1,2,3$, we have above determined a
  $\delta>0$ such that $\iPar(\vpsi),\zPar(\vpsi)\in \Pdom_{j-1}$,
  when $\vpsi\in U_{\delta}(\Pdom_j)$. Thus we can now give the
  relation between the distances $\delta_{\Mf}$ ($\delta_{P}$), the
  minimal distance between the manifolds and $\delta$ introduced above
  (See Section~\ref{sec:main}, Figure~\ref{fig:domain}). This relation
  is $\delta_{\Mf}\leq c\delta$, $\delta_{P}\leq c\delta$.  For some
  constant $c>0$, which partly is determined by the size of $C$ in part
  (ii) and (iii) above.
\end{remark}
\begin{remark}
  If we choose in Proposition~\ref{prop:decomp} an even smaller
  distance, $\eps<\delta$, and consider the tubular neighborhood
  $U_{\eps}(\Pdom_3)$, then part $(ii)$ and $(iii)$ above hold with
  $\delta$ replaced by $\eps$.
\end{remark}

The proof of Proposition~\ref{prop:decomp} needs four intermediate
results and is given in the end of this section. We define the
function $G:\Half\times \oPdom\rightarrow \RR^8$ by 
\begin{equation}\label{eq:G}
G_j(\vpsi,\Par) := \symp{\vpsi - \vsolp}{\vz_{j,\Par}},\ \ j=1,\ldots, 8,
\end{equation}
where $\vz_{j,\Par}$ is the $j$:th tangent vector in $\TMo$ to the
soliton manifold at the point $\vsolp$, see the ordering given by the
list \eqref{eq:zj}. We will consider solutions to the equation $G=0$
that are close to a solution $(\vpsi,\Par)\mapsto
(\vsol_{\iniPar},\iniPar)$.  We introduce the notation $\iniPar$ to
distinguish an arbitrary parameter $\Par$ from the {\it center
  position}, $\iniPar$, of the ball where we solve the equation $G=0$.
We have the first result:
\begin{lemma}\label{lem:ball}
  Suppose Assumption~\ref{ass:ker} is satisfied. Then, for every
  center position $\iniPar\in\oPdom$ there are balls
  $b_{R_1}(\vsol_{\iniPar})$ and $B_{\rho_1}^8(\iniPar)$ in
  $\tilde{\Gamma}$ and $\oPdom$, respectively, with centers
  $\vsol_{\iniPar}$, $\iniPar$ and radii $R_1$, $\rho_1$ and a unique
  $\C{1}(b_{R_1},B^8_{\rho_1})$ map, such that
  $G(\vpsi,\iPar(\vpsi))=0$ for all $\vpsi\in
  b_{R_1}(\vsol_{\iniPar})$.  Both $R_1$ and $\rho_1$ depend on the
  center position $\iniPar$.
\end{lemma}
\begin{proof}
  We use an implicit function theorem to solve the equation $G=0$.  We
  need to show that $(a)$ $G$ is $\C{1}$, $(b)$
  $G(\vsol_{\iniPar},\iniPar)=0$ and $(c)$ $\left. \partial_{\Par}
    G(\vsol_{\iniPar},\Par) \right|_{\Par=\iniPar}$ is invertible.

  $(a)$ $G$ is $\C{1}$ in $\vpsi$ since it is linear in $\vpsi$. $G$
  is $\C{1}$ in $\Par$ since both $\vsolp$ and $\vz_{j,\Par}$ are
  $\C{1}$ in $\Par$ see Proposition~\ref{prop:sol};

  $(b)$ follows from the definition of $G$;

  $(c)$ calculate
  \begin{equation}\label{eq:cinv}
    \partial_{\Par_k} G_j(\vpsi,\Par) = -\symp{\partial_{\Par_k}\vsolp}
    {\vz_{j,\Par}} + \symp{\vpsi-\vsolp}{\partial_{\Par_k}\vz_{j,\Par}}
\end{equation}
at $(\vpsi,\Par)=(\vsol_{\iniPar},\iniPar)$
\begin{equation}\label{eq:inv}
  \left .\partial_{\Par_k} G_j(\vpsi,\iniPar)\right|_{\vpsi=\vsol_{\iniPar}} = -\symp{\partial_{\Par_k}\vsolp}{z_{\Par,j}}.
\end{equation}
Choose coordinate axis such that $\vel$ is parallel to $x_3$ then
\begin{equation}
\left .\partial_{\Par_k} G_j(\vpsi,\iniPar)\right|_{\vpsi=\vsol_{\iniPar}} 
= -\jOM,
\end{equation}
with $\jOM$ as in Section~\ref{sec:J}.  Thus, by
Proposition~\ref{prop:J}, $\det \jOM>\kappa>0$ and we have shown
$(c)$.

All assumptions in the implicit function theorem are thus satisfied
and, therefore there are open neighborhoods $W\subset \tilde{\Gamma}$
and $V\subset \oPdom$ around $\vsolp$ and $\Par$ respectively and an
unique $\C{1}$-function, $\iPar:W\rightarrow V$ such that for all
$\vpsi\in W$ $G(\vpsi,\iPar(\vpsi))=0$. Now choose $R_1$ sufficiently
small so that $b_{R_1}\subset W$ and $\rho_1$ sufficiently large so
that $V\subset B^8_{\rho_1}$. By possibly reducing $R_1$ further (and
consequently $\rho_1$ by continuity), we find that
$B^8_{\rho_1}\subset \oPdom$.
\end{proof}

To single out a tubular neighborhood $U_\delta$ around the soliton
manifold of constant `width' $\delta$, we examine in the next two
corollaries how the radii in Lemma~\ref{lem:ball} depend on the
parameters. First, by using symmetries of $G$ we have:
\begin{corollary}\label{cor:dep}
  For every $\iniPar\in\oPdom$, the radii of Lemma~\ref{lem:ball} 
  only depend of $\freq$ and $\vel$.
\end{corollary}
\begin{proof}
  The function $G$ is invariant under translation in the sense that if
  a parameter $y$ of $\Par$ maps to $y+a$ and $\vpsi(x)\mapsto
  \vpsi(x-a)$ then the value of $G$ is unchanged. This implies that
  the balls $b_{R_1}$ and $B_{\rho_1}^8$ are independent of which
  position $y$ they are calculated for.

  Analogously, $G$ is phase invariant in the sense that $\phase\mapsto
  \phase + \gamma$ and $\vpsi\mapsto \lexp{-\gamma J}\vpsi$ leave $G$
  unchanged. Thus $R_1=R_1(\vel,\freq)$ and
  $\rho_1=\rho_1(\vel,\freq)$.
\end{proof}

To achieve uniform radii in Lemma~\ref{lem:ball}, we have the following
result:
\begin{corollary}\label{cor:uniform}
  There are a number $r_2'$, $0<r_2'<r_1$, and a non-empty open interval
  $I_2'\Subset I_1$ such that if the $\Par\in \Pdom(r_2',I_2')$, then the
  result of Lemma~\ref{lem:ball} holds with uniform radii.
  Furthermore, for sufficiently small $\rho_2$ there exists $c>0$
  such that
\begin{equation}\label{eq:masterR}
  R_2\leq c\rho_2.
\end{equation}
Uniform here implies that $R_2$, $\rho_2$ (and $c$) only depend on
$r_1$, $r_2'$, $I_1$ and $I_2'$.
\end{corollary}
Let $\tPdom':=\Pdom(r_2',I_2')$. The proof of this corollary is somewhat
tedious and is placed in Appendix~\ref{app:ift}. It relies on two
observations. First, the ground state and maps thereof are well
defined and have a $\C{1}$ dependence on the parameters on the whole
manifold $\MfS$. This allows us to extract uniform radii away from the
boundaries. Second, by choice of $r_2'$ and $I_2$, the only constrained
directions, we are a fixed distance away from the boundaries of $\MfS$
and $\oPdom$, respectively. Thus we can find a uniform radii on this
smaller set.

\begin{figure}[htbp]
\psfrag{b}{$b_{\dio}$}
\psfrag{p}{$\vpsi$}
\psfrag{TM}{$\set{T}_{\vsol_{\Par^{(0)}}}\MfS$}
\psfrag{x}{$\vsol_{\iniPar}$}
\psfrag{x0}{$\vsol_{\Par^{(0)}}$}
\psfrag{xp}{$\vsol_{\iPar(\vpsi)}$}
\psfrag{T}{$\tilde{\set{T}}_{\vsol_{\iPar(\vpsi)}}\MfS$}
\psfrag{M}{$\MfS$}
\psfrag{R}{$\dio$}
\psfrag{z}{$\iniPar$}
\psfrag{z0}{$\Par^{(0)}$}
\psfrag{zp}{$\iPar(\vpsi)$}
\psfrag{Z}{$\oPdom$}
\psfrag{s}{$\Par\mapsto \vsolp$}
\psfrag{r}{$B^8_{c\dio}$}
\psfrag{G}{$\tilde{\Gamma}$}
   \centering
   \centerline{\includegraphics[scale=.75]{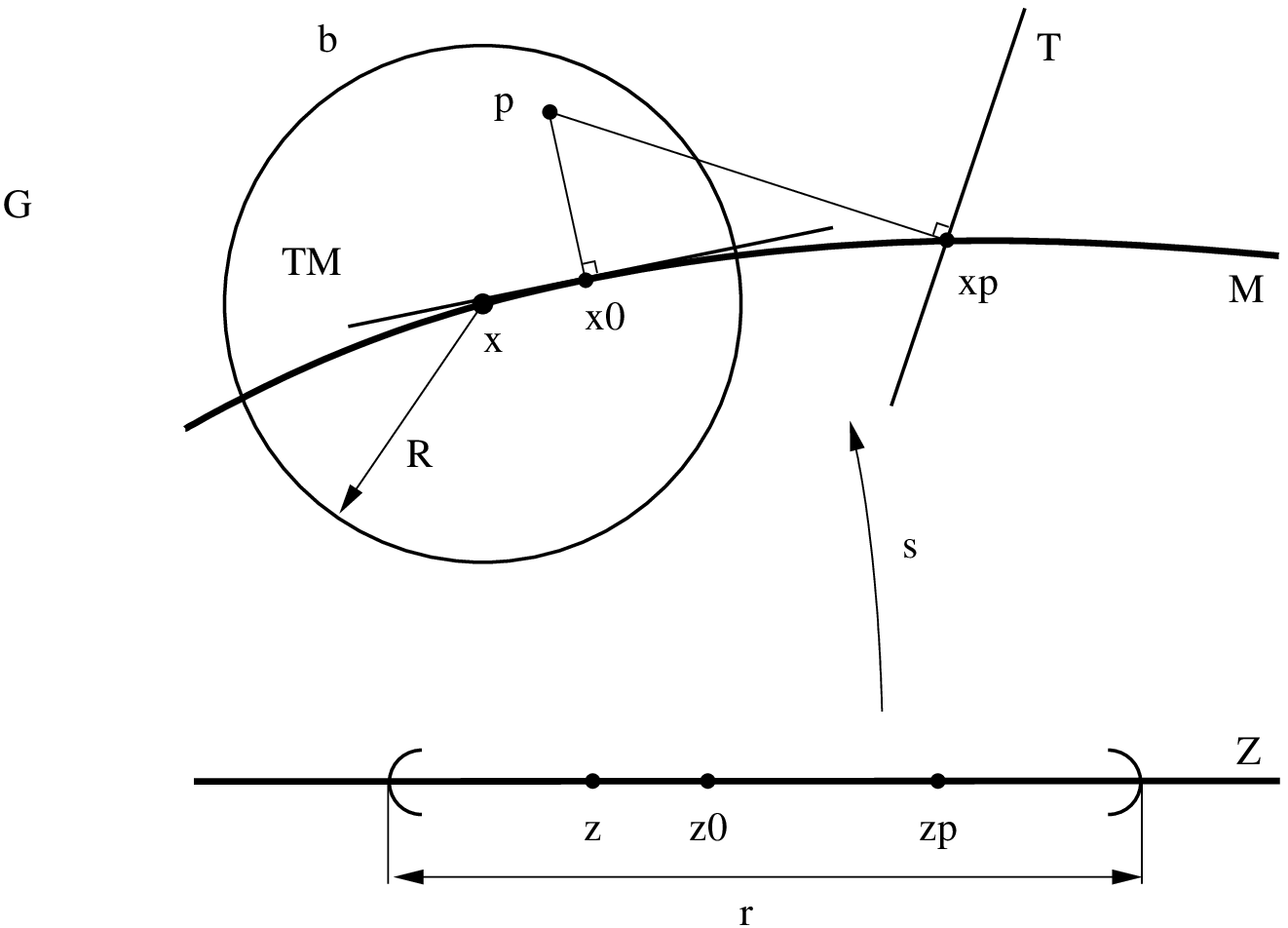}}
   \parbox{\linewidth}{
     \caption{The above figure displays the sets $\oPdom$ and $\MfS$,
       as well as $\set{T}_{\vsol_{\Par^{(0)}}}\MfS$ and the
       symplectically orthogonal plane
       $\tilde{\set{T}}_{\vsol_{\iPar(\vpsi)}}\MfS$. Furthermore, we
       schematically sketch the orthogonal and symplectically
       orthogonal decomposition of $\vpsi$.}
\label{fig:ift}}
\end{figure}

The next lemma is captured by Figure~\ref{fig:ift}.
\begin{lemma}\label{lem:on}
  Suppose that Assumption~\ref{ass:ker} is satisfied. There are 
  numbers $0<\dio<R_2$ and $0<r_2<r_2'$ as well as a non-empty open
  interval $I_2\Subset I_2'$ such that the following holds. If $\vpsi\in
  b_{\dio}(\vsol_{\iniPar})$, $\iniPar\in \Pdom(r_2,I_2)$, then
  there exists a unique $\zPar\in \oPdom$ such that
\begin{enumerate}
\item $\zsol$ minimizes $\nrmX{\vpsi-\vsolp}$;
\item $\nrmX{\soli-\zsol}\leq C\dio$;
\item $\Rnrm{\iPar(\vpsi)-\zPar}\leq C\dio$.
\end{enumerate}
Here $C$ depends only on $r_1$, $r_2$, $r_2'$, $I_1$ and $I_2$, $I_2'$.
\end{lemma}
\begin{proof}
  First note that $\tilde{\Gamma}$ is a Hilbert space. There exists
  $0<r_1''<r_0$ and a non-empty open interval $I_1''\Subset I_0$, with
  corresponding domain $\Pdom_1'':=\Pdom(r_1'',I_1'')$ and
  $\Mf_1'':=\Mf(\Pdom''_1)$ such that for $\vpsi$ that in
  $\nrmX{\cdot}$-norm are sufficiently close to $\Mf_1''$ the orthogonal
  projection of $\vpsi$ onto $\Mf_0$ exists and is unique.  To see
  this, define $f_{\vpsi}(\Par):=\nrmX{\vpsi-\vsolp}^2$.  Observe that
  $f_{\vpsi}\in\C{2}(\Pdom_0,\RR)$, and that $\Pdom_0$ is a convex
  open set. To see that $f_{\vpsi}$ has a minimum, it suffices to show
  that (i) there exists a $\Par^{(0)}=\Par^{(0)}(\vpsi)$ such that
  $f'_{\vpsi}(\Par^{(0)})=0$ and (ii) $f''_{\vpsi}(\Par)>0$, for all
  $\Par\in \Pdom_0$. To show (ii), we calculate
\begin{equation}\label{eq:onproj}
  (f''_{\vpsi}(\Par))_{jk} = \dotX{\vpsi-\vsolp}{\partial_{\Par_j}\vz_{k,\Par}} + 
  \dotX{\vz_{j,\Par}}{\vz_{k,\Par}},
\end{equation}
where $\vz_{k,\Par}:=\partial_{\Par_k}\vsolp$.  The last term,
$(f''_{\vsolp}(\Par))_{jk}=\dotX{\vz_{j,\Par}}{\vz_{k,\Par}}$
simplifies to $\dotX{\vz_{j,\vel,\freq}}{\vz_{k,\vel,\freq}}$, since
phase and translation vanish in the integral. By the reflection
symmetries of $\vz_{k,\vel,\freq}$, see
Proposition~\ref{prop:sol}\ref{sol:1}, together with the fact that
both $|x|$ and $\sqrt{1-\Laplace}$ commutes with rotation and
reflection, we find that $\det f''_{\vsolvn}(\Par)>0$ for $|\vel|\leq
r_1''$, for some $0<r_1''<r_0$ and $\freq\in I_1''$, and $I_1''$ open
an non-empty, such that $I_1''\Subset I_0$. To extend the positivity
of $f_{\vsolvn}''$ to positivity of $f_{\vpsi}''$, we see that the
first term in~\eqref{eq:onproj} can be made small, when
$\nrmX{\vpsi-\vsolp}$ is small. Thus, for $\vpsi$ sufficiently close
to $\Mf(\Pdom(r_1'',I_1''))$ we have $f''_{\vpsi}>0$ and we have proved (ii).
To show (i), we note that $f'_{\vsolp}(\Par)=0$, $f''_{\vsolp}>0$ and
$f'_{\vpsi}\in \C{1}$ thus we can use the implicit function theorem to
find a small neighborhood of $\vsolp$ where there is a $\C{1}$-map
$\Par^{(0)}(\vpsi)$ such that $f_{\vpsi}(\Par^{(0)})=0$.  Furthermore,
this map $\Par^{(0)}$ is unique and we have shown (i). Thus close to
the manifold there exists a unique minimizer.

Now repeating the minimization procedure and replace $\Pdom_{0}$ with
the smaller domain $\tPdom'$ we find that there is a orthogonal
minimizer on $\tPdom''$, where $\tPdom'':=\Pdom(r_2'',I_2'')$ for some
$0<r_2''<r_2'$ and $I_2''$ non-empty and open, $I_2''\Subset I_2'$.
Let $\Mf_2'':=\Mf(\tPdom'')$.

The projection on $\Mf_2''$ is denoted by $\zsol$ and the inverse of
the map $\Par\mapsto \vsolp$ is the coordinate map of the manifold and
uniquely defines $\zPar$. The corresponding radii $R_3$, $\rho_3$ of
balls in $\tilde{\Gamma}$ and $\RR^8$ respectively depends on the
centrum point $\vsol_{\Par^{(c)}}$ respectively $\Par^{(c)}$ around
which the implicit function theorem above is constructed. Analogously
to the proof of Corollary~\ref{cor:uniform} we can reduce the domain
of center positions, $\Par^{(c)}\in \Pdom(r_2,I_2)$, for some
$0<r_2<r_2''$, $I_2\Subset I_2''$ and thus find uniform radii $R_4$,
$\rho_4$ where the decomposition is valid. Moreover, by continuity of
the the map $\Par^{(0)}(\vpsi)$ we have $R_4=c\rho_4$ for some $c>0$. Let $\tPdom:=\Pdom(r_2,I_2)$.

We now show part 2 and 3. Let $\dio=\min(R_4,R_2)$ (where $R_2$ is the
radius where the symplectic decomposition is valid). For $\vpsi\in
b_{\dio}(\vsol_{\Par^{(c)}})$, $\Par^{(c)}\in \tPdom$ both the
orthogonal and the symplectic projection is well defined. We note that
$\zsol\in b_{\dio}(\vsol_{\Par^{(c)}})$ and that
$\nrmX{\soli-\vsol_{\Par^{(c)}}}\leq C'\dio$ thus
\begin{equation}
  \nrmX{\zsol-\soli}\leq C\dio.
\end{equation}
Furthermore, since $\zPar\in B_{c'\dio}^8(\Par^{(c)})$ and $\iPar(\vpsi)\in
B_{c''\dio}^8(\Par^{(c)})$, it follows that $\Rnrm{\zPar -
  \iPar(\vpsi)}\leq C\dio$.
\end{proof}

\begin{proof}[Proof of Proposition~\ref{prop:decomp}]
  We first note that $r_2<r_2'$, $I_2\Subset I_2'$, with corresponding
  $\tPdom=\Pdom(r_2,I_2)$.  With $\dio>0$ as chosen in
  Lemma~\ref{lem:on} we note that $U_{\dio}(\Pdom)$, is the union of
  balls $b_{\dio}(\Isol)$ over all central points in $\iniPar\in
  \tPdom$. By Lemma~\ref{lem:on}, there exists a point $\zPar\in
  \tPdom$ such that $\zsol$ is the orthogonal projection of $\vpsi$
  onto $\Mf_{1}$.  Part (2) and (3) of this lemma ensure that (ii) and
  (iii) of the Proposition~\ref{prop:decomp} is satisfied with
  $\delta=\dio$.

  Lemma~\ref{lem:ball} on the smaller balls, with radii $\dio$ and
  $\rho_2=c\dio$ chosen as above, shows that in each such ball there
  exists a unique $\C{1}$ map $\iPar$, with the property that
  \begin{equation}
    \symp{\vpsi-\soli}{\vz} = 0
  \end{equation}
  for all $\vz\in \set{T}_{\vsol_{\iPar(\vpsi)}}\MfS$.  As shown
  above, $\iPar$ is unique in each ball, hence it is unique in
  $U_{\delta}(\tPdom)$, with $\delta=\dio$.

  That the decomposition exists for $\vpsi$ in an even smaller set,
  $\Mf(\Pdom_3)$, with $\Pdom_3=\Pdom(r_3,I_3)$, $0<r_3<r_2$, $I_3$
  non-empty, open interval, $I_3\Subset I_2$, is clear. To show that
  the resulting decomposition map $\iPar \in \Pdom_2$ rather than in
  $\Pdom_1$ and verify Part (ii) and (iii) of this proposition for
  this smaller set, is done by repeating the above lemmas and
  corollaries with $\Pdom_1$ replaced with $\Pdom_2$. The resulting
  radius $\delta>0$, with $\vpsi\in b_{\delta}(\vsol_{\Par^{(c)}})$
  will be smaller or equal to $\dio$ constructed above.

  Due to the uniform radii constructed above, we find that the
  constant $C$ in the Part~(ii) and (iii) of the proposition depend
  only on the $r$'s and the $I$'s constructed above and not on the points
  $\iPar$ or $\Par^{(0)}$. We have proved the proposition.
\end{proof}

\section{Dynamics in a moving frame}
\label{sec:dyn}

In this section we apply the unique decomposition provided by
Proposition~\ref{prop:decomp} to a class of solutions
of~\eqref{eq:SRH} and find the resulting equations for the decomposed
parts. Another way to see this decomposition is that we make the
change of variables $\vpsi\mapsto (\Par,\vw)$ and derive the equation
of motion for this set of variables. As mentioned above, $\vw$ can be
seen as a {\it perturbation} to a solitary wave parameterized by
$\Par$.

For the decomposition of a solution $\vpsi$ to Eq.~\eqref{eq:SRH} to
exist, we require that $\vpsi$ to remain in the tubular neighborhood
$U_{\delta}(\tPdom)$ for some interval of times. This is ensured by
the requirement that the corresponding initial condition $\vpsi_0$
belongs to $U_{\delta}(\Pdom_3)$. The decomposition is defined by the
unique function $\iPar(\vpsi)$, with $\iPar=(y,\vel,\phase,\freq)$
that solves the equation $G(\vpsi,\Par)=0$, where
$G(\vpsi,\Par):=\symp{\vpsi-\vsolp}{\vz_{\cdot,\Par}}$, and the
relation
\begin{equation}\label{eq:dec}
  \vpsi(x,t)= \lexp{-\phase(t) J}\big(\vsol_{\vel(t),\freq(t)}(x-y(t))
   +\vw(x-y(t),t)\big).
\end{equation}
Thus, the existence of $\iPar$ ensures that $\symp{\vw}{\vz}=0$ for all
$\vz$ in $\set{T}_{\vsolvn}\MfT$. Here
\begin{equation}
\lexp{-\phase J}=
\begin{pmatrix}
\cos \phase & -\sin \phase \\ 
\sin\phase & \cos \phase
\end{pmatrix}.
\end{equation}
The solution $\vpsi$ depends on time, and consequently so does
$\Par(t):=\iPar(\vpsi(\cdot,t))$ and $\vw$. 

By substituting the decomposition~\eqref{eq:dec} into the
Eq.~\eqref{eq:srh} (which is the Hamiltonian formulation of
\eqref{eq:SRH}) and upon applying `projections' of the symplectic form
we have the result:
\begin{proposition}\label{prop:dyn}
  Let $U_\delta(\Pdom_3)$ be defined as above. Suppose that
  Assumption~\ref{ass:ker} is satisfied. Let $\vpsi(\cdot,t)$ be a
  solution to~\eqref{eq:srh} with initial condition $\vpsi_0\in
  U_\delta(\Pdom_3)$ and let $\vw(\cdot,t)$ and $\Par(t)=$
  $(y(t),\vel(t),\phase(t),\freq(t))$ be the decomposed parameters
  corresponding to $\vpsi(\cdot,t)$.  Furthermore, let the external
  potential $V$ satisfy~\eqref{eq:Vass} for some $\eps>0$.  Then, 
\begin{enumerate}
\item [(i)] the parameters $\Par=(y,\vel,\phase,\freq)$ satisfy the {\em
    modulation equations}
\begin{equation}\label{eq:aeqs}
  \alpha_j + \Nn(\vsolvn)\sum_{\nu=1}^3
  (\OMvos)^{-1}_{j\nu}\partial_{y_{\nu}}V(y) = Y_j
\end{equation}
where
\begin{equation} \label{eq:alpha}
\alpha := (\vel-\dot y,\dot \vel,\freq - \dot \phase - V(y),\dot
\freq),
\end{equation}
and the perturbation terms $Y_j$ are given as the right-hand side
of~\eqref{eq:Yterms} and satisfies the estimate
\begin{equation}\label{eq:e_alpha}
  |Y| \leq  C(\nrmH{\vw}^3 + \nrmH{\vw}^2
  + \eps^2 + |\alpha|\nrm{\vw}).
\end{equation}
Furthermore, we have $|\alpha|\leq C\eps + |Y|$.  Here $\OMvos$ is the
matrix~\eqref{eq:explOM} and $\Nnum(\freq,\vel)=\Nn(\vsolvn)$. The
constant $C$ depends only on $r_1$, $r_2$, $I_1$ and $I_2$, see
Proposition~\ref{prop:J} and~\ref{prop:decomp}.
\item [(ii)] Furthermore, the perturbation $\vw$ satisfies the
  equation of motion
\begin{multline}\label{eq:vw}
\dot\vw =J(\Lv\vw + \NL(\vw) + (V_y-V(y))\vw + \VR\vsolvn)
\\-\Big((\vel-\dot y) \cdot \nabla_x (\vsolvn + \vw)
+\dot \vel \cdot \nabla_{\vel} \vsolvn
+\dot\freq\partial_\freq \vsolvn \\
+(\freq -\dot \phase-V(y)) J(\vsolvn+\vw) 
- J x\cdot \nabla_x V(y)\vsolvn  \Big),
\end{multline}
where $\Lv$ is defined in ~\eqref{eq:Lv}, 
\begin{equation}\label{eq:NL}
-\NL(\vw) = (\frac{1}{|x|}*|\vw|^2)\vsolvn + 
\big(\frac{2}{|x|}*(\vsolvn\cdot \vw)\big)\vw + 
(\frac{1}{|x|}*|\vw|^2)\vw
\end{equation}
and
\begin{equation}\label{eq:VR}
  \VR(x):= V_y(x) - V(y) - x\cdot \nabla_y V(y),
\end{equation}
where $V_y(x)=V(x+y)$.
\end{enumerate}
\end{proposition}
\begin{remark}\label{rem:dyn}
  With the explicit form of $\OMvos$ in Corollary~\ref{cor:OM} we
  rewrite Eq.~\eqref{eq:aeqs} and~\eqref{eq:alpha} as
\begin{align}
\gamma\dot\vel + \nabla_y V(y) = \Nn(\vsolvn)^{-1}(Y_4,Y_5,Y_6)^T && 
\dot\freq -\pn^{-1}\pf^{T}\gamma^{-1}\nabla_y V(y) = Y_8 \label{eq:alpha1} \\
\dot y - \vel = -(Y_1,Y_2,Y_3) && \dot\phase -\freq + V(y) = -Y_7,
\label{eq:alpha2}
\end{align}
where $\Nnum:=\Nn(\vsolvn)$, $\pfj = \partial_{\vel_j}\Nn(\vsolvn)$,
$\pn=\partial_{\freq}\Nn(\vsolvn)$,
$\pv_{jk}:=\dotp{\partial_{\vel_j}\vsolvn}{\Lv\partial_{\vel_k}\vsolvn}$
and $\gamma_{jk}:=\Nnum^{-1}(\pv_{jk} + \pn^{-1}\pfj\pfk)$. See also
Corollary~\ref{cor:dyn} below.
\end{remark}
\begin{remark}\label{rem:dv}
  The expression \eqref{eq:vw} is equivalent to \eqref{eq:srh} in the
  moving frame and with the decomposition~\eqref{eq:dec} inserted.
  This equation does contain the information about
  Eqns.~\eqref{eq:alpha1}--\eqref{eq:alpha2}. We can of course remove
  this information from~\eqref{eq:vw} by a `projection'.  But since we
  do not explicitly need this form of Eq.~\eqref{eq:vw}, we have
  refrained from writing out this expression.
\end{remark}

\begin{proof}
  By Proposition~\ref{prop:gamma} and $\vpsi_0\in U_\delta(\Pdom_3)$,
  the solution $\vpsi$ to Eq.~\eqref{eq:srh} satisfies $\vpsi\in
  U_{\delta}(\Pdom_2)$ for some positive times, and the decomposition
  of $\vpsi$ into $(\Par,\vw)$ exists and is unique. For such times we
  express~\eqref{eq:srh} in terms of $\Par,\vw$, with $\Par\in
  \oPdom$. First, we calculate the time derivative of~\eqref{eq:dec}:
\begin{multline}\label{eq:AL}
d_t \vpsi = \lexp{-\phase J}\Big(-
\dot y \cdot \big(\nabla_x \vsolvn(x-y) + \nabla_x \vw(x-y,t)\big)
+\dot \vel \cdot \nabla_{\vel} \vsolvn(x-y) \\ +
\dot \freq \partial_\freq \vsolvn(x-y) 
- \dot \phase J \big(\vsolvn(x-y)+\vw(x-y,t)\big) + \partial_t \vw(x-y,t)\Big).
\end{multline}
We denote the decomposition of $\vpsi$ inserted into the right-hand side of
\eqref{eq:srh} by $A$. That is
\begin{equation}
  A:=J\HV'(\lexp{-\phase J}(\vsolvn(x-y)+\vw(x-y,t))).
\end{equation}
We expand $\HV'$ around $\vsolvn$, with the relations
$\EE'(\vsolvn)=\Hn_{V\equiv 0}'(\vsolvn)+\freq\solvn -\vel\cdot \nabla_x
J\vsolvn$, $\Lv:=\EE''(\vsolvn)$ (for its explicit form see~\eqref{eq:Lv})
and $\EE'(\vsolp)=0$ we find
\begin{multline}\label{eq:AR} 
A = \lexp{-\phase J}J
  \big(\Lv \vw - \freq (\vsolvn + \vw)
  + \vel\cdot \nabla_x J(\vsolvn + \vw)+\NL(\vw)\big)(x-y,t) + \\
  (V(x)-V(y)+V(y))(\vsolvn(x-y)+\vw(x-y,t)),
\end{multline}
where
\begin{equation}
  \NL(\vw):= \HV'(\vsolvn + \vw) - \HV'(\vsolvn) - \HV''(\vsolvn)\vw.
\end{equation}
We insert the explicit form of $\HV$ into the expression for $\NL$
above, simplification gives the result~\eqref{eq:NL}.

The expressions~\eqref{eq:AL} and \eqref{eq:AR} are the right and
left-hand side of~\eqref{eq:srh}. Both sides have a common phase which
we cancel.  Furthermore, both sides also have a common spacial
translation $x\mapsto x-y(t)$, which we remove. That is, we consider
the equation in a moving frame.  Thus, we can rewrite~\eqref{eq:srh}
into the form
\begin{multline}\label{eq:dvw}
  (\vel- \dot y) \cdot \nabla_x ( \vsolvn + \vw)
  +\dot \vel \cdot \nabla_{\vel} \vsolvn +(\freq-\dot \phase - V(y)) J
  \big(\vsolvn+\vw\big) + \dot \freq
  \partial_\freq \vsolvn  \\ -\nabla_y V(y)\cdot x J\solvn 
  + \partial_t \vw = J \big(\Lv \vw +\NL(\vw)
  + (V_y-V(y))\vw) + \VR\vsolvn \big),
\end{multline}
after collecting terms of similar types. Here $V_y(x):=V(x+y)$, and
$\vsolvn$ and $\vw$ are evaluated at $x$ and $x,t$ respectively. Furthermore,
$\VR$ is defined as
\begin{equation}
  \VR(x):= V_y(x) - V(y) - x\cdot \nabla_y V(y).
\end{equation}
Thus~\eqref{eq:dvw} is the desired equation~\eqref{eq:vw} with the
terms somewhat rearranged.  We have showed part (ii) of the
proposition. See also Remark~\ref{rem:dv}.

To show part (i), let $\vz_k\in \set{T}_{\vsolvn}\MfS$ where
$\{\vz_k\}$ are ordered as in~\eqref{eq:zj}, and apply the symplectic
form $\symp{\vz_k}{\cdot}$ to~\eqref{eq:dvw}, then:
\begin{multline}\label{eq:proj}
\symp{\vz_k}{(\vel-\dot y) \cdot \nabla_x (\vsolvn + \vw)
+\dot \vel \cdot \nabla_{\vel} \vsolvn
+(\freq -\dot \phase-V(y)) J(\vsolvn+\vw) +\dot\freq\partial_\freq \vsolvn \\
- \nabla_y V(y)\cdot x J\vsolvn  +
\partial_t\vw}
\\=\symp{\vz_k}{J\Lv \vw + J\NL(\vw)  + J(V_y-V(y))\vw + J\VR\vsolvn}.
\end{multline}
Denote the right-hand side of this equation with $B_k$, we claim that
the term $\symp{\vz_k}{J\Lv\vw}$ vanishes for all $\vz_k\in
\set{T}_{\vsolvn}\MfS$.  To show this claim, first note the identity
$\symp{\vz_k}{J\Lv\vw}=\dotp{\Lv\vz_k}{\vw}$, where we used that $\Lv$
is symmetric.  Secondly, $\Lv\vz_k$ either is zero, or $J\vz_{k'}$ for
some $k'$; see eqns.~\eqref{eq:Lz1} and~\eqref{eq:Lz2}. In the first
case, we have showed the claim, in the latter case recall that $\vw$
satisfies the decomposition conditions~\eqref{eq:decomp}. Thus we have
$\symp{\vz_{k'}}{\vw}=0$ and we have shown the claim. The consequence
is
\begin{equation}\label{eq:Bk}
B_k =\symp{\vz_k}{J\NL(\vw) \\ + J(V_y-V(y))\vw + J\VR\vsolvn}.
\end{equation}
To estimate the nonlinear term in $B_k$, we use the
Hardy-Littlewood-Sobolev inequality~\cite{Lieb+LossII} and a Sobolev
embedding theorem. The remaining terms involve the potential and
Taylor expansions of it, we bound these terms by using the fundamental
theorem of calculus, and eqn.~\eqref{eq:Vass}. We find
\begin{equation}\label{eq:BJ}
  B_k\leq C(\nrmH{\vw}^3+\nrmH{\vw}^2 + \eps^2).
\end{equation}

We now return to~\eqref{eq:proj}, let the infinitesimal generators,
$K$, and their coefficients, $\alpha$, be defined by
\begin{equation}\label{eq:K}
  K:=(\nabla_x, \nabla_{\vel}, J,\partial_\freq), \ \alpha:=(v-\dot y,\dot \vel,\freq-\dot\phase-V(y),\dot\freq).
\end{equation}
We keep the above notation $B=(B_1,\ldots,B_8)$ to represent the
right-hand side of~\eqref{eq:proj}, which by~\eqref{eq:BJ} is a
perturbation for sufficiently small $\vw$ and $\eps$. With the
observations that
$\symp{\vz_k}{\nabla_{x}\vw}=-\symp{\nabla_x\vz_k}{\vw}$,
$\symp{\vz_k}{J\vw}=-\symp{J\vz_k}{\vw}$, $\symp{\vz_k}{\vw}=0$ and
\begin{equation}
0=\partial_{t}\symp{\vz_k}{\vw} = \dot{\vel}\cdot\symp{\nabla_\vel \vz_k}{\vw}+\dot\freq\symp{\partial_{\freq}\vz_k}{\vw} + \symp{\vz_k}{\vw}.
\end{equation}
we re-write~\eqref{eq:proj} as
\begin{equation}
  \sum_j \big((\OMvos)_{kj} - \symp{K_j\vz_k}{\vw}\big)\alpha_j -
  \symp{\vz_k}{\nabla_y V(y)\cdot xJ\vsolvn}  = B_k,
\end{equation}
where $(\OMvos)_{kj}=\symp{\vz_k}{K_j\vsolvn}$ is as in 
Corollary~\ref{cor:OM}. Solving for the leading term in $\alpha$ we
find
\begin{equation}\label{eq:Yterms}
\alpha_j - \sum_k(\OMvos)^{-1}_{jk}\symp{\vz_k}{\nabla_y V(y)\cdot xJ\vsolvn}
 = \sum_k \big(\OMvos)^{-1}_{jk}(B_k + \sum_l \symp{K_l\vz_k}{\vw}\alpha_l \big).
\end{equation}
Denote the right-hand side with $Y_j$, then the uniform lower bound on
$\OMvos$, given by Proposition~\ref{prop:J}, yields that
\begin{equation}
  |Y| \leq C(\eps^2 + \nrm{\vw}|\alpha| + \nrmH{\vw}^2 + \nrmH{\vw}^3
  ).
\end{equation}
and $|\alpha|\leq C\eps + |Y|$. The constant $C$, in both cases,
depends only on $r_1$, $r_2$, $I_1$ and $I_2$.  Since $\vz_k =
K_k\vsolvn$ we find that
\begin{equation}
  \symp{\vz_k}{x\cdot\nabla_y V(y)} = 
  -\delta_{k\nu}\nabla_{y_\nu}V(y)\Nn(\vsolvn),
\end{equation}
for $\nu=1,2,3$. Thus 
\begin{equation}
\alpha_j + \Nn(\vsolvn)\sum_{\nu=1}^3 (\OMvos^{-1})_{j\nu}\nabla_{y_{\nu}}V(y) = 
Y_j,
\end{equation}
and we have proved the proposition.
\end{proof}
We have derived a set of ordinary differential
equations~\eqref{eq:aeqs}. The right-hand side remains small by the
main theorem, provided the decomposition exists. Standard ODE theory shows
that the solution to~\eqref{eq:aeqs} is well defined as long as the
decomposition is well defined.  This agrees with what we expect from
the global well-posedness of the solution $\vpsi$ to \eqref{eq:srh}.

We close this section with yet another form of~\eqref{eq:alpha1}:
\begin{corollary}\label{cor:dyn}
  With the change of variables $(\freq,\vel)\mapsto(\Nn,\Pn)$, through
  $\Nn=\Nn(\vsolvn)$ and $\Pn=\Pn(\vsolvn)$, defined in \eqref{eq:Nn}
  and \eqref{eq:Pn}, the equations for $\dot\vel$ and $\dot\freq$,
  (Eqns.\eqref{eq:alpha1}) take the form
\begin{align}
  d_t \Pn + \Nn\nabla_y V(y) &= (X_{1},X_2,X_3)^T,
\label{eq:dP} \\
d_t \Nn &= X_{8}, \label{eq:dN} 
\end{align}
where each of $X_{j}$, $j=1,2,3,8$ are related to
$Y_j$ above by $X_j=\sum_k(\OMvos)_{jk} Y_k$.
\end{corollary}

\begin{proof}
  Insert $\vz_k=(\nabla_{x}\vsolvn,J\vsolvn)$ into~\eqref{eq:proj} and
  simplify to obtain
\begin{align}
  d_t \Pn(\vsolvn) + \Nn(\vsolvn)\nabla_y V(y) &= (X_1,X_2
  ,X_3)^{T} \\
  d_t \Nn(\vsolvn) & = X_8, 
\end{align}
where
\begin{equation}\label{eq:Xs}
  X_{j} = \symp{K_j \vsolvn}{J(\NL(\vw)+(V_y-V(y))\vw+\VR\vsolvn)}
  + \sum_{k=1}^8\alpha_k \symp{K_k K_j\vsolvn}{\vw},
\end{equation}
in which $K$ is defined by~\eqref{eq:K}.
The change of variables $(\freq,\vel)\mapsto (\Nn,\Pn)$
gives~\eqref{eq:dP} and~\eqref{eq:dN}. To see the relation between $X$
and $Y$, see equation~\eqref{eq:Yterms}.
\end{proof}

\begin{proposition} \label{prop:gamma}
If $\vpsi_0 \in \tilde{\Gamma}$, then the solution of \eqref{eq:SRH} satisfies
\[
\vpsi \in \C{0}\big ( [0,T); \tilde{\Gamma} \big ),
\]
where $T \in (0,\infty]$ is the maximal time of existence. Furthermore, we have that $T= \infty$ holds whenever $\Nn(\vpsi) < \Nc$, for some universal constant $\Nc > 2/\pi$. 
\end{proposition}

\begin{proof}
  For $\tilde{\Gamma} \subset \Half$ replaced by $\Half$, the claim
  follows from the well-posedness results for (\ref{eq:SRH}) proven in
  \cite{Lenzmann2005a} (where also more general $V$'s are treated).

  It remains to show that $t \mapsto |x|^{1/2} \vpsi(t)$ is a
  continuous map from $[0,T)$ into $\Lp{2}$. First, we notice that
  $\vpsi_0 \in \tilde{\Gamma}$ implies that $\vpsi(t) \in
  \tilde{\Gamma}$ for all $0 \leq t < T$. This claim follows in
  particular from a direct adaption of
  \cite{Frohlich+Lenzmann2005}[Lemma A], yielding the formula
\begin{equation} \label{eq:heisen}
\dotp{\vpsi(t)}{|x| \vpsi(t)} = \dotp{\vpsi(0)}{|x| \vpsi(t)} + \int_0^t \dotp{\vpsi(s)}{J [|x|,\sqrt{-\Delta +m^2}] \vpsi(s)} \, \diff s .
\end{equation}  
Here the commutator $[|x|,\sqrt{-\Delta+m^2}]$ is a bounded operator
on $\Lp{2}$; see also \cite{Frohlich+Lenzmann2005} for this.

Moreover, equation \eqref{eq:heisen} shows in particular that $t
\mapsto \| |x|^{1/2} \vpsi(t)Ê\|_2$ is continuous. Assume now that
$t_n \rightarrow t_* \in [0,T)$ is a sequence of times. Then $u_n:=
|x|^{1/2} \vpsi(t_n)$ is a bounded sequence in $\Lp{2}$. By possibly
passing to a subsequence, we have that $u_n$ converges weakly to some
$u_*$ with $\liminf_{n \rightarrow \infty} \nrm{u_n} \geq \| u_* \|_2$. But
since $t \mapsto \| |x|^{1/2} \vpsi(t) \|_2$ is continuous, we have
that $\lim_{n \rightarrow \infty} \| u_n \|_2 = \| u_* \|_2$ holds.
Thus we conclude that $u_n$ actually converges strongly to $u_*$ in
$\Lp{2}$, showing that $t \mapsto |x|^{1/2} \vpsi(t)$ is a continuous
map from $[0,T)$ into $\Lp{2}$.
\end{proof}

\section{Weighted dynamics}
\label{sec:xdyn}

Let $\vpsi$ be a solution to~\eqref{eq:srh} with initial condition
$\vpsi_0\in U_{\eps}(\Pdom_3)$. Then, under Assumption~\ref{ass:ker},
there is, for some positive time, a unique decomposition of $\vpsi$
into $\Par,\vw$ (Proposition~\ref{prop:decomp}).  Furthermore, if we
add that $V$ satisfies~\eqref{eq:Vass} for some $\eps>0$, we find the
equations of motion for $\vw$ in Proposition~\ref{prop:dyn}. Let
$\pos(t):=\dotp{\vw(\cdot,t)|x| \lexp{-\delta_0|x|}}{\vw(\cdot,t)}$.
In this section we use the equations of motions for $\vw$ to show that
the weighted expectation value $\po(t)=\po_{\delta_0=0}(t)$ is well
defined and small.

We have the result:
\begin{proposition}\label{prop:x}
  Let $U_{\eps}(\Pdom_3)$ be defined as above. Let the
  Assumption~\ref{ass:ker} be satisfied, and let $\vpsi$ be a solution
  to~\eqref{eq:srh} with initial condition $\vpsi_0 \in
  U_{\eps}(\Pdom_3)$. Denote its decomposition by $(\Par,\vw)$. Let the
  external potential $V$ satisfy~\eqref{eq:Vass}, for some small
  parameter $\eps>0$.  Then, for times such that the decomposition is
  unique there is a constant $0<c<\infty$ depends only on $r_j$,
  $I_j$, $j=1,2,3$ such that
\begin{multline}\label{eq:x}
  \sup_{s\leq t}\pos(s) \leq \pos(0) + c t\sup_{s\leq t} \big(\pos(s)
  (\nrmH{\vw(s)}+\nrmH{\vw(s)}^2) \\ +
  (\eps+|\alpha(s)|)\nrm{\vw(s)}+
  \nrmH{\vw(s)}^2+\nrmH{\vw(s)}^3\big),
\end{multline}
where $\alpha$ is defined in~\eqref{eq:alpha}.
\end{proposition}
Let $\po(t):=\dotp{\vw(\cdot,t)|x|}{\vw(\cdot,t)}$. From the above
result and the assumption on the initial condition in the main
Theorem~\ref{thm:main} we have
\begin{corollary}\label{cor:x}
  Assume, in addition to the assumptions of Proposition~\ref{prop:x},
  that $\nrmX{\vpsi_0 -\vsol_{\Par_0}}\leq \eps< \delta$. Then there
  is a constant $0<C<\infty$ depending only on $I_j$ and $r_j$ for
  $j=1,2,3$ such that
\begin{equation}\label{eq:xx}
  \sup_{s\leq t}\po(s) \leq C\eps + \sup_{s\leq t}\nrmH{\vw(s)},
\end{equation}
for positive times $t$ such that $\vpsi\in U_\delta(\oPdom)$ and such that time,
$t$, satisfies the inequality
\begin{equation}\label{eq:t1}
  t\leq \frac{1}{2c}\frac{1}{\eps + \sup_{s\leq t}(|\alpha(s)|+
    \nrmH{\vw(\cdot,s)}+\nrmH{\vw(\cdot,s)}^2)},
\end{equation}
with constant $c$ as in Proposition~\ref{prop:x}.
\end{corollary}
\begin{proof}[Proof of Corollary~\ref{cor:x}]
  By assumption $\nrmX{\vpsi_0-\vsol_{\Par_0}}\leq \eps$, thus by the
  definition of $\nrmX{\cdot}$,
  $\nrm{(\eps|x|^{1/2})(\vpsi_0-\vsol_{\Par_0})}\leq \eps$.  Hence
\begin{equation}
\nrmX{\vpsi_0 - \vsol_{\iPar(\vpsi_0)}} \leq \eps + \nrmX{\vsol_{\iPar(\vpsi_0)}-\vsol_{\Par_0}}.
\end{equation}
Part (ii) in Proposition~\ref{prop:decomp} yields that
$\nrmX{\vsol_{\Par_0}-\vsol_{\iPar(\vpsi_0)}}\leq C\eps$. Thus $\pos(0)\leq
\po(0)\leq C\eps$.

As we consider times, {\it the decomposition time}, such that $\vpsi\in
U_\delta(\tPdom)$ (and hence $\Par(t)\in \oPdom$), we have that the
result in Proposition~\ref{prop:x} holds.  Let $t$ be such that
it is smaller than the minimum of the decomposition time and the times
such that~\eqref{eq:t1} holds.  For such times, estimate \eqref{eq:x}
simplifies to
\begin{equation}
  \sup_{s\leq t}\pos(s) \leq C\eps + \frac{1}{2} \nrmH{\vw}.
\end{equation}
The right-hand side is independent of $\delta_0$, we can thus take the
limit to find the result~\eqref{eq:xx}.
\end{proof}
\begin{proof}[Proof of Proposition~\ref{prop:x}]
Notice that 
\begin{equation}
\pos(t) = \pos(0) + \int_0^t d_s \pos(s)\diff s.
\end{equation}
Since $\pos$ is positive for all times, we find
\begin{equation}
\pos(t) \leq \pos(0) + t\sup_{s\leq t}|d_s \pos(s)|.
\end{equation}
The right-hand side is independent under the map $t\rightarrow s$,
$\sum_{s\leq t}$. Thus we find
\begin{equation}
\sup_{s\leq t} \pos(s) \leq \pos(0) + 
t\sup_{s\leq t}|d_s \pos(s)|.
\end{equation}

To bound this we need to estimate $d_s\pos(s)$. As mentioned in the
introduction to this section we have assumed $\vpsi_0\in
U_{\eps}(\Pdom_3)$, thus for some times (to be determined) $\vpsi\in
U_\delta(\tPdom)$. For such $\vpsi$, there is, under
Assumption~\ref{ass:ker} and by Proposition~\ref{prop:decomp}, a unique
decomposition of $\vpsi\mapsto (\vw,\Par)$. 
Proposition~\ref{prop:dyn} yields the equation of motion for $\vw$
in~\eqref{eq:vw}, that is
\begin{multline}
\partial_t \vw 
= J\big(\Lv \vw +\NL(\vw) + (V_y-V(y))(\vw + \vsolvn) \big) 
+(\dot y-v) \cdot \nabla \big(\vsolvn + \vw\big) \\
-\dot v \cdot \nabla_v \vsolvn
-\dot{\freq}\partial_\freq \vsolvn 
-(\freq -\dot \phase-V(y)) J(\vsolvn+\vw) ,
\end{multline}
where $\Lv$ is the $4\times 4$ matrix operator introduced
in~\eqref{eq:Lv}.  We repeat the explicit form of $\Lv$ for clarity.
\begin{align}
  L_{11}\xi_1&:=\sqrt{-\Laplace + m^2}\xi_1 +(-m+\freq)\xi_1 -
  \frac{1}{|x|}*|\vsolvn|^2\xi_1 -
  (\frac{2}{|x|}*(\xi_1 \rsol))\rsol, \\
  L_{12}\xi_2&:=v\cdot \nabla \xi_2-(\frac{2}{|x|}*(\xi_2 \isol))\rsol, \\
  L_{21}\xi_1&:=-v\cdot \nabla \xi_1 -(\frac{2}{|x|}*(\xi_1 \rsol))\isol, \\
  L_{22}\xi_2&:=\sqrt{-\Laplace + m^2}\xi_2 +(-m + \freq)\xi_2 -
  (\frac{1}{|x|}*|\vsolvn|^2)\xi_2 -( \frac{2}{|x|}*(\xi_2
  \isol))\isol,
\end{align}
where we have used the notation $\vsolvn = (\rsol,\isol)^{T}$.  We
regularize $|x|$ by
\begin{equation}
  \fe(x) := |x|\lexp{-\delta_0 |x|}.
\end{equation}
This is a bounded function, and $2\pos:=\dotp{\vw}{\fe\vw}$ is well
defined, since $\vw\in \Ltwo$ and $\fe\in \Lp{\infty}$.

The time derivative of $\pos$ can be expressed as
\begin{equation}
  2d_t \pos = \dotp{\partial_t\vw}{\fe\vw} + \dotp{\vw}{\fe\partial_t\vw}
  =2\dotp{\partial_t \vw}{\fe\vw}.
\end{equation}
Inserting the above equation for $\vw$, we find
\begin{multline}
  d_t \pos = \dotp{J\big(\Lv \vw +\NL(\vw) + (V_y-V(y))(\vw + \vsolvn)
    \big)}{\fe\vw}
   \\
  +\dotp{-\dot v \cdot \nabla_v \vsolvn -\dot{\freq}\partial_\freq
    \vsolvn -(\freq -\dot \phase-V(y)) J(\vsolvn+\vw) }{\fe\vw} \\ 
+\dotp{(\dot y-v) \cdot \nabla(\vsolvn + \vw)}{\fe\vw}.
\end{multline}
To simplify this expression, we note that $\dotp{Ja}{r(x)a} = 0$ for
all bounded scalar functions $r(x)$.  Similarly $2\dotp{\nabla
  a}{r a}= -\dotp{a}{a\nabla r}$, for any scalar function $r(x)$. Thus
\begin{multline}
d_t \pos = 
\dotp{J\big(\Lv \vw +\NL(\vw) \big)}{\fe\vw} 
+\dotp{J (V_y-V(y))\vsolvn}{\fe\vw} 
+\\ (\dot y-v) \cdot (
\dotp{\nabla \vsolvn}{\fe\vw}-\frac{1}{2}\dotp{\vw}{\vw\nabla\fe}
)
\\
-(\freq -\dot \phase-V(y))\dotp{ J\vsolvn }{\fe\vw} 
-\dotp{\dot v \cdot \nabla_v \vsolvn
+\dot{\freq}\partial_\freq \vsolvn}{\fe\vw}.
\end{multline}

To estimate $d_t \pos$, we begin with recalling the definition of
$\alpha$ in eqn.~\eqref{eq:alpha}, thus all the terms terms $\dot y
-\vel$, $\dot \vel$, etc. are bounded by $|\alpha|$.  Furthermore, we
note that $|\nabla\fe| \leq 1$. Thus
\begin{multline}
  |d_t \pos| \leq |\dotp{J\Lv \vw}{\fe\vw}|+|\dotp{\NL(\vw)}{\fe\vw}|
  + C\eps\nrm{|x|\fe \vsolvn}\nrm{\vw}\\
  +|\alpha|(\nrm{\fe\nabla \vsolvn}+\nrm{\fe J\vsolvn}
  +\nrm{\fe\partial_\freq \vsolvn}+\nrm{\fe\nabla_v \vsolvn}) \nrm{\vw}
  +\frac{1}{2}\nrm{\vw}^2.
\end{multline}
By Proposition~\ref{prop:sol}, we know that all terms of the form
$\nrm{\fe \vz_j}$ satisfy $\nrm{\fe \vz_j} \leq \nrm{|x| \vz_j}\leq
C$, where $\vz_j$ is of the form $K\vsolvn$ and
$K\in\{|x|,\nabla_x,J,\partial_\freq,\nabla_{\vel}\}$, and $C$ being
independent of $\delta_0$. Hence
\begin{equation}\label{eq:xQ}
|d_t \pos| \leq 
|\dotp{J\Lv \vw}{\fe\vw}| +|\dotp{J\NL(\vw)}{\fe\vw}| +
 C (\eps+|\alpha|)\nrm{\vw}+\frac{1}{2}\nrm{\vw}^2.
\end{equation}

To estimate the term $\dotp{\Lv \vw}{\fe\vw}$, we write down this
expression in detail
\begin{multline}
\dotp{J\Lv \vw}{\fe \vw} = 
\dotp{\begin{pmatrix} 0 & 1 \\ -1 & 0 \end{pmatrix}
\begin{pmatrix} L_{11} & L_{12} \\ L_{21} & L_{22}\end{pmatrix}\vw}{\fe\vw}
=\dotp{\begin{pmatrix} L_{21} & L_{22} \\
 -L_{11} & -L_{12}\end{pmatrix}\vw}{\fe\vw} \\ 
=\dotp{L_{21}\rw + L_{22}\iw}{\fe\rw} - \dotp{L_{11}\rw +L_{12}\iw }{\fe\iw} .
\end{multline}
Inserting the explicit expressions for the operator $\Lv$
yields
\begin{multline}
\dotp{L_{12}\iw}{\fe\iw} = 
\dotp{v\cdot \nabla\iw}{\fe \iw} - 
\dotp{\frac{2}{|x|}*(\isol \iw)\rsol}{\fe\iw}
\\ =-\frac{v}{2}\cdot \dotp{\iw}{\iw\nabla \fe} - 
\dotp{\frac{2}{|x|}*(\isol \iw)}{\fe\rsol\iw}.
\end{multline}
Once again we observe that $\sup_x |\nabla \fe|\leq 1$ holds,
independent of $\delta_0$. Using the Hardy-Littlewood-Sobolev
inequality we thus find
\begin{multline}
|\dotp{L_{12}\iw}{\fe\iw}| \leq 2|\vel|\nrm{\iw}^2
+ C\nrmLp{12/5}{\isol\iw} \nrmLp{12/5}{\fe\rsol\iw}
\\\leq 2|\vel|\nrm{\iw}^2+C\nrmLp{12/5}{\iw}^2
(\nrmLp{\infty}{\isol}^{12/5}\nrmLp{\infty}{\fe\rsol}^{12/5})^{5/12}.
\end{multline}
Observe that $\fe\leq |x|$, $\sup ||x|\vsolvn|<C$ and that
$\nrmLp{12/5}{\vw}\leq C\nrmH{\vw}$. We find that
\begin{equation}
|\dotp{L_{12}\iw}{\fe\iw}| \leq 
2|\vel|\nrm{\iw}^2 + C\nrmH{\iw}^2.
\end{equation}

For the $L_{21}$-term we have
\begin{multline}
\dotp{L_{21}\rw}{\fe\rw} = 
\dotp{-v\cdot \nabla\rw}{\fe \rw} - 
\dotp{\frac{2}{|x|}*(\rsol \rw)\isol}{\fe\rw}
\\=\frac{v}{2}\cdot \dotp{\rw}{\rw\nabla \fe} - 
\dotp{\frac{2}{|x|}*(\rsol \rw)}{\fe\isol\rw}.
\end{multline}
Similar to the estimate of $L_{12}$, we obtain that
\begin{equation}
|\dotp{L_{21}\rw}{\fe\rw}| \leq 
2|\vel|\nrm{\rw}^2 + C\nrmH{\rw}^2.
\end{equation}

The last two terms yield 
\begin{multline}
\dotp{L_{22}\iw}{\fe\rw}-\dotp{L_{11}\rw }{\fe\iw} = \\
\dotp{\sqrt{-\Laplace + m^2}\iw +(-m + \freq)\iw -
  (\frac{1}{|x|}*|\vsolvn|^2)\iw -( \frac{2}{|x|}*(\iw
  \isol))\isol}{\fe\rw} \\ - 
\dotp{\sqrt{-\Laplace + m^2}\rw +(-m+\freq)\rw -
  (\frac{1}{|x|}*|\vsolvn|^2)\rw -
  (\frac{2}{|x|}*(\rw \rsol))\rsol}{\fe\iw} \\
= \dotp{[\fe,\sqrt{-\Laplace + m^2}]\iw}{\rw}  -\dotp{( \frac{2}{|x|}*(\iw
  \isol))\isol}{\fe\rw}  \\ + 
\dotp{  (\frac{2}{|x|}*(\rw \rsol))\rsol}{\fe\iw},
\end{multline}
where $[A,B]=AB-BA$.  The last two terms are both bounded by
$C\nrmH{\vw}^2$, analogous to the estimate for $L_{12}$. For the first
term we use~\cite[Lemma A.3]{Frohlich+Jonsson+Lenzmann2005}, see also
Stein~\cite{Stein1993}, that shows
\begin{equation}
|\dotp{[\fe,\sqrt{-\Laplace+m^2}]\iw}{\rw}| \leq C\nrm{\vw}^2,
\end{equation}
with $C$ independent of $\delta_0$.  Thus we find, using that $|v|\leq
1$,
\begin{equation}\label{eq:xL}
|\dotp{J\Lv\vw}{\fe\vw}|\leq C\nrmH{\vw}^2.
\end{equation}

The last term to estimate is $\dotp{\NL(\vw)}{\fe\vw}$. To this end,
we recall from Proposition~\ref{prop:dyn} that
\begin{equation}
-\NL(\vw) = \frac{1}{|x|}*|\vw|^2\vsolvn + 
\frac{2}{|x|}*(\vsolvn\cdot \vw)\vw + 
\frac{1}{|x|}*|\vw|^2\vw.
\end{equation}
In this case we cannot use the Hardy-Littlewood-Sobolev estimate.  But
instead we can use the Kato~\cite[\S V.5.4, eq.  (5.33)]{Kato1995}
inequality:
\begin{equation}\label{eq:Kato}
  \int_{\RR^3} \frac{1}{|y|}|\vw(y)\cdot \vu(y)| \diff y 
  \leq C \nrmH{\vu}\nrmH{\vw};
\end{equation}
see \eg~\cite{Herbst1977}.  We estimate
$\dotp{\NL(\vw)}{\fe\vw}$ as follows
\begin{multline}
  |\dotp{\NL(\vw)}{\fe\vw}|\leq \nrmLp{1}{\fe |\vw|^2} 
  \sup_{x} \int_{\RR^3} (\frac{1}{|y|}(|\vw(x+y)|^2 
  + |\vw(y+x)\cdot\vsolvn(y+x)|))\diff y 
  \\ + \nrmLp{1}{\fe\vsolvn\cdot \vw}
  \sup_x\int_{\RR^3}\frac{1}{|y|}|\vw(y+x)|^2 \diff y.
\end{multline}
Using~\eqref{eq:Kato} we find
\begin{equation}
  |\dotp{\NL(\vw)}{\fe\vw}|\leq \nrmLp{1}{\fe |\vw|^2} 
  (\nrmH{\vw}^2 + \nrmH{\vw}\nrmH{\vsolvn}) 
   + \nrmLp{1}{\fe\vsolvn\vw}\nrmH{\vw}^2.
\end{equation}
Note that $\nrmLp{1}{\fe|\vw|^2} = \pos(t)$ and
$\nrmLp{1}{\fe\vsolvn\cdot \vw}\leq C\nrm{\vw}$, where $C$ is
independent of $\delta_0$.  Thus
\begin{equation}\label{eq:xN}
|\dotp{\NL(\vw)}{\fe\vw}| \leq 
c\nrmH{\vw}^3
+c\nrmH{\vw}(1+\nrmH{\vw})\pos(t),
\end{equation}
where $c$ is independent of $\delta_0$.

Inserting the results in \eqref{eq:xL} and \eqref{eq:xN} into
\eqref{eq:xQ} yields
\begin{equation}
|d_t \pos| \leq 
 C\nrmH{\vw}^2 +c\nrmH{\vw}^3
+c\nrmH{\vw}(1+\nrmH{\vw})\pos + 
 C (\eps+|\alpha|)\nrm{\vw}.
\end{equation}
This concludes the proof of the proposition.
\end{proof}

\section{Estimates of the Lyapunov functional from below}
\label{sec:low}

In this section, we define a Lyapunov
functional $\Lyp=\Lyp(t)$ as
\begin{equation}
\Lyp := \hL(\vpsi) - \hL(\vsolp).
\end{equation}
Here $\hL$ is defined by
\begin{equation}
\hL(\vpsi) := (\freq-V(y)) \Nn(\vpsi) 
- \frac{1}{2}\vel\cdot\dotp{\nabla J \vpsi}{\vpsi} + \HV(\vpsi)
\end{equation}
where $\Nn(\vpsi):=\frac{1}{2}\nrm{\vpsi}^2$, and
\begin{equation}
\HV(\vpsi):=\frac{1}{2}\dotp{\vpsi}{(\sqrt{-\Laplace + m^2}-m)\vpsi} +
\frac{1}{2}\dotp{V\vpsi}{\vpsi} -\dotp{\frac{1}{4|x|}*|\vpsi|^2}{|\vpsi|^2}.
\end{equation}
The function $\hL$ is a linear combination of conserved and almost
conserved quantities, $\HV$, $\Nn$, and $\Pn$.

The parameters $\Par=(y,\vel,\phase,\freq)$ above are chosen such that
if $\vpsi$ is decomposed then $\Par=\iPar(\vpsi)$ (see
Proposition~\ref{prop:decomp}, for the construction of $\iPar$).  In
this section we show that this Lyapunov functional is coercive up to
small corrections.  This will be used to bound the perturbations $\vw$
from above.

We recall the notation and a result shown in previous sections: If
$\vpsi\in U_{\delta}(\tPdom)$ then, under Assumption~\ref{ass:ker},
there exists a unique decomposition of $\vpsi\rightarrow
(\Par=(y,\vel,\phase,\freq),\vw)$ by $\Par\in \oPdom$ (see
Proposition~\ref{prop:decomp}).

We have the result:
\begin{proposition}\label{prop:low}
  Let $\vpsi\in U_\delta(\tPdom)$ and let Assumption~\ref{ass:ker} be
  satisfied. Denote the unique decomposition of $\vpsi$ by
  $(\Par,\vw)$.  Let the external potential $V$
  satisfy~\eqref{eq:Vass} for some number $\eps>0$ and let
  $\rho=\rho(r_1,I_1)>0$, be defined as in
  Proposition~\ref{prop:Lpos}.  Then
\begin{equation}
  \Lyp \geq \frac{7}{8}\rho \nrmH{\vw}^2 -C\eps\po- C \eps^2
  - C\nrmH{\vw}^4,
\end{equation}
with $\po:=\dotp{\vw}{|x|\vw}$.
\end{proposition}
\begin{remark}
  The major limitation of $\eps$ appears here. The lower bound,
  $\rho$, depends on the distance from zero to the start of the
  essential spectrum and is hence of size $\ell_{sol}^{-1}$, whereas
  the upper bound (next section) is given in terms of gradients of the
  potential, initial distances both parameterized by $\eps$ together
  with the so far unknown size of the perturbation $\vw$.  Thus the
  requirement that $\eps = \ell_{sol}/\ell_{pot}\ll 1$ arises here.
\end{remark}
\begin{proof}
  Using the decomposition (Proposition~\ref{prop:decomp}) of $\vpsi$
  into $\Par,\vw$, with $(y,\vel,\phase,\freq)=\Par:=\iPar(\vpsi)$ we
  can write $\vpsi$ as $U_\delta(\tPdom)\ni\vpsi(x,\cdot) = \lexp{-J
    \phase}(\vsol_{\vel,\freq}(x-y) + \vw(x-y,\cdot))$. Inserting this
  into $\Lyp$ gives
\begin{multline}\label{eq:F}
\Lyp = \hL(\vpsi)-\hL(\vsolp) = \dotp{\hL'(\vsolvn(\cdot-y))}{\vw(\cdot -y)}\\ + 
\frac{1}{2}\dotp{\hL''(\vsolvn(\cdot -y))\vw(\cdot -y)}{\vw(\cdot-y)} + 
\Rest(\vw,\vsolvn)\\ = A + B + \Rest(\vw,\vsolvn).
\end{multline}
Here we define $A$ as 
\begin{equation}\label{eq:AAA}
A:= \dotp{\hL'(\vsolvn(\cdot-y))}{\vw(\cdot -y)},
\end{equation}
and $B$ as
\begin{equation}\label{eq:BBB}
B:=\frac{1}{2}\dotp{\hL''(\vsolvn(\cdot -y))\vw(\cdot -y)}{\vw(\cdot-y)}.
\end{equation}
The remainder, $\Rest(\vw,\vsolvn)$ is defined as
\begin{multline}
\Rest(\vw,\vsolvn) := \frac{1}{4}\dotp{\frac{1}{|x|}*(|\vsolvn+\vw|^2)}{|\vsolvn+\vw|^2} - 
  \frac{1}{4}\dotp{\frac{1}{|x|}*(|\vsolvn|^2)}{|\vsolvn|^2} \\-
  \dotp{\frac{1}{|x|}*(|\vsolvn|^2)}{\vsolvn\cdot \vw} - 
  \frac{1}{2}\dotp{\frac{1}{|x|}*(|\vsolvn|^2)}{|\vw|^2} \\ -
  \dotp{\frac{1}{|x|}*(\vsolvn\cdot \vw)}{\vsolvn\cdot \vw},
\end{multline}
$\Rest$ can also be defined directly from~\eqref{eq:F} as the
remainder of the their given Taylor expansion of $\hL(\vpsi)$ around
$\vsolvn$ to second order. Thus the rest term contains only the Taylor
expansion of the nonlinear term in $\hL$ which is what is
written out above in detail. By expansion of the polynomials, $\Rest$
simplifies to
\begin{equation}\label{eq:rest}
\Rest(\vw,\vsolvn) = -\dotp{\frac{1}{|x|}*|\vw|^2}{\vsolvn\cdot \vw} -
\frac{1}{4}\dotp{\frac{1}{|x|}*|\vw|^2}{|\vw|^2}.
\end{equation}

We now proceed to estimate the terms $A$, $B$ and $\Rest$. We begin
with $\Rest$.  The Hardy-Littlewood-Sobolev
inequality yields
\begin{equation}
|\Rest(\vw,\vsolvn)| \leq c ( \nrmLp{12/5}{\vw}^4 + \nrmLp{12/5}{\vw}^3).
\end{equation}
From the Sobolev inequality we have $\nrmLp{12/5}{f}\leq
C\nrmHp{1/4}{f}\leq C\nrmH{f}$.
Hence,
\begin{equation}
|\Rest(\vw,\vsolvn)| \leq c' ( \nrmH{\vw}^4 + \nrmH{\vw}^3).
\end{equation}
Cauchy's inequality in the form $2ab \leq \eta a^2 + \eta^{-1}b^2$ finally
yields that
\begin{equation}\label{eq:estR}
|\Rest(\vw,\vsolvn)| \leq C \nrmH{\vw}^4 
+ \frac{\rho}{16}\nrmH{\vw}^2.
\end{equation}

To estimate $A$ as defined in~\eqref{eq:AAA}, we relate it to 
$\EE'$. The functional $\EE$ is defined in~\eqref{eq:EE} as
\begin{equation}
\EE(\vpsi):= \freq \Nn(\vpsi) 
- \frac{1}{2}\vel\cdot\dotp{\nabla J \vpsi}{\vpsi} + \Hn_{V=0}(\vpsi).
\end{equation}
All groundstates $\vsolvn$ satisfy~\eqref{eq:EEp}, that is
\begin{equation}
\EE'(\vsolvn)=0.
\end{equation}
We write out the terms in $A$ explicitly and identify
$\EE'(\vsolvn)$. This gives
\begin{multline}\label{eq:Aid}
  A=\dotp{\hL'(\vsolp(\cdot-y))}{\vw(\cdot-y)} = (\freq-V(y))
  \dotp{\vsolvn}{\vw} - \vel\cdot\dotp{\nabla J \vsolvn}{\vw} \\+
  \dotp{\Hn'_{V=0}(\vsolvn)}{\vw} + \dotp{V_y\vsolvn}{\vw} =
  \dotp{\EE'(\vsolvn)}{\vw} +\dotp{(V_y-V(y))\vsolvn}{\vw}.
\end{multline}
Using that $\EE'(\vsolvn)=0$ and $\dotp{\vsolvn}{\vw}=0$, together
with that $V$ satisfies~\eqref{eq:Vass} for some small $\eps$, we find that
\begin{equation}\label{eq:A}
|A| =|\dotp{(V_y-V(y))\vsolvn}{\vw}|\leq c_1\eps\nrm{\vw}. 
\end{equation}

Next we estimate $B$ defined in~\eqref{eq:BBB}. We also rewrite $B$ in
terms of $\Lv:=\EE''(\vsolvn)$. A calculation, similar to the one
above, shows the relation
\begin{equation}\label{eq:Lrel}
\dotp{\vw(\cdot-y))}{\hL''(\vsolvn(\cdot - y))\vw(\cdot-y)} =  
\dotp{\vw}{\Lv \vw} + \dotp{\vw}{(V_y-V(y))\vw}
\end{equation}
Denote 
\begin{equation}
B_1:=\dotp{\vw}{(V_y-V(y))\vw}.
\end{equation}

On the space where $w$ is symplectically orthogonal to $\vTM$, we have 
\begin{equation}\label{eq:estL}
\dotp{\vw}{\Lv \vw} \geq \rho \nrmH{\vw}^2 
\end{equation}
which is shown in Appendix~\ref{app:Lpos}. Note that $\rho$ here and
in Proposition~\ref{prop:Lpos} depends only on $r_1$ and $I_1$. 

To bound $B_1$, we expand $V_y$ around $y$ to obtain
\begin{equation}
|B_1|\leq C\eps \nrm{|x|^{1/2}\vw}^2. 
\end{equation}
Hence $B$ obeys the lower bound
\begin{equation}\label{eq:B}
B \geq \rho \nrmH{\vw}^2 
- C\eps\nrm{|x|^{1/2}\vw}^2.
\end{equation}
To complete the proof, we use estimates \eqref{eq:estR}, \eqref{eq:A}
and \eqref{eq:B} to obtain
\begin{equation}
\Lyp \geq \rho \nrmH{\vw}^2 - 
C\eps\nrm{|x|^{1/2}\vw}^2 -c_1\eps\nrm{\vw} 
-  C  \nrmH{\vw}^4 - \frac{\rho}{16}\nrmH{\vw}^2.
\end{equation}
By using $c_1 \eps \nrm{\vw}\leq 4c_1^2 \eps^2/\rho +
(\rho/16)\nrm{\vw}^2$ in the above equation we conclude the proof.
\end{proof}

\section{Estimates of the Lyapunov functional from above}
\label{sec:up}

In this section we show that the above defined Lyapunov functional
$\Lyp$ is almost conserved, to cubic order in terms of small
quantities. First, we recall that, mass, energy are
conserved and that the momentum satisfies the Ehrenfest identity \ie
\begin{equation}\label{eq:cons}
d_t \Nn = 0, \ d_t \dotp{J\nabla \vpsi}{\vpsi} = -\dotp{\vpsi}{(\nabla V)\vpsi}
, \ d_t \HV = 0.
\end{equation} 
The conservation laws are proved in~\cite{Lenzmann2005a} and for
Ehrenfest's lemma see the comment after~\eqref{eq:Ehr}. Once again,
recall that the Lyapunov-Schmidt functional is defined as
\begin{equation}
\Lyp:=\hL(\vpsi) - \hL(\vsolp),
\end{equation}
with $\Par=(y,\vel,\phase,\freq)=\iPar(\vpsi)$, provided that $\iPar$
exists (see Proposition~\ref{prop:decomp}), and where
\begin{equation}
\hL(\vpsi) := (\freq-V(y)) \Nn(\vpsi) 
- \frac{1}{2}\vel\cdot\dotp{\nabla J \vpsi}{\vpsi} + \HV(\vpsi).
\end{equation}

We can now state the following result.
\begin{proposition}\label{prop:high}
  Let Assumption~\ref{ass:ker} be satisfied and let $\vpsi$ be a
  solution to~\eqref{eq:srh}, with initial condition $\vpsi_0\in
  U_\eps(\Pdom_3)$, $\eps\leq \delta$. The decomposition of $\vpsi$, which exists for
  some times, is denoted by $(\Par=(y,\vel,\phase, \freq),\vw)$.  Let
  $\alpha:=(\dot y-\vel,\dot\vel, \freq- \dot\phase-V(y),\dot\freq)$,
  see eqn.~\eqref{eq:alpha}, and let the external potential $V$
  satisfy~\eqref{eq:Vass} for some small parameter $\eps>0$.
  Finally, let $\Lyp(t)$ be defined as above. Then
\begin{equation}
|\Lyp(t)| \leq  |\Lyp(0)|+C t\sup_{s\leq t} \big(
(\eps + |\alpha(s)|)
(\eps^2 + \nrmH{\vw(\cdot,s)}^2) 
+ \nrmH{\vw(\cdot,s)}^3(1+\nrmH{\vw(\cdot,s)}^2)\big),
\end{equation}
where $C$ depends only on $r_j$, and $I_j$, $j=1,2,3$.
\end{proposition}
Using Proposition~\ref{prop:decomp}, we have the following corollary.
\begin{corollary}\label{cor:high}
  In addition to the assumptions in
  Proposition~\ref{prop:high}, we assume that
  $\nrmX{\vpsi_0-\vsol_{\Par_0}}\leq \eps<\delta$, for $\eps$ sufficiently
  small. Then
\begin{equation}
|\Lyp(t)| \leq  C\eps^2+C t\sup_{s\leq t}\big( 
(\eps + |\alpha(s)|)
(\eps^2 + \nrmH{\vw(\cdot,s)}^2) 
+ \nrmH{\vw(\cdot,s)}^3(1+\nrmH{\vw(\cdot,s)}^2)\big),
\end{equation}
\end{corollary}
\begin{proof}
Expanding $\Lyp(0)=\left.\hL(\vpsi)-\hL(\vsolp)\right|_{t=0}$, around
the soliton yields
\begin{multline}
|\Lyp(0)|\leq |\dotp{(V_y-V(y))\vsolvn}{\vw}| + |\dotp{\Lvz \vw}{\vw}|
+ |\dotp{(V_y-V(y))\vw}{\vw}| \\+ |\Rest(\vw,\vsolz)| \big|_{t=0}.
\end{multline}
Here $\Rest$ is defined in~\eqref{eq:rest}. Using estimate
\eqref{eq:estR} yields
\begin{equation}
|\Lyp(0)|\leq C(\eps^2 + (1+\eps)\nrmH{\vw}^2 + 
\nrmH{\vw}^4)\big|_{t=0}
\end{equation}
To estimate $\nrmH{\vw}\big|_{t=0}$, we recall
from Proposition~\ref{prop:decomp}(ii) with $\vpsi_0 \in
U_{\eps}(\Pdom_3)$ that
\begin{equation}
  \nrmH{\vw}\big|_{t=0} = \nrmH{\vpsi_0-\vsol_{\iPar(\vpsi_0)}}
  \leq \nrmH{\vpsi _0- \vsol_{\Par_0}} + \nrmH{\vsol_{\Par_0}
  -\vsol_{\iPar(\vpsi_0)}} \leq C\eps.
\end{equation}
Here $C$ depends only on $I_j$ and $r_j$, $j=1,2,3$.  Since
$\eps^4\leq C\eps^2$, we find that
\begin{equation}
|\Lyp(0)|\leq C\eps^2,
\end{equation}
whenever $\eps$ is sufficiently small.
\end{proof}

\begin{proof}[Proof of Proposition~\ref{prop:high}]
  The proof of this proposition is a straightforward calculation.
  Using that
\begin{equation}
\Lyp(t)-\Lyp(0) = \int_0^t d_s \Lyp(s)\diff s,
\end{equation}
we find 
\begin{equation}
|\Lyp(t)|\leq |\Lyp(0)| + t \sup_{s\leq t} |d_s \Lyp(s)|.
\end{equation}
Thus the desired result corresponds to controlling $|d_t \Lyp|$ in
terms which are of third order or higher in $\eps$, $|\alpha|$ and
$\nrmH{\vw}$.

By Proposition~\ref{prop:gamma} and $\vpsi_0\in U_\delta(\Pdom_3)$
there is some positive time such that the solution $\vpsi$
to~\eqref{eq:srh} satisfies $\vpsi\in U_\delta(\Pdom_2)$
to~\eqref{eq:srh} and the it has a unique decomposition into
$\Par,\vw$.

We now calculate the time derivative of $\hL(\vpsi)$ for a solution
$\vpsi$ to~\eqref{eq:srh}. By~\eqref{eq:cons}, we find
\begin{equation}\label{eq:dtH1}
d_t \hL(\vpsi) = \Nn(\vpsi)d_t (\freq 
- V(y))- 
\frac{1}{2}\dot \vel\cdot \dotp{J\nabla \vpsi}{\vpsi} + \frac{1}{2}\vel\cdot \dotp{\vpsi}{\nabla V \vpsi}. 
\end{equation}
Here $\freq$, $\vel$ and $y$ are taken from the decomposition of
$\vpsi$.

Next, we rewrite it in terms of $\EE$ functional for the (boosted)
solitary waves
\begin{multline}
  \hL(\vsolvn(\cdot - y)) = \freq \Nn(\vsolvn) - \frac{1}{2}\vel \cdot
  \dotp{J \nabla \vsolvn}{\vsolvn} + \Hn_{V=0}(\vsolvn) \\ +
  \dotp{\vsolvn (V_y- V(y))}{\vsolvn}= \EE(\vsolvn) + \dotp{\VR
    \vsolvn}{\vsolvn},
\end{multline}
where we used that the symmetry properties of $\vsolvn$ to conclude
that $|\vsolvn|^2$ is even in all directions and hence $\dotp{\vsolvn
  x_j}{\vsolvn}=0$, for $j=1,2,3$.  Thus
\begin{equation}\label{eq:dtH2}
d_t \hL(\vsolvn(\cdot -y )) = \dot \freq \Nn(\solvn)
- \frac{1}{2}\dot \vel\cdot \dotp{J\nabla \vsolvn}{\vsolvn}
+ d_t \dotp{\VR\vsolvn}{\vsolvn},
\end{equation}
where we used that $\EE'(\vsolvn)=0$.

Subtracting \eqref{eq:dtH2} from \eqref{eq:dtH1}, using
the orthogonality relations $\dotp{\vsolvn}{\vw}=0$ and
$\dotp{J\nabla \vsolvn}{\vw}=0$ gives 
\begin{multline}\label{eq:middle}
d_t \Lyp = \frac{1}{2}\nrm{\vw}^2\dot \freq 
- \frac{1}{2}\dot \vel \cdot \dotp{J\nabla \vw}{\vw}
+\frac{1}{2}\vel \cdot \dotp{\vpsi}{\nabla V \vpsi}
-\dot y \cdot \nabla V(y) \Nn(\vpsi) \\- d_t \dotp{\VR\vsolvn}{\vsolvn} 
\end{multline}
The first two terms are of cubical order, quadratic in $\nrmH{\vw}$
and linear in $\alpha$; recall the definition of $\alpha$
in~\eqref{eq:alpha}.  The last term is also of third order or higher.
Indeed, let
\begin{equation}
A_2:=d_t \dotp{\VR\vsolvn}{\vsolvn}.
\end{equation}
Then 
\begin{equation}
|A_2| = |\dot{y}\cdot 
\dotp{\vsolvn\nabla_y \VR}{\vsolvn} + 
\dotp{\VR \vsolvn}{\dot \freq \partial_\freq \vsolvn 
+ \dot \vel \cdot \nabla_v \vsolvn}|.
\end{equation}
To bound $A_2$, we recall that $|\vel|\leq 1$ and that $\dot y_j =
\vel_j - Y_j(\Par,\vw)$, $j=1,2,3$ (see \eqref{eq:alpha1}) together
with equation~\eqref{eq:e_alpha}. This gives the estimate for $Y_{j}$;
see Proposition~\ref{prop:dyn}.  We thus find
\begin{equation}
  |A_2|\leq C \eps^2 \big(|\alpha|+\eps |Y|)\big).
\end{equation}

The middle two terms of~\eqref{eq:middle} are also of at least cubic
order, Indeed, let
\begin{equation}
  A_1:= \frac{1}{2}\vel \cdot \dotp{\vpsi}{\nabla V \vpsi} -
  \dot y \cdot \nabla V(y)\Nn(\vpsi).
\end{equation}
Decomposing $\vpsi$ gives
\begin{multline}
A_1 = \frac{1}{2}\vel \cdot \Big(
\dotp{\vsolvn}{\nabla V_y \vsolvn} 
+2\dotp{\vw}{\nabla V_y \vsolvn}
+\dotp{\vw}{\nabla V_y \vw} \Big)\\-
\frac{1}{2}\dot y \cdot \nabla V(y)(\nrm{\vw}^2+\nrm{\vsolvn}^2).
\end{multline}
Using the orthogonality relation $\dotp{\vw}{\vsolvn}=0$, and
$\dotp{\vsolvn}{x_j \vsolvn}=0$, we find
\begin{multline}
A_1 = \frac{1}{2}\vel \cdot \Big(
\dotp{\vsolvn}{\nabla_y \VR \vsolvn} 
+\dotp{\vw}{(\nabla V_y-\nabla V(y)) \vw} \Big) \\ 
+\frac{1}{2}(\vel-\dot y) \cdot \nabla V(y) (
\nrm{\vw}^2+\nrm{\vsolvn}^2)
+\vel\cdot \dotp{\vw}{(\nabla V_y-\nabla V(y)) \vsolvn}
\end{multline}
Hence,
\begin{equation}
|A_1|\leq C \eps(\eps^2 + \nrm{\vw}^2 + |Y|(1+\nrmH{\vw}^2)),
\end{equation}
with $Y$ defined as in Proposition~\ref{prop:dyn}. From this, we infer
\begin{equation}
|d_t \Lyp|\leq C|\alpha|\nrmH{\vw}^2 
+ |A_1| + |A_2|.
\end{equation}
Inserting the above estimates for $A_1$ and $A_2$ gives
\begin{equation}
|d_t \Lyp|\leq C(\eps+|\alpha|)(\nrmH{\vw}^2+\eps^2) + 
C\eps|Y|(1+\eps^2 +\nrmH{\vw}^2) 
\end{equation}
Note that $\eps\leq C$, inserting the bound of $|Y|$, given
in~\eqref{eq:e_alpha}, we simplify the above result to obtain
\begin{equation}
|d_t \Lyp|\leq C(\eps+|\alpha|)(\nrmH{\vw}^2+\eps^2) + 
C\nrmH{\vw}^3(1+\nrmH{\vw}^2)
\end{equation}
For some constant $0<C<\infty$ depending only on $r_j$ and $I_j$, $j=1,2,3$.
\end{proof}

\section{Proof of Theorem~\ref{thm:main}}
\label{sec:main}

In this section we use the lower and upper bound on the Lyapunov
functional together with the modulation equations to bound
$\nrmX{\vw}$ and $|\alpha|$.
\begin{proof}[Proof of Theorem~\ref{thm:main}]
  The Theorem~\ref{thm:main} assumes that Assumption~\ref{ass:ker} is
  satisfied and that the external potential satisfies~\eqref{eq:Vass}
  for some $\eps>0$. Furthermore, we require that the initial
  condition $\vpsi_0$ satisfies the inequality
\begin{equation}
\nrmX{\vpsi_0-\vsol_{\Par_0}}\leq \eps,
\end{equation}
for some $\Par_0\in\Pdom_3$. By Proposition~\ref{prop:gamma}
$\vpsi(\cdot,t)\in U_{\delta}(\tPdom)$, for some $\delta$ and up to
some time $T_\delta$.  Here $U_\delta$ is constructed in
Proposition~\ref{prop:decomp}. Thus all the assumptions for
Corollary~\ref{cor:high} are satisfied, and we obtain
\begin{equation}
|\Lyp(t)|\leq C\eps^2 + C t\sup_{s\leq t} f(s),
\end{equation}
where 
\begin{equation}
  f(s) = (\eps + |\alpha(s)|)
  (\eps^2 + \nrmH{\vw(\cdot,s)}^2) 
  + \nrmH{\vw(\cdot,s)}^3(1+\nrmH{\vw(\cdot,s)}^2).
\end{equation}
For times $t\leq T_\delta$, we can invoke Proposition~\ref{prop:low} and
Corollary~\ref{cor:high} to find
\begin{equation}
  \frac{7}{8}\rho\nrmH{\vw(\cdot,t)}^2 \leq C\eps^2 + 
  C \sup_{s\leq t} (t f(s)+\eps\nrm{|x|^{1/2}\vw(\cdot,s)}^2+\nrmH{\vw(\cdot,s)}^4).
\end{equation}
Thus, for all such times $t\leq T_\delta$ we have the above
inequality, and since the right-hand side is independent under
$t\rightarrow s'$, $\sup_{s'\leq t}$, for $t\leq T_\delta$ we can also
apply this to the left hand side. This gives us
\begin{equation}\label{eq:ineq}
  \frac{7}{8}\rho \sup_{s\leq t} 
  \nrmH{\vw(\cdot,s)}^2 \leq C\eps^2 + 
  C \sup_{s\leq t} (tf(s)+\eps \nrm{|x|^{1/2}\vw(\cdot,s)}^2+\nrmH{\vw(\cdot,s)}^4).
\end{equation}
Consider the inequality
\begin{equation}\label{eq:t2}
t\leq \frac{\rho}{8C}\frac{1}{\eps 
+ \sup_{s\leq t}(|\alpha(s)|+\nrmH{\vw(\cdot,s)})}.
\end{equation}
This inequality implicitly defines a maximal time, $T_2$, dependent on
$\eps$, the size of $|\alpha|$ and $\nrmH{\vw}$, such that when $t\leq
T_2$ the inequality holds. We now choose the minimal time of $T_2$ and
$T_\delta$. Since this minimal time is necessarily smaller than
the right-hand side of the inequality~\eqref{eq:t2}, we can use this
inequality to re-write~\eqref{eq:ineq} as
\begin{multline}\label{eq:77}
\frac{7}{8}\rho \sup_{s\leq t} \nrmH{\vw(\cdot,s)}^2 \leq  
C\eps^2 + \sup_{s\leq t}\Big( \frac{\rho}{8}\big(\eps^2 + 2\nrmH{\vw(\cdot,s)}^2 +\nrmH{\vw(\cdot,s)}^4\big)\\ +C\big(\eps\nrm{|x|^{1/2}\vw(\cdot,s)}^2
+\nrmH{\vw(\cdot,s)}^4\big)\Big).
\end{multline}

Let $T_1$ be the maximal time such that for $t\leq T_1$
Eqn.~\eqref{eq:t1} in Corollary~\ref{cor:x} holds.  By choosing the
minimal of the three times $T_\delta$, $T_1$ and $T_2$
we can apply the result in Corollary~\ref{cor:x}.
That is, we use
\begin{equation}\label{eq:xup}
  \sup_{s\leq
    t}\nrm{|x|^{1/2}\vw(\cdot,s)}^2\leq C\eps + \sup_{s\leq t}\nrmH{\vw(\cdot,s)},
\end{equation}
in~\eqref{eq:77} above. We find, for this minimal time, 
\begin{equation}
\frac{1}{2}\rho \sup_{s\leq t} \nrmH{\vw(\cdot,s)}^2 \leq  
C(\eps^2 +  \sup_{s\leq t}\nrmH{\vw(\cdot,s)}^4).
\end{equation}
Recalling that the initial condition is small enough we simplify the
inequality to find
\begin{equation}\label{eq:wup}
\nrmH{\vw(\cdot,t)}^2\leq \sup_{s\leq t}\nrmH{\vw(\cdot,s)}^2 \leq C\eps^2
\end{equation}
We now use the definition~\eqref{eq:nrmX} to find from~\eqref{eq:wup}
and~\eqref{eq:xup} that
\begin{equation}
\nrmX{\vw}\leq C\eps.
\end{equation}

We insert the result~\eqref{eq:wup} into the modulation equations,
\eqref{eq:aeqs}, we find that $|\alpha|$, as defined in
Proposition~\ref{prop:dyn}, satisfies the inequality
\begin{equation}
\sup_{t\leq s}|\alpha(s)| \leq C' \sup_{s\leq t}
\big(|\alpha(s)|\eps+2\eps \big)
\end{equation}
Choosing $\eps$ sufficiently small, \ie $C'\eps \leq 1/2$, leads to
\begin{equation}\label{eq:aup}
\sup_{t\leq s}|\alpha(s)| \leq c' \eps.
\end{equation}
This inserted into~\eqref{eq:e_alpha} gives that $|Y_j|\leq C\eps^2$,
for all $j$ and hence the finite dimensional modulation equations are
bounded by $C\eps^2$.  We insert the above upper bounds on
$\nrm{\vw}$, $|\alpha|$, into the inequalities~\eqref{eq:t1}
and~\eqref{eq:t2} that determines the times $T_1$ and $T_2$, both
inequalities simplify to
\begin{equation}
t\leq \frac{c}{\eps}.
\end{equation}
By possibly reducing the constant $c$ we find that $c/\eps<T_\delta$, and
we have proved the theorem.
\end{proof}

\section*{Acknowledgements}

Lars Jonsson and J\"urg Fr\"ohlich are grateful to the Swiss National Foundation
(NF-Project 20-105493).

\appendix

\section{Proof of Proposition~\ref{prop:sol}\ref{sol:spec}}
\label{app:spec}

Here we prove Proposition~\ref{prop:sol}\ref{sol:spec}.
In~\cite[App. C]{Frohlich+Jonsson+Lenzmann2005} we showed that the
essential spectrum starts at $\freq - \freq_l(\vel)$, where
$\freq_l(\vel)=(1-\sqrt{1-\vel^2})m$. For the remaining claims we
have:
\begin{proposition}\label{prop:specEE}
  Suppose Assumption \ref{ass:ker} is satisfied for the frequency
  $\freq_0$.  There is a neighborhood, $W\subset \RR^2$, around
  $(0,\freq_0)$ such that 
\begin{equation}
  \dim \Ker{\Lv} = 4,
\end{equation}
for all $(|\vel|,\freq)\in W$.  Furthermore, $\Lv$ has exactly one
negative eigenvalue, and around zero there is a gap to the next
spectral point.
\end{proposition}
\begin{proof}
  To prove the proposition, we begin with the point
  $(\vel,\freq)=(0,\freq_0)$, here $\vsolw=(\solw,0)$ and $\Lv$
  reduces to $\Lzz=\diag(\LzOz,\LzTz)$. That Assumption \ref{ass:ker} implies that $\dim \Ker \Lzz = 4$ has already been shown in the proof of Proposition \ref{prop:exists}.
  
  For general velocities, $\vel\neq 0$, and frequencies, let
  $\set{K}:=\Ker{\Lv}$ and let $k=\dim \set{K}$.
  Equations~\eqref{eq:Lz1} show that
\begin{equation}
J\vsolvn, \partial_j \vsolvn \in \set{K}, 
\end{equation}
and consequently $k\geq 4$, to show that $k=4$, we use Kato's
perturbation of the spectrum: Define the operator $A:=\Lv-\Lzz = -\vel
\cdot \nabla J +
\En_{0,\freq}''(\vsolvn)-\En_{0,\freq_0}''(\vsol_{\freq_0},0)$, it is
$\Lzz$-bounded;
\begin{equation}
\nrm{A\vu}\leq c_{|\vel|,\freq-\freq_0}\nrm{\vu} + C_{|\vel|,\freq-\freq_0} \nrm{\Lzz \vu},
\end{equation}
where both constants approach zero as both $|\vel|$ and
$|\freq-\freq_0|$ approach zero, which follows from the fact
$\nrmHp{1}{\vsolvn-\vsol_{\freq_0}}\leq (|v|+|\freq-\freq_0|)C$, for
small enough $|\vel|$ and $|\freq-\freq_0|$.

Denote the spectral distance in $\sigma(\Lzz)$ from zero to nearest
spectral point $d$ and consider the inequality:
\begin{equation}
c_{|\vel|,\freq-\freq_0} + C_{\vel,\freq-\freq_0} d\leq d/2
\end{equation}
For neighborhoods $W$ such that the above inequality is satisfied for
all $(|\vel|,\freq)\in W$, \cite[\S.V.4.3]{Kato1995} states that
within the circle with center zero and radius $d/2$ there are exactly
four (repeated) eigenvalues of $\Lv$ (since $\Lzz$ has a degeneracy
four zero eigenvalue). Thus $k=\dim K\leq 4$ consequently $k=4$.
Furthermore, $\sigma(\Lv)$ has a spectral gap of at least $d/2$ from
zero to the next spectral point. The circle thus separates the
spectrum into three parts.

The function $\vsol_{\freq_0}$ is a minimizer with one constraint,
thus its corresponding Hessian $\Lzz$, can have at most one negative
eigenvalue, see \eg~\cite{FGJS-I}. But by
\begin{equation}
\dotp{\partial_{\freq}\vsolw}{\Lz\partial_\freq \vsolw} = -\Nn'(\vsolw)<0.
\end{equation}
it has at least one negative eigenvalue. Thus it has exactly one
negative eigenvalue. The above separation of the spectrum together
with the fact that the eigenvalues (of self-adjoint operators) are
constrained to the real axis ensure that $\Lv$ has exactly one
negative eigenvalue.
\end{proof}

\section{Positivity of $\dotp{\Lv \solv}{\solv}$}
\label{app:Lpos}

In this section we show that $\dotp{\vw}{\Lv\vw}\geq \rho\nrm{\vw}^2$
for $\symp{\vw}{\vz}=0$ for all $\vz\in \TMo$. From Proposition~\ref{prop:sol}
we know that $\Lv$ has one negative eigenvalue. 
We have the following result
\begin{proposition}\label{prop:Lpos}
  Under Assumption~\ref{ass:ker} and with $|\vel|< r_1<1$ and
  $\freq\in I_1$ there is a $\rho>0$ dependent only on $I_1$ and $r_1$
  defined in Proposition~\ref{prop:J} such that if $\symp{\vw}{\vz}=0$
  for all $\vz\in \TMo$, then
\begin{equation}
\dotp{\vw}{\Lv\vw}\geq \rho \nrmH{\vw}^2.
\end{equation}
\end{proposition}
We follow the proof of Proposition D.1 in~\cite{FGJS-I} with necessary
modifications to the pseudo-relativistic Hartree equation. But we
repeat the proof here for completeness. We break the proposition into
three steps.
\begin{lemma}[Step 1] 
  Let $X_1:=\{\vw\in\Half: \nrm{\vw}=1, \dotp{\vw}{\vsolvn}=0 \}$,
  and $|\vel|\leq r_1$, $\freq\in I_1$.  Then
\begin{equation}\label{eq:min1}
\inf_{\vw\in X_1} \dotp{\vw}{\Lv\vw}=0.
\end{equation} 
\end{lemma}
\begin{proof}
  Let $a:=\inf_{\vw\in X_1} \dotp{\vw}{\Lv\vw}$. Clearly $\nu\leq a
  \leq 0$, where $\nu<0$ is the negative eigenvalue of $\Lv$.  That
  $a\leq 0$ is clear as $\vw = J\vsolvn/\nrm{\vsolvn}\in X_1$ yields
  $\dotp{\vw}{\Lv \vw} = 0$.  Moreover $a \neq \nu$. Indeed if
  $a=\nu$, then the (local) minimizer, $\TheNewVec{\phi}$, of
  \eqref{eq:min1} would be an eigen-function of $\Lv$ corresponding to
  the smallest eigenvalue $\nu$ and $\TheNewVec{\phi}\in X_1$ and
  $\vsolvn\bot \TheNewVec{\phi}$. Now, since $\vsolvn \bot \Ker{\Lv}$
  and since $\nu$ is the only negative eigenvalue, we conclude that
  $\vsolvn$ is in the spectral subspace of $\Lv$ corresponding to the
  interval $[\delta,\infty)$ for some $\delta>0$. Therefore
  $\Lv^{-1}\vsolvn$ is well defined and
  $\dotp{\vsolvn}{\Lv^{-1}\vsolvn}>0$. On the other hand the equation
  $\Lv\partial_\freq \vsolvn = - \vsolvn$ implies that
\begin{equation}\label{eq:D3}
\dotp{\vsolvn}{\Lv^{-1}\vsolvn} = - N'(\vsolvn)< 0 
\end{equation}
which contradicts $\dotp{\vsolvn}{\Lv^{-1}\vsolvn}>0$. Hence $a =
\nu$ is impossible.

To show that $a=0$ we use the Euler-Lagrange equations
corresponding to \eqref{eq:min1}
\begin{equation}
\Lv \vw = a \vw + b \vsolvn,
\end{equation}
where $a$ and $b$ are Lagrange multipliers corresponding to
$\nrm{\vw}=1$ and $\dotp{\vsolvn}{\vw}=0$ respectively. Assume
$\nu<a<0$. If $b=0$, then $a$ would be a negative eigenvalue in
$(\nu,0)$ which contradicts that $\nu$ is the only negative
eigenvalue. Thus $b\neq 0$. Given $\nu<a<0$, we can solve the 
Euler-Lagrange equation as
\begin{equation}
\vw = b (\Lv - a)^{-1}\vsolvn.
\end{equation}
The inner product of the equation above with $\vsolvn$, together with
the orthogonality relation $\dotp{\vsolvn}{\vw}=0$, and $b\neq 0$,
give
\begin{equation}\label{eq:zeroq}
0 = \dotp{\vsolvn}{(\Lv+|a|)^{-1}\vsolvn} = : q(|a|)
\end{equation}
$q(\lambda)$ is analytic in $\lambda\in (0,|\nu|)$, and hence 
differentiable. Moreover it is monotonically decreasing, since 
\begin{equation}
q'(\lambda) = -\dotp{\vsolvn}{(\Lv+\lambda)^{-2} \vsolvn} = 
-\nrm{(\Lv + \lambda)^{-1}\vsolvn}^2 < 0.
\end{equation}
Furthermore by \eqref{eq:D3} $q(0)=\dotp{\vsolvn}{\Lv^{-1}\vsolvn}<0$. Thus
$q(|a|)\neq 0$, for $a \in (\nu,0)$, which contradicts 
\eqref{eq:zeroq}. Hence $a=0$
\end{proof}

\begin{lemma}[Step 2]
Let $X:=\{\vw\in \Half:\nrm{\vw}=1, \omega(\vw,z)=0, \forall z\in \TMo\}$.
Then
\begin{equation}\label{eq:LL9}
\inf_{\vw \in X}\dotp{\vw}{\Lv\vw}>0
\end{equation}
\end{lemma}
\begin{proof}
The Euler-Lagrange equation corresponding to \eqref{eq:LL9} is
\begin{equation}
\Lv \vw = a \vw + \sum_k \gamma_k J \vz_k
\end{equation} 
where $\{\vz_k\}$ is a basis for $\TMo$. Here $a$ and $\{\gamma_k\}$
are the Lagrange multipliers corresponding to the constraints
$\nrm{\vw}=1$ and $\omega(\vw,\vz_{k})=0$ $\forall k$ respectively.
Note that $a=\dotp{\vw}{\LL \vw}$, and that $\set{X}\subset
\set{X}_1$, hence $a\geq 0$.  Assume that $a=0$, and that
$\gamma_j\neq 0$ for some $j$.  Then, by Corollary ~\ref{cor:nondeg},
there exists a $\vz=\sum_{j,l} \gamma_j (\OMvos^{-1})_{jl}\vz_l \in
\TMo$ such that
\begin{equation}
\dotp{\vz}{\LL \vw} = \sum_{j,k,l} \gamma_j (\OMvos^{-1})_{jl}(\OMvos)_{lk}\gamma_k = \sum_j |\gamma_j|^2>0,
\end{equation}
which contradicts $\dotp{\vz}{\Lv \vw}=\dotp{\Lv \vz}{\vw}=0$.  Here
we have used that $\vz=\sum_j b_j\vz_j$
and $\vz_j$ is either a zero-eigenfunction or an associated zero-mode
for $\Lv$.  Thus either $a>0$ or $a=0$ and $\gamma_j=0$. Consider the latter
case.  In this case
\begin{equation}
\Lv \vw = 0.
\end{equation}
which implies that $\vw\in \Ker{\Lv}$. Since $\Ker{\Lv}\subset \TMo$,
the relation $\omega(\vw,\vz)=0$ for all $\vz\in
\TMo$ contradicts the non-degeneracy of $\OMvos$ on $\MfS$ 
(see Corollary~\ref{cor:nondeg}).
Thus $a>0$.
\end{proof} 

\begin{proof}[Step 3. End of Proof.]
Equation~\eqref{eq:LL9} implies that there exists a $\rho'>0$
such that
\begin{equation}
   \dotp{\vw}{\Lv \vw}\geq \rho' \nrm{\vw}^2\; , \label{eq:cL}
\end{equation}
for some $\rho'=\rho'(\freq,\vel)$ and all $\vw \in X$.  To improve the coercivity from $\Ltwo$
to $\Half$, we let $0<\delta<1$, and estimate $\dotp{\vw}{\Lv
  \vw}$ using \eqref{eq:cL} as
\begin{equation}\label{eq:ett}
    (1-\delta)\rho'\nrm{\vw}^2 + 
        \delta \dotp{\vw}{\Lv \vw}
    \leq 
        \dotp{\vw}{\Lv \vw} \; .
\end{equation}
Upon using the explicit form of $\Lv$ we find that
\begin{equation}\label{eq:tvaa}
    \dotp{\vw}{\Lv \vw} \geq
          \dotp{\vw}{\sqrt{-\Laplace+m^2} \vw} - \vel\cdot \dotp{J\nabla \vw}{\vw}- C_\freq\nrm{\vw}^2,
\end{equation}
where 
\begin{equation}
      C_\freq \leq  |m-\freq| + C(3,1)\nrmLp{3}{\vsolvn}^2+\sup_x  |\frac{1}{|x|}*|\vsolvn|^2|.
\end{equation}
Here we have used Kato's inequality (see~\eqref{eq:Kato}) and the
Hardy-Littlewood-Sobolev inequality with sharp constant $C(3,1)$ see
\eg in~\cite[Thm.  4.3]{Lieb+LossII}.  The two
estimates~\eqref{eq:ett}, \eqref{eq:tvaa} with $\delta := \rho'(1+
\rho' + C_\freq)^{-1}$ imply
\begin{equation}
   \dotp{\vw}{\Lv \vw} \geq  \rho\nrmH{\vw}^2 \; ,
\end{equation}
where $0<\rho = \inf_{\freq\in
  I_1,\vel<r_1}(1-|\vel|)\rho'(1+\rho'+C_\freq)^{-1}$.  Thus for
$\freq\in I_1$, $|\vel|<r_1$, we find that $\rho$ depends only on
$I_1$ and $r_1$.  This concludes the proof of
Proposition~\ref{prop:Lpos}.
\end{proof}

\section{Proof of Corollary~\ref{cor:uniform}}
\label{app:ift}

In this appendix we prove Corollary~\ref{cor:uniform} by using the
proof of an implicit function theorem as proven in~\cite[Thm.
10.2.1]{Dieudonne1969} and~\cite[Thm 4E]{Zeidler109}. From these
proofs we find that it suffices to consider three restrictions of the
radii. They are 1) distance to the boundary 2) sufficient conditions
for contraction 3) differentiability of $\iPar$. We follow closely the
proof of the implicit function theorem in~\cite[Thm.
10.2.1]{Dieudonne1969}, applied to this case and with bounds expressed
in terms of known quantities \eg maps of ground states.

  Let $p\in [1,\infty]$ and consider the $p$-metric on $\RR^8$, 
\begin{equation}
  |\Par-\Par'|_{(p)}:=(\sum_j |\Par_j-\Par_j'|^p)^{1/p}.
\end{equation}
These metrics are all equivalent, and it does not matter for the
result which of these we use. But to make a definite choice we use the
$p=2$ metric as the default one, and denote the distance
$\Rnrm{\cdot}:=|\cdot|_{(2)}$. To obtain uniform bounds, we
use $\Inrm{\cdot}$ and $\Dnrm{\cdot}$ defined by
\begin{equation}
  \Inrm{\OMvos} := \sup_{\freq\in I_2',\Rnrm{\vel}< r_2'} \sup_{\Rnrm{\Par'}=1}
  \Rnrm{\sum_k (\OMvos)_{\cdot k}\Par_k'} 
\end{equation}
and
\begin{equation}
  |\OMvos|_{1,\infty} := |\partial_{\freq}\OMvos|_{\infty} + 
  \sum_{l=1}^3 |\partial_{\vel_l}\OMvos|_{\infty}
\end{equation}

To elucidate our proof we make a change of notation to emphasize the
difference between the center of the specific balls $B^8_{\rho_2}$
and $b_{R_2}$ from any point on the soliton manifold $\MfS$: The
center of the balls are denoted by $\iniPar=(\iy,\iv,\iphase,\ifreq)$
and $\vsol_{\iniPar}$ respectively, whereas $\Par$ (or $\vsolp$) is any
point in $\tPdom'$ (or in $\MfT'$).

The size of the radii clearly depend on how close to the boundary of
the soliton manifold (parameter space) the point $\vsol_{\iniPar}$
($\iniPar$) is located. In order to obtain uniform radii, consider
$(r_2',I_2')$ such that $0<r_2'<r_1$ and $I_2'$ a non-empty, open
interval with $I_2'\Subset I_1$.  By the implicit function theorem
there are balls {\it in} the domain $\oPdom$, $\MfS$ such that
Lemma~\ref{lem:ball} holds. The symplectic manifold $\MfS$ is well
defined up to and including its boundary and by the argument before,
the only boundary directions are the $\vel$ and $\freq$ directions.
Thus the radii have to satisfy the inequalities
\begin{align}\label{eq:lim1}
R_2< \inf_{\freq\in \partial I_1, |\vel|=r_1, \ifreq\in \partial I_2', |\iv|=r_2'}
   \nrmX{\vsolvn-\vsol_{\iv,\ifreq}},\\
\rho_2< \min(|r_1-r_2'|,\min_{\freq\in \partial I_1,\ifreq\in \partial I_2'}|\freq-\ifreq|),
\end{align}
which depend only on the choice of $r_2'$ and $I_2'$.

Now given a point on the manifold $\vsol_{\iniPar}\in \MfT'$, with its
corresponding ball $b_{R_1}$. The second restriction on the radii is
the contraction restriction which we consider by recasting the
equation $G(\vpsi,\Par)=0$ into a contraction equation:
$g(\vpsi,\Par)=\Par$ where
\begin{equation}\label{eq:con}
  g_j(\vpsi,\Par):=\Par_j + \sum_{k}((\iOM)^{-1})_{jk}G_k(\vpsi,\Par).
\end{equation}
Here we used that $\jOM=\OM_{\vsol_{\iv,\ifreq}}$ and
$\iOM:=\OM_{\vsol_{\iv,\ifreq}}$. To apply the contraction
theorem~\cite[Thm.  10.1.1]{Dieudonne1969} to $g(\vpsi,\Par)=\Par$, it
suffices to show that, with $\vpsi\in b_{R_1}$ and
$\Par^{(1)},\Par^{(2)}\in B^8_{\rho_1}$ that a)
\begin{equation}\label{eq:aa}
  \Rnrm{g(\vpsi,\Par^{(1)})-g(\vpsi,\Par^{(2)})}\leq 
  \frac{1}{2}\Rnrm{\Par^{(1)}-\Par^{(2)}}.
\end{equation}
and b)
\begin{equation}\label{eq:bb}
  \Rnrm{g(\vpsi,\iniPar)-\iniPar)} \leq  \frac{\rho_2}{2}.
\end{equation}
Then existence, uniqueness and continuity of $\iPar$ are ensured for
$\vpsi\in b_{R_1}(\vsol_{\iniPar})$.

To translate these two constraints into uniform bounds on the radii
$R_2\leq R_1$ and $\rho_2\leq \rho_1$, we estimate a `Taylor series
remainder' or, equivalently, use a mean value theorem. First assume
that $\vpsi\in b_{R_2}(\vsol_{\iniPar})$ and $\Par^{(1)},\Par^{(2)}\in
B^8_{R_2}(\iniPar)$. The left-hand side of eq.~\eqref{eq:aa} with
\eqref{eq:con} inserted is bounded from above by
\begin{equation}
  A_1:=\Inrm{\OMvos^{-1}} \Rnrm{G(\vpsi,\tPar{1})-G(\vpsi,\tPar{2})+\iOM 
  (\tPar{1}-\tPar{2})},
\end{equation}
where $\Inrm{\cdot}$ was defined above. Let the vector $f$ be defined
through its elements
\begin{equation}
  f_j(\vpsi,\Par'):= G_j(\vpsi,\Par')
  -\partial_{\Par_\cdot}G_j(\vsol_{\iniPar},\iniPar)\cdot\Par'.
\end{equation}
The difference $f(\vpsi,\Par^{(1)})-f(\vpsi,\Par^{(2)})$, which up to a
constant describes $A_1$, is estimated by the mean value
theorem~\cite[Thm 8.6.2]{Dieudonne1969}, yielding the upper bound on
$A_1$ as
\begin{align}
  A_1 &\leq \Rnrm{\Par^{(1)}-\Par^{(2)}} \Inrm{\OMvos^{-1}}
  \sup_{\Par\in b_{R_2},\vpsi\in V_{\rho_2}} \Rnrm{\iOM - \OM_{\vsolp}
    +
    \dotp{\vpsi-\vsolp}{J^{-1}\partial_{\Par_\cdot}\vz_{\cdot,\Par}}}
  \nonumber \\ &\leq \Inrm{\OMvos^{-1}}(\rho_2 \Dnrm{\OMvos} + R_2
  \dNrm{\vz_{\Par}})\Rnrm{\Par^{(1)}-\Par^{(2)}}.
  \label{eq:lim2}
\end{align}
Here $\Dnrm{\cdot}$ is defined above, and to define
$\dNrm{\vz_{\Par}}$, let the matrix $Y$ by its elements
$Y_{jk}:=\nrmX{\partial_{\Par_k}\vz_{j,\Par}}$, and let the vector $V$
be the elements $V_k = \nrmX{\vz_{k,\Par}}$. We then have
\begin{equation}
 \zNrm{\vz_{\Par}}:=\Inrm{V},
\end{equation}
and
\begin{equation}
  \dNrm{\vz_{\Par}} := \Inrm{Y} +\zNrm{\vz_{\Par}}.
\end{equation}
Constants, norms, in~\eqref{eq:lim2}, involving $\OMvos$ and its
inverse are independent of $\Par$ due to the supremum over $\freq$ and
$\vel$.  The numbers $\nrmX{\partial_{\Par_k}\vz_{j,\Par}}$ are
independent of translation and phase, since the integral over space
and its absolute value removes all appearances of phase and
translation. Hence $\dNrm{\vz_{\Par}}$ is independent of $\Par$ and
depends only on $r_2'$, $I_2'$. Thus the allowed radii $\rho_2$ and
$R_2$ in~\eqref{eq:lim2}, so that we obtain the constant a half as
required in~\eqref{eq:aa}, only depend on $r_2'$ and $I_2'$. We
say that~\eqref{eq:aa} is {\it uniformly} satisfied on $\tPdom'$.

Similarly, consider~\eqref{eq:bb} with~\eqref{eq:con} inserted, its
left-hand side is
\begin{equation}
  \Rnrm{(\iOM)^{-1}G(\vpsi,\iniPar)}.
\end{equation}
For $\vpsi\in b_{R_2}$, we find
\begin{equation}\label{eq:lim3}
  \Rnrm{g(\vpsi,\iniPar)-\iniPar} \leq 
  |\OMvos^{-1}|_{\infty}\zNrm{\vz_{\freq,\vel}}R_2\leq \frac{\rho_2}{2},
\end{equation}
for $R_2$ small enough.  Thus~\eqref{eq:bb} is also uniformly
satisfied on $\tPdom'$ for $R_2$ small enough.

The third constraint on the radii $R_2$ and $\rho_2$ are the
conditions to ensure differentiability of $\iPar$ in $b_{R_2}$.  To
find these constraints we use the differentiability of $G$ and the
chain rule. To this end, we consider the Taylor expansion of
$G(\vpsi+\vw,\iPar(\vpsi+\vw))$ around the solution
$(\vpsi,\iPar(\vpsi))$:
\begin{multline}\label{eq:diffG}
  A_2:=\Rnrm{G(\vpsi+\vw,\iPar(\vpsi+\vw)) - G(\vpsi,\iPar(\vpsi))
  - \dotp{\vw}{J\vz_{\cdot,\iPar(\vpsi)}}\\  - 
  \partial_{\Par_\cdot}G(\vpsi,\iPar(\vpsi))\cdot
  (\iPar(\vpsi+\vw)-\iPar(\vpsi))},
\end{multline}
which by the construction of $\iPar$ for $\vpsi+\vw\in b_{R_2}$
reduces to
\begin{equation}\label{eq:diffG2}
A_2=\Rnrm{\dotp{\vw}{J\vz_{\cdot,\iPar(\vpsi)}} + 
  \partial_{\Par_\cdot}G(\vpsi,\iPar(\vpsi))\cdot 
  (\iPar(\vpsi+\vw)-\iPar(\vpsi))}
\end{equation}
Differentiability of $G$ ensures a relation between the upper bound of
$A_2$ and the radii. Let $\tau:=\iPar(\vpsi+\vw)-\iPar(\vpsi)$, and
let
\begin{equation}
  f_j^{1}(\vw,\tau'):=G_j(\vpsi+\vw,\iPar(\vpsi)+\tau') 
  - \dotp{\vw}{J\vz_{j,\Par}} 
  - \partial_{\iPar_{\cdot}(\vpsi)}G_j(\vpsi,\iPar(\vpsi))\cdot \tau'.
\end{equation}
$A_2$ is simply $\Rnrm{f^1(\vw,\tau)-f^1(0,0)}$ and by the mean value 
theorem for we have
\begin{multline}\label{eq:lim4}
  A_2\leq 
 \Rnrm{\tau}(\Rnrm{\OM_{\vsol_{\iPar(\vpsi)+\tau}}-\OM_{\vsol_{\iPar(\vpsi)}}} 
 + \Rnrm{\dotp{\vpsi+\vw-\vsol_{\iPar(\vpsi)+\tau}}{J\partial_{\Par_\cdot}\vz_{\cdot,\iPar(\vpsi)+\tau}}
 -\dotp{\vpsi-\vsol_{\iPar(\vpsi)}}{J\partial_{\Par_\cdot}\vz_{\cdot,\iPar(\vpsi)}}}) 
\\ + \nrmX{\vw} \Rnrm{\vz_{\iPar(\vpsi)+\tau}-\vz_{\iPar(\vpsi)}} 
\leq   \Rnrm{\tau}(2\dNrm{\vz} R_2 
  + \Dnrm{\OMvos}\rho_2) +\nrmX{\vw} \dNrm{\vz_{\Par}}\rho_2 ,
\end{multline}
where we used that $|\tau|\leq \rho_2$. The differentiability of $G$
can now be expressed as follows: For every $K_2>0$, there exist,
by~\eqref{eq:lim4}, radii $R_2>0$ and $\rho_2>0$, such that
\begin{equation}\label{eq:K2}
A_2\leq K_2(|\tau|+\nrmX{\vw}).
\end{equation}
The explicit calculation~\eqref{eq:lim4} shows that \eqref{eq:K2} can
be satisfied uniformly on $\tPdom'$.

To convert the differentiability of $G$ into differentiability of
$\iPar$, we use that $\partial_{\Par} G$ is invertible at
$\vsol_{\iniPar},\iniPar$. We recall that
\begin{equation}
  \partial_{\Par_{k}}G(\vpsi,\Par) = -\OMvos 
+ \symp{\vpsi-\vsolp}{J\partial_{\Par_k}\vz_{j,\Par}}.
\end{equation}
For all $\Par'\neq 0$ we have
\begin{equation}
  \Rnrm{\partial_{\Par_{\cdot}} G(\vpsi,\iPar(\vpsi))\cdot\Par'} 
    \geq \Rnrm{\OMvos \Par'} 
  - \dNrm{\vz_{\Par}}R_2\Rnrm{\Par'},
\end{equation}
and since $\OMvos$ is invertible we have 
\begin{equation}\label{eq:lim5}
  \Rnrm{\partial_{\Par_{\cdot}} G(\vpsi,\iPar(\vpsi))\cdot\Par'} 
    \geq (\Inrm{\OMvos^{-1}}^{-1} 
  - \dNrm{\vz_{\Par}}R_2)\Rnrm{\Par'} \geq \frac{1}{2}\Rnrm{\Par'}\Inrm{\OMvos}>0,
\end{equation}
by choice of $R_2$. For such $R_2$ is $\partial_{\Par}G$ uniformly
invertible in each ball $b_{R_2}$.  If we possibly
reduce $\rho_2$ and $R_2$ further, we may assume that
\begin{equation}\label{eq:lim6}
  K_2 \Inrm{\partial_{\Par}G(\vpsi,\iPar(\vpsi))^{-1}}\leq \frac{1}{2},
\end{equation}
and hence by~\eqref{eq:diffG2} and \eqref{eq:lim4} we then have
\begin{equation}\label{eq:chain}
  \Rnrm{\partial_{\Par}G^{-1}(\vpsi,\iPar(\vpsi))\dotp{J\vz_{\cdot,\Par}}{\vw} + \tau} \leq \frac{1}{2}(\Rnrm{\tau}+\nrmX{\vw}).
\end{equation}
The triangle inequality leads to
\begin{equation}
  \Rnrm{\tau} \leq 
  (1+2\Inrm{\partial_{\Par}G^{-1}(\vpsi,\iPar(\vpsi))}
  \zNrm{\vz_{\Par}})\nrmX{\vw}.
\end{equation}
Insert this inequality into~\eqref{eq:chain} to obtain
\begin{multline}
  \Rnrm{\tau + D_2G^{-1}(\vpsi,\iPar(\vpsi))\dotp{D_1
      G(\vpsi,\iPar(\vpsi))}{\vw}}\\ \leq
  2K_2(1+\Inrm{D_2G^{-1}(\vpsi,\iPar(\vpsi))}\zNrm{\vz_{\Par}})\nrmX{\vw},
\end{multline}
and thus $\iPar$ is differentiable. Furthermore, we have shown
that the above constraints can be chosen uniformly in $\Par$ on
$\tPdom'$.

If we now go back and study the restrictions on the radii
\eqref{eq:lim1}, \eqref{eq:lim2}, \eqref{eq:lim3}, \eqref{eq:lim4},
\eqref{eq:lim5} and \eqref{eq:lim6}, we notice that the constraint
area can, for a sufficiently small number $\tilde{\rho}_2$ such that
$\rho_2\leq \tilde{\rho}_2$, be chosen to be all points in the
triangle $0<R_2\leq c\rho_2$, $\rho_2\leq \tilde{\rho}_2$, which
concludes the proof of the corollary. \qed

\newcommand{\SortNoop}[1]{}

$ $\\
{\sc
J\"urg Fr\"ohlich\\
Institute for Theoretical Physics\\
ETH Z\"urich\\
CH-8093 Z\"urich, Switzerland.\\
{\it email:} \verb+juerg@itp.phys.ethz.ch+\\
\\
B. Lars G. Jonsson\\
Institute for Theoretical Physics\\
ETH Z\"urich\\
CH-8093 Z\"urich, Switzerland.}\\
\\
Alternative address:\\
{\sc School of Electrical Engineering\\
Electromagnetic engineering\\
Royal Institute of Technology (KTH)\\
SE-100 44 Stockholm.\\
{\it email:} \verb+jonsson@itp.phys.ethz.ch+\\
\\
Enno Lenzmann\\
Department of Mathematics, HG G 33.1\\
ETH Z\"urich\\
CH-8092 Z\"urich, Switzerland.\\
{\it email:} \verb+lenzmann@math.ethz.ch+
}


\begin{thebibliography}{10}

\bibitem{BP92}
V.~S. Buslaev and G.~S. Perel'man.
\newblock Scattering for the nonlinear {S}chr{\"o}dinger equation: states that
  are close to a soliton.
\newblock {\em Algebra i Analiz}, 4(6):63--102, 1992,
  \\MR1199635\footnote{http://www.ams.org/mathscinet-getitem?mr=MR1199635}.

\bibitem{Buslaev+Sulem2002}
V.~S. Buslaev and C.~Sulem.
\newblock On asymptotic stability of solitary waves for nonlinear
  {S}chr{\"o}dinger equations.
\newblock {\em Ann. IHP. Anal. Nonl.}, 20:419--475, 2003, doi:
  10.1016/S0294-1449(02)00018-5
\footnote{http://dx.doi.org/10.1016/S0294-1449(02)00018-5}.

\bibitem{Dejak+Jonsson2005}
S.~I. Dejak and B.~L.~G. Jonsson.
\newblock Long-time dynamics of variable coefficient {mKdV} solitary waves.
\newblock {\em Accepted by J. Math. Phys.}, 2006,
  ArXiv:math-ph/0503016\footnote{http://arxiv.org/abs/math-ph/0503016}.

\bibitem{Dejak+Sigal2006}
S.~I. Dejak and I.~M. Sigal.
\newblock Long-time dynamics of {KdV} solitary waves over a variable bottom.
\newblock {\em Comm. Pure Appl. Math.}, 59(6):869--905, 2006, doi:
  10.1002/cpa.20120.

\bibitem{Demanet+Schlag2005}
L.~Demanet and W.~Schlag.
\newblock Numerical verification of a gap condition for linearized {NLS}.
\newblock {\em Nonlinearity}, 19(4):829--852, 2006,
  doi:10.1088/0951-7715/19/4/004.

\bibitem{Dieudonne1969}
J.~Dieudonn\'e.
\newblock {\em Foundations of Modern Analysis}, volume 10-I of {\em Pure and
  Applied Mathematics}.
\newblock Academic Press, New York, third edition, 1969, MR0349288.
\newblock Enlarged and corrected printing.

\bibitem{Elgart+Schlein2005}
A.~Elgart and B.~Schlein.
\newblock Mean field dynamics of boson stars.
\newblock {\em Comm. Pure Appl. Math.}, 2006, doi: 10.1002/cpa.20134.
\newblock Published online.

\bibitem{FGJS-I}
J.~Fr\"ohlich, S.~Gustafson, B.~L.~G. Jonsson, and I.~M. Sigal.
\newblock Solitary wave dynamics in an external potential.
\newblock {\em Comm. Math. Phys.}, 250(3):613--642, 2004, doi:
  10.1007/s00220-004-1128-1.

\bibitem{Frohlich+Jonsson+Lenzmann2005}
J.~Fr\"ohlich, B.~L.~G. Jonsson, and E.~Lenzmann.
\newblock Boson stars as solitary waves.
\newblock Submitted, 2005, ArXiv:math-ph/0512040.

\bibitem{Frohlich+Jonsson+Lenzmann2006b}
J.~Fr\"ohlich, B.~L.~G. Jonsson, and E.~Lenzmann.
\newblock The kernel condition for boson stars.
\newblock In preparation, 2006.

\bibitem{Frohlich+Lenzmann2004}
J.~Fr{\"o}hlich and E.~Lenzmann.
\newblock Mean-field limit of quantum {B}ose gases and nonlinear {H}artree
  equation.
\newblock In {\em S\'emin. \'Equ. D\'eriv. Partielles. 2003--2004}, pages
  XIX--1--26. \'Ecole Polytech., Palaiseau, 2004, ArXiv:math-ph/0409019.

\bibitem{Frohlich+Lenzmann2005}
J.~Fr\"ohlich and E.~Lenzmann.
\newblock Blow-up for nonlinear wave equations describing boson stars.
\newblock {\em To appear in Comm. Pure Appl. Math.}, 2005, ArXiv:math.AP/0511003.

\bibitem{Frohlich+Tsai+Yau2002}
J.~Fr\"{o}hlich, T.-P. Tsai, and H.-T. Yau.
\newblock On the point-particle ({N}ewtonian) limit of the non-linear {H}artree
  equation.
\newblock {\em Comm. Math. Phys.}, 225(2):223--274, 2002,
  doi:10.1007/s002200100579.

\bibitem{Herbst1977}
I.~W. Herbst.
\newblock Spectral theory of the operator $(p^2+m^2)^{1/2}-ze^2/r$.
\newblock {\em Comm. Math. Phys.}, 53(3):285--294, 1977,
  doi:10.1007/BF01609852.

\bibitem{Hislop2000}
P.~D. Hislop.
\newblock Exponential decay of two-body eigenfunctions: a review.
\newblock In {\em Proceedings of the Symposium on Mathematical Physics and
  Quantum Field Theory (Berkeley, CA, 1999)}, volume~4 of {\em Electron. J.
  Differ. Equ. Conf.}, pages 265--288 (electronic), San Marcos, TX, 2000.
  Southwest Texas State Univ., MR1785381.

\bibitem{FGJS-II}
B.~L.~G. Jonsson, J.~Fr\"ohlich, S.~Gustafson, and I.~M. Sigal.
\newblock Long time motion of {NLS} solitary waves in a confining potential.
\newblock {\em To appear in Ann. Henri Poincare}, 2006, ArXiv:math-ph/0503009.

\bibitem{Kato1995}
T.~Kato.
\newblock {\em Perturbation theory for linear operators}.
\newblock Classics in Mathematics. Springer-Verlag, Berlin, 1995, MR1335452.
\newblock Reprint of the 1980 edition.

\bibitem{Kaup1976}
D.~J. Kaup.
\newblock A perturbation expansion for the {Z}akharov-{S}habat inverse
  scattering transform.
\newblock {\em SIAM J. Appl. Math.}, 31(1):121--133, 1976, doi:10.1137/0131013.

\bibitem{Kodama+Ablowitz1981}
Y.~Kodama and M.~J. Ablowitz.
\newblock Perturbations of solitons and solitary waves.
\newblock {\em Stud. Appl. Math.}, 64(3):225--245, 1981, MR615541.

\bibitem{Lenzmann2005a}
E.~Lenzmann.
\newblock Well-posedness for semi-relativistic {H}artree equations of critical
  type.
\newblock {\em To appear in Math. Phys. Anal. Geom.}, 2006,
  ArXiv:math.AP/0505456.

\bibitem{Lieb+LossII}
E.~H. Lieb and M.~Loss.
\newblock {\em Analysis}, volume~14 of {\em Graduate Studies in Mathematics}.
\newblock American Mathematical Society, Providence, RI, second edition, 2001,
  MR1817225.

\bibitem{Lieb+Thirring1984}
E.~H. Lieb and W.~E. Thirring.
\newblock Gravitational collapse in quantum mechanics with relativistic kinetic
  energy.
\newblock {\em Ann. Physics}, 155(2):494--512, 1984,
  doi:10.1016/0003-4916(84)90010-1.

\bibitem{Lieb+Yau1987}
E.~H. Lieb and H.-T. Yau.
\newblock The {C}handrasekhar theory of stellar collapse as the limit of
  quantum mechanics.
\newblock {\em Comm. Math. Phys}, 112(1):147--174, 1987,
  \\doi:10.1007/BF01217684.

\bibitem{McLaughlin+Scott1978}
D.~W. McLaughlin and A.~C. Scott.
\newblock Perturbation analysis of fluxon dynamics.
\newblock {\em Phys. Rev. A}, 18(4):1652--1680, 1978,
  doi:10.1103/PhysRevA.18.1652.

\bibitem{Merle2005}
F.~Merle and P.~Raphael.
\newblock The blow-up dynamic and upper bound on the blow-up rate for critical
  nonlinear {S}chr\"odinger equation.
\newblock {\em Ann. of Math. (2)}, 161(1):157--222, 2005,
  euclid.annm/1111509197
\footnote{http://projecteuclid.org/getRecord?id=euclid.annm/1111509197}.

\bibitem{Messer1981}
J.~Messer.
\newblock {\em Temperature dependent Thomas-Fermi theory}, volume 147 of {\em
  Lect. Notes Phys.}
\newblock Springer, Berlin, 1981, doi:10.1007/3-540-10875-0.

\bibitem{Ruffini+Bonazzola1969}
R.~Ruffini and S.~Bonazzola.
\newblock Systems of self-graviting particles in general relativity and the
  concept of an equation of state.
\newblock {\em Phys. Rev. II}, 187(5):1767--1783, 1969,
  doi:10.1103/PhysRev.187.1767.

\bibitem{Slaggie+Wichmann1962}
E.~L. Slaggie and E.~H. Wichmann.
\newblock Asymptotic properties of the wave function for a bound
  nonrelativistic three-body system.
\newblock {\em J. Math. Phys.}, 3:946--968, 1962, doi: 10.1063/1.1724311.

\bibitem{Stein1993}
E.~M. Stein.
\newblock {\em Harmonic analysis: real-variable methods, orthogonality, and
  oscillatory integrals}, volume~43 of {\em Princeton Mathematical Series}.
\newblock Princeton University Press, 1993, MR 1232192.
\newblock With the assistance of Timothy S. Murphy, Monographs in Harmonic
  Analysis, III.

\bibitem{Stuart1992}
D.~M.~A. Stuart.
\newblock Perturbation theory for kinks.
\newblock {\em Comm. Math. Phys.}, 149(3):433--462, 1992,
  doi:10.1007/BF02096938.

\bibitem{Stuart2001}
D.~M.~A. Stuart.
\newblock Modulational approach to stability of non-topological solitons in
  semilinear wave equations.
\newblock {\em J. Math. Pure. Appl.}, 80(1):51--83, 2001, doi:
  10.1016/S0021-7824(00)01189-2.

\bibitem{Stuart2004b}
D.~M.~A. Stuart.
\newblock Geodesics and the {E}instein nonlinear wave system.
\newblock {\em J. Math. Pure. Appl.}, 83(5):541--587, 2004,
  doi:10.1016/j.matpur.2003.09.009.

\bibitem{Thirring1983}
W.~Thirring.
\newblock Bosonic black holes.
\newblock {\em Phys. Lett. B}, 127(1--2):27--29, 1983, doi:
  10.1016/0370-2693(83)91623-4.
  
\bibitem{Weder1974}
R.~Weder.
\newblock Spectral properties of one-body relativistic spin-zero
              {H}amiltonians.
\newblock {\em Ann. Inst. H. Poincar\'e Sect. A (N.S.)}, 20:211--220, 1974.
  
\bibitem{Weinstein1985}
M.~I. Weinstein.
\newblock Modulational stability of ground states of nonlinear
  {S}chr\"{o}dinger equations.
\newblock {\em SIAM J. Math. Anal.}, 16(3):472--491, 1985, doi:10.1137/0516034.

\bibitem{Zeidler109}
E.~Zeidler.
\newblock {\em Applied Functional Analysis, Main Principles and Their
  Applications}, volume 109 of {\em Applied Mathematical Sciences}.
\newblock Springer-Verlag, New York, 1995, MR1347692.

\end{thebibliography}
\end{document}